\documentclass[11pt,a4paper]{article}
\usepackage{jcappub}
\pdfoutput=1
\usepackage{amsmath}
\usepackage{mathrsfs}
\usepackage{enumerate}
\usepackage{amssymb,mathabx}
\usepackage{graphicx}
\usepackage{multirow}
\usepackage{xcolor}
\usepackage{bm}
\usepackage[normalem]{ulem}
\usepackage{bracket}
\usepackage{booktabs}

\def\m@th{\mathsurround=0pt }
\def\eqalign#1{\null\,\vcenter{\openup1\jot \m@th
 \ialign{\strut\hfil$\displaystyle{##}$&$\displaystyle{{}##}$\hfil
 \crcr#1\crcr}}\,}

\begin{document}

\title{Supergravity, $\alpha$-attractors and primordial non-Gaussianity}

\author[1,2,3]{Nicola Bartolo,}
\author[1]{Domenico Matteo Bianco,}
\author[4,5]{Raul Jimenez,}
\author[1,2,3,6]{Sabino Matarrese,}
\author[4,5]{Licia Verde}

\affiliation[1]{Dipartimento di Fisica e Astronomia ``G. Galilei'', Universit\`a degli Studi di Padova, via Marzolo 8, I-35131, Padova, Italy}
\affiliation[2]{INFN, Sezione di Padova, via Marzolo 8, I-35131, Padova, Italy}
\affiliation[3]{INAF-Osservatorio Astronomico di Padova, Vicolo dell'Osservatorio 5, I-35122 Padova, Italy}
\affiliation[4]{ICCUB, University of Barcelona, Marti i Franques 1, Barcelona, 08028, Spain.}
\affiliation[5]{ICREA, Pg. Lluis Companys 23, Barcelona, 08010, Spain.} 
\affiliation[6]{Gran Sasso Science Institute, Viale F. Crispi 7, I-67100 L'Aquila, Italy}

\emailAdd{nicola.bartolo@pd.infn.it; raul.jimenez@icc.ub.edu; sabino.matarrese@pd.infn.it; liciaverde@icc.ub.edu}

\abstract{We compute in detail how deviations from Einstein gravity at the inflation energy scale could appear as non-Gaussian features in the sky. To illustrate this we use multi-field $\alpha-$attractor models in the framework of supergravity to realise inflation. We find no obvious obstacle for having choices of model's parameters that generate non-Gaussian features of the equilateral and local type at the $\cal O$(1) level in the $f_{\rm NL}$ 
non-Gaussianity parameter, thus being potentially detectable in future cosmological surveys. This non-Gaussianity has its origin in either the non-canonical kinetic term (which, in turn, is an immediate consequence of assuming an hyperbolic geometry of the moduli space), the interactions of the fields in the potential or the $\alpha-$parameter, or a combination of these three. This opens up the exciting possibility of constraining the law of gravity at energy scales close to the Planck one.}
 
\maketitle

\section{Introduction}

It was shown in Ref.~\cite{MG} that modifications of Einstein gravity at the energy scale of inflation could seed observable non-Gaussian (NG) features in the density field of matter. The argument is simple: any modification to gravity can be seen as an extra field that  interacts with the inflaton. If the interaction is non-negligible, then non-Gaussian features may be generated during the slow-roll phase. While in most  inflationary models Einstein gravity is assumed to be  the correct description of gravity,  it is reasonable to consider  that the assumption may break down at very high energies:  for example there could be  a true modification of General Relativity (GR) or quantum effects  may become relevant. Indeed, in the leading theory to unify the four forces of nature, string theory~\cite{strings}, the law of gravity is modified at high energies. It is well known~\cite{Gangui,Acquaviva:2002ud,Maldacena:2002vr} that in the simplest, standard inflationary models, without significant modifications of GR, the primordial non-Gaussianity produced is small and only measurable with long-term future surveys \cite{Pillepich:2006fj,Cooray:2006km,Kamion}, and potentially with CMB spectral distortions~\cite{Pajer:2012vz,Ganc:2012ae,Emami:2015xqa,Bartolo:2015fqz}. 
Departures from Einstein gravity during inflation have been considered before for specific models (see e.g. \cite{Berkin:1991nm,Mollerach:1991qx,Minimal,MinimalII,Maldacena:2011nz,Soda:2011am,Kallosh:2013wya,Kallosh:2013xya,Kallosh:2013hoa}), starting indeed from the first inflationary model, the Starobinsky model~\cite{Starobinsky:1980te}, but the non-Gaussianity produced in these cases is also generally too small to be measurable i.e., of the order of the slow-roll parameters~\footnote{However see,  e.g.,~\cite{Choudhury:2014uxa,Ellis:2014opa,Kawai:2014gqa,Arkani-Hamed:2015bza,Baumann:2015xxa,Hetz:2016ics,Lee:2016vti,Bartolo:2017szm}, for other related works on the subject covering various aspects of high-energy scale inflation.}. 
In Ref.~\cite{MG} we considered general inflation models arising from the most general Lagrangian containing all generally covariant terms up to two derivatives built with the metric and one scalar field --assumed to be the inflaton. It was found that under fairly general conditions, modifications of Einstein gravity in this context lead to a quasi-local non-Gaussianity, with amplitude, parameterised by the $f_{\rm NL}$ parameter,  with values $\sim  O(-1)$. This level of non-Gaussianity, and the fact that it is so similar to the local type, makes the signature potentially measurable in upcoming galaxy surveys, see e.g.,  \cite{Carbone, Karagiannis,Castorina:2018zfk} and refs therein.

This opens up an exciting window to peer into the physics of  the highest energies at which gravity maybe modified. Therefore here we present a more detailed and quantitative study of the conditions for inflation and how modifications of gravity can be measured or characterised from their observational signatures. 

We  focus on  exact computations of the bispectrum generated by the interacting fields and abandon approximations as much as possible. In this vein, we want to make robust predictions for models that are expected to modify Einstein General Relativity at high energies that can be probed at the inflationary epoch. We focus on supergravity~\cite{supergravity1,supergravity2,supergravityreviews} models of inflation. They emerge in a low-energy limit of string theory which remains a most promising candidate to unify the laws of nature. String theory introduces modifications of Einstein gravity and brings along additional fields. We will use for our computations $\alpha-$attractor models~\cite{Ferrara:2013rsa,Kallosh:2013yoa,Kallosh:2015lwa,Roest:2015qya,DiMarco:2017sqo,DiMarco:2017zek,Kallosh:2017wku} (see also~\cite{Kallosh:2013hoa,Kallosh:2013daa} and~\cite{Kallosh:2013wya,Kallosh:2013xya}) which are a wide class of solutions of supergravity models. In this context there are two aspects which are relevant for the main results of this paper. The first is a non-trivial geometry in field space, which is typical of supergravity models and brings along non-trivial kinetic terms (see, e.g.,~\cite{Kallosh:2015zsa}). For $\alpha-$attractor models the field metric acquires a specific form, see Eq.~(\ref{eq:two-alpha-model-2}) and ~(\ref{eq:metric-f-matrix}), which, as we shall see, can play a crucial role as far as primordial non-Gaussianity is concerned. The second aspect is that this very same form can essentially arise from a construction of $\alpha-$attractor models that has its origin in models of gravity where GR is modified, see, e.g., Eq.~(\ref{eq:lagrangian-two-1}). Indeed these types of scalar dynamics afford an alternative interpretation via modifications of the Einstein term, in supergravity extensions of the Starobinsky construction~\cite{Cecotti:1987sa}.

Non-Gaussianity is an invaluable tool to learn about the physics of inflation, as well as the late-time universe (e.g. Ref.~\cite{Gangui,Acquaviva:2002ud,Maldacena:2002vr,MF,Komatsu:2009kd}).  While the local type of  primordial non-Gaussianity produces a prominent and promising signature on the large-scale structure power spectrum~\cite{Dalal,HaloBiasI,HaloBiasII}, the bispectrum is the statistics that carries the full (shape) information, hence is most suitable to detect the specific signature of the effect. For any practical application, the measured, late-time bispectrum contains a combination of primordial, gravitational, clustering and tracers signal. Separating these effects is an active area of research that will not be touched upon here. Nevertheless, it is broadly agreed in the community that a local primordial non-Gaussianity of $|f_{\rm NL}| \sim1$ is detectable from forthcoming surveys (e.g. \cite{Carbone,Karagiannis,Castorina:2018zfk} and references therein) and $|f_{\rm NL}| \sim 0.01$ will be detected in ultimate surveys \cite{Kamion}; for this reason, and also accounting for the present constraints on primordial non-Gaussianity from the {\it Planck} data~\cite{Ade:2015ava}, we pay particular attention to models that can in principle produce (local) non-Gaussian signatures at or above $|f_{\rm NL}| \sim1$. 

This paper is organized as follows: In \S \ref{sec:2} we first motivate the so-called $\alpha-$attractor models in supergravity and review them in an appendix. We then take a general (two-field) Lagrangian for these models and compute the conditions under which inflation takes place. Having found the general conditions for inflation, we compute the bispectrum generated by the interaction of the fields (\S \ref{sec:2}) and clarify under which conditions it leads to measurable non-Gaussian signatures. The impatient reader will find our main results in Eq.~\eqref{eq:chen-term-enhancement-alpha}, together with Eq.(\ref{fNL3}), the discussion in~\ref{sec:NSF-generalizations} and Table~\ref{tab:bispectrum-contributions}.
We conclude in \S \ref{sec:conclusions}. In Appendix~\ref{AppA} we recall some basics of single-field models of $\alpha$-attractors to introduce the typical multi-field models studied in this paper. In  Appendix~\ref{sec:setup-of-in-in}  we write down the building-blocks for the computation of the primordial non-Gaussianity in the models investigated. In Appendix~\ref{sec:non-canonical-kinetic-slow-roll} we give some details about slow-roll parameters in the case of multifield models of inflation with a non-canonical kinetic term. 

 \section{Inflation models and modifications of gravity}
 \label{sec:2}
We wish to study in a general framework how modifications of Einstein gravity can be measured at the scale of inflation, which is the highest energy probe we currently have in the Universe (Inflation is supposed to occur at energies as high as $\sim 10^{15-16} \, \text{GeV}$). 

Supergravity, as a low-energy limit of string theory, underlies the corresponding attempts to unify the four fundamental forces
of nature at high energies based on underlying symmetries. We will use a general framework
to describe inflationary models that has emerged from conformal symmetries embedded in theories of 
supergravity.

The $\alpha$-attractor models~\cite{Ferrara:2013rsa,Kallosh:2013yoa,Kallosh:2015lwa,Roest:2015qya,DiMarco:2017sqo,DiMarco:2017zek,Kallosh:2017wku} arise in the context of supergravity~\cite{Scalisi1, Yamaguchi:2011kg,Ferrara:2013rsa,Ellis:2013nxa}, along with the attempt to build an inflationary mechanism (e.g., \cite{Carrasco, Antoniadis:2014oya,Ferrara:2014kva, Dall'Agata:2014oka}).  Important simplifications are obtained, in this context, when supersymmetry is non-linearly realized (for a review see~\cite{Ferrara:2015cwa}), as it occurs in String Theory, in ``brane supersymmetry breaking" ~\cite{S,ADS,A,Mourad:2017rrl}.

The Planck~\cite{Ade:2015lrj} and WMAP~\cite{0067-0049-208-2-20} results   are consistent with standard, single-field models of inflation, in particular, the original Starobinsky model~\cite{Starobinsky:1980te} is  among those that are allowed by current constraints. Several different models of inflation, despite their different origin, make very similar cosmological predictions and provide a good fit to observations. This, apparently surprising, coincidence led to the discovery of  ``cosmological attractors" models. The first class of inflationary attractor models discovered was the conformal attractors~\cite{Kallosh:2013hoa,Kallosh:2013daa}, which were soon generalized to $\alpha$-attractors~\cite{Kallosh:2013yoa,Ferrara:2013rsa}. Another class of models, dubbed $\xi$-attractors, describes inflation through a non-minimal coupling to gravity~\cite{Kallosh:2013tua,Giudice:2014toa,Pallis:2013yda}. In addition to those, there are models with two attractor points~\cite{Kallosh:2014rga,Kallosh:2014laa,Mosk:2014cba} that can cover large regions of the tensor-to-scalar ratio vs primordial tilt parameter space ($r-n_s$)  (in particular they span the currently allowed region).~\footnote{However note the recent work~\cite{Dias:2018pgj} that elaborates on some potential issues on realising concrete $\alpha$-attractor models of inflation.}

In what follows, we review the ideas on which $\alpha$-attractors are based, focusing from the start on the the extension of these models to the multi-field case which will be especially important to understand how non-Gaussian features are directly related to modifications of gravity. The action we start from, Eq.~(\ref{eq:two-alpha-model-2}), derives indeed from a model of gravity, Eq.~(\ref{eq:lagrangian-two-1}) with modifications reflected by the presence of scalar fields.  We recall the construction that leads to Eq.~(\ref{eq:two-alpha-model-2}) in Appendix A. There we provide a summary of some details about inflationary alpha-attractors, starting from the action in the Jordan frame, introducing the concept of spontaneously broken conformal symmetry and recalling  why and to what extent the predictions of these models can be considered to be universal. In particular we recall the construction.   

 Our numerical simulations of the background dynamics of the inflaton are based on the \textit{Mathematica} code developed by Ref.~\cite{Dias:2015rca}, which we have suitably modified. 
 \subsection{Specific cases}

We will start with a toy model from Ref.~\cite{Kallosh:2013hoa} and generalise it to two fields. We will show later that this model is the basis for a more general analysis.

Our starting point is the Lagrangian~\cite{Kallosh:2013daa}
\begin{equation}
\mathcal{L} = \sqrt{-g} \biggl[\frac{R}{2} - \frac{1}{2} \partial_\mu \varphi \partial^\mu \varphi - 3 \sinh^2 \frac{\varphi}{\sqrt{6}} \, \partial_\mu \theta \partial^\mu \theta - V \left( \tanh (\varphi / \sqrt{6}) \cos\theta , \tanh (\varphi / \sqrt{6}) \sin\theta \right) \biggl],
\label{eq:two-alpha-model-2}
\end{equation}
where $V$ is the potential term. Note that $\theta$ exhibits a non-canonical kinetic term. If we interpret the fields as classical variables, $\theta$ would be an angle and $\varphi$ would be related to the radial distance from the origin in polar coordinates.

We begin by studying in detail  two examples of  the two-field models described by this Lagrangian --Eq. (\ref{eq:two-alpha-model-2})--. 
In particular, we  compute the power spectrum and the bispectrum of curvature perturbations and show under what conditions observable non-Gaussianities can be generated. These specific cases will then be generalized later.

In order to be as general as possible within a Lagrangian of the type~\eqref{eq:two-alpha-model-2}, we are going to focus on
\begin{equation}
\mathcal{L} = a^3(t) \biggl[\frac{R}{2} - \frac{1}{2} \partial_\mu \varphi \partial^\mu \varphi - \frac{f(\varphi)}{2} \partial_\mu \theta \partial^\mu \theta - V(\varphi,\theta) \biggl],
\label{eq:tgh-lagrangian}
\end{equation}
where we have defined
\begin{equation}
f(\varphi) \equiv 6 \sinh^2 \frac{\varphi}{\sqrt{6}} \hspace{2pt} .
\label{eq:function-f}
\end{equation}
In this way, the following equations can be generalized without effort to a different non-canonical kinetic term. In addition to this, we have assumed that the background spacetime is Friedman-Robertson-Walker. 

In this case, using the general treatment of multi-field models of inflation~\cite{Lahiri:2005xj,Peterson:2010np,vandeBruck:2014ata} (shortly summarized in Appendix C) the field space metric (see Eq.~\ref{eq:S-multi}) is given by
\begin{equation}
G_{IJ} = 
 \begin{pmatrix}
  1 & 0  \\
  0 & f(\varphi)
 \end{pmatrix}
 \, , 
 \label{eq:metric-f-matrix}
\end{equation}
which yields the equations of motion for the fields
\begin{align}
& \ddot{\varphi} + 3 H \dot{\varphi} -\frac{1}{a^2} \nabla^2 \varphi + V_\varphi +  \frac{f^\prime}{2} \partial_\mu \theta \partial^\mu \theta  = 0 \hspace{1pt}, \label{eq:W-eq-motion-phi} \\
& \ddot{\theta} + 3 H \dot{\theta} - \frac{1}{a^2} \nabla^2 \theta  + \frac{1}{f} V_\theta  - \frac{f^\prime}{f} \partial_\mu \varphi \partial^\mu \theta= 0 \hspace{1pt}, 
\label{eq:W-eq-motion-theta}
\end{align}
where
\begin{equation}
V_{\chi} \equiv \frac{\partial V}{\partial \chi} \hspace{2pt}  .
\end{equation} 
and $\chi$ denotes the field either $\theta$ or $\phi$.
The last terms on the LHS in both equations come from the non-canonical kinetic term for $\theta$. Notice that both the non-canonical kinetic term and the potential are responsible for interactions between the fields. The previous equations are decoupled if, and only if, the potential has the form $V = V(\varphi) + U(\theta)$ and $f \equiv \text{const}$. Hereafter, we will denote the derivatives of the potential with respect to the fields without the comma, for convenience.

As usual, we split the fields into a background term and a perturbation as
\begin{subequations}
\begin{align}
\varphi (t,\textbf{x}) &= \varphi_0(t) + \delta \varphi (t,\textbf{x}) \hspace{1pt}, \\
\theta (t,\textbf{x}) &= \theta_0(t) + \delta \theta (t,\textbf{x}) \hspace{1pt} .
\end{align}
\end{subequations}
The background equations of motion are
\begin{align}
& \ddot{\varphi}_0 + 3 H \dot{\varphi}_0  - \frac{f^\prime}{2} \dot{\theta}^2_0  + V_\varphi = 0 \hspace{1pt}, \label{eq:W-eq-background-phi} \\
& \ddot{\theta}_0 + 3 H \dot{\theta}_0 + \frac{f^\prime}{f} \dot{\varphi}_0 \dot{\theta}_0 + \frac{1}{f} V_\theta = 0 \hspace{1pt}.
\label{eq:W-eq-background-theta}
\end{align}
The non-trivial field space metric introduces a new potential term $- \frac{f^\prime}{2} \dot{\theta}^2_0 $ dependent on the angular velocity $\dot{\theta}_0$ into the first equation. Notice that this term is always negative and hence it forces the field $\varphi$ to accelerate. Conversely, in the second one, the non-trivial field-space metric introduces the term $\frac{f^\prime}{f} \dot{\varphi}_0 \dot{\theta}_0$ dependent on $\dot{\varphi}_0$. The sign of this term depends on the sign of the velocity $\dot{\varphi}_0$ and hence it can accelerate or decelerate $\theta$. 

The first Friedmann equation reads
\begin{equation}
H^2 = \frac{1}{3} \Bigl( \frac{1}{2} \dot{\varphi}_0^2 + \frac{f}{2} \dot{\theta}_0^2 + V \Bigl) ,
\end{equation}
with
\begin{equation}
\begin{split}
\rho = \frac{1}{2} \dot{\varphi}_0^2 + \frac{f}{2} \dot{\theta}_0^2 + V \hspace{1pt}, \\
P = \frac{1}{2} \dot{\varphi}_0^2 + \frac{f}{2} \dot{\theta}_0^2 - V \hspace{1pt}.
\end{split}
\end{equation}
In what follows, we will adopt the shortcut notation where we will drop the subscript ``0" that characterises the background fields. The complete field  will appear only in equations~\eqref{eq:in-in-pi-momenta-W} and~\eqref{eq:in-in-hamiltonian-W}.

The equations for the fluctuations, up to second order, read
\begin{align}
&\ddot{\delta\varphi} + 3 H \dot{\delta\varphi} - \frac{1}{a^2} \nabla^2 \delta\varphi - \frac{\dot{\theta}^2}{2} \Bigl( f^{\prime\prime} \delta \varphi + \frac{1}{2} f^{\prime\prime\prime} \delta\varphi^2 \Bigl) - f^\prime \Bigl( \dot{\theta} \dot{\delta\theta} - \frac{1}{2} \partial_\mu \delta\theta \partial^\mu \delta\theta \Bigl) - f^{\prime\prime} \dot{\theta} \delta\varphi \dot{\delta\theta}  \nonumber \\
&+ V_{\varphi\varphi} \delta\varphi + V_{\varphi\theta} \delta\theta + \frac{1}{2} V_{\varphi\varphi\varphi} \delta\varphi^2 + V_{\varphi\varphi\theta} \delta\varphi\delta\theta + \frac{1}{2} V_{\varphi\theta\theta} \delta\theta^2 = 0 \hspace{1pt}, \label{eq:W-perturbations-phi}\\
&\ddot{\delta\theta} + 3 H \dot{\delta\theta}- \frac{1}{a^2} \nabla^2 \delta\theta + \frac{f^\prime}{f} \Bigl( \dot{\varphi} \dot{\delta\theta} + \dot{\theta}\dot{\delta\varphi} \Bigl) - \frac{f^\prime}{f} \partial_\mu \delta\varphi \partial^\mu \delta\theta + \biggl( \frac{f^\prime}{f} \biggl)^{\hspace{-2pt} \prime} \Bigl(\dot{\varphi} \dot{\theta} + \dot{\theta} \dot{\delta\varphi} + \dot{\varphi} \dot{\delta\theta}   \Bigl)  \delta\varphi  \nonumber\\
&+ \biggl( \frac{f^\prime}{f} \biggl)^{\hspace{-2pt} \prime\prime} \dot{\varphi} \dot{\theta} \delta\varphi^2 + \frac{1}{f} \biggl( V_{\theta\theta} \delta\theta +V_{\theta\varphi} \delta\varphi + \frac{1}{2} V_{\theta\theta\theta} \delta\theta^2 + V_{\theta\theta\varphi} \delta\varphi \delta\theta + \frac{1}{2} V_{\theta\varphi\varphi} \delta\varphi^2 \biggl)  \nonumber \\
&+ \biggl( \frac{1}{f} \biggl)^{\hspace{-2pt} \prime} \Bigl( V_\theta + V_{\theta\theta} \delta\theta + V_{\theta\varphi} \delta\varphi \Bigl) \delta\varphi + \frac{1}{2} \biggl( \frac{1}{f} \biggl)^{\hspace{-2pt} \prime \prime} V_\theta \delta\varphi^2 = 0 \hspace{1pt} . \label{eq:W-perturbations-theta}
\end{align}
Notice that all the terms with derivatives of $f$ would vanish if the kinetic term of $\theta$ was canonical.

\subsection{Viable initial conditions for inflation}
\label{sec:W-model-initials}

Acceptable initial conditions for the fields should lead to an inflationary period lasting at least $50-70$ $e$-folds (for a review of the problem of initial conditions for inflation, including, $\alpha$-attractor models, see~\cite{Linde:2017pwt}). Recall that the number of e-folds $N$ is defined by $dN =  H dt$. Initial conditions are given at $N=0$ and inflation may start some time later. To study the inflationary dynamics, it is useful to rewrite the background equations~\eqref{eq:W-eq-background-phi} and~\eqref{eq:W-eq-background-theta} in terms of the $e$-folds number $N$.

Let us denote $\varphi'=d\varphi/dN$ i.e. prime for the fields denote derivative with respect to $N$, but $f^\prime$ stands for $df / d\varphi$.
Defining 
\begin{align}
\omega_\theta &\equiv - \frac{f^\prime}{2} \theta^{\prime 2}  \label{eq:omega-theta} , \\
\omega_\varphi &\equiv  \frac{f^\prime}{f} \theta^\prime \varphi^\prime \label{eq:omega-phi} ,
\end{align}
 we can write
\begin{eqnarray}
&&\varphi^{\prime\prime} + \omega_\theta + (3 - \varepsilon ) \left( \varphi^\prime + \frac{V_\varphi}{V} \right) = 0 \hspace{1pt}, \label{eq:phi-background-W-1} \\
&&\theta^{\prime\prime} + \omega_\varphi + (3- \varepsilon) \left( \theta^\prime + \frac{1}{f}  \frac{V_\theta}{V} \right) = 0 \hspace{1pt} , \label{eq:theta-background-W-1} \\
&&H^2 = \frac{V}{3 - \varepsilon} \label{eq:H-background-z} \hspace{2pt} , \\
&&\varepsilon \equiv - \frac{\dot{H}}{H^2}  = \frac{1}{2} {\varphi^\prime}^2 + \frac{f}{2} {\theta^\prime}^2 \, . \label{eqsback4}
\end{eqnarray}

The background equations of motion in the slow-roll and slow-turn approximation reduce to 
\begin{subequations}
\begin{align}
&\varphi^\prime +  \frac{V_\varphi}{V}  = 0 \hspace{1pt}, \label{eq:SRST-W-phi}\\
&\theta^\prime + \frac{1}{f} \frac{V_\theta}{V}  = 0 \hspace{1pt}, \label{eq:SRST-W-theta}\\
&H^2 = \frac{V}{3}  \hspace{2pt}. \label{eq:SRST-W-H}  
\end{align}
\label{eq:SRST-W}
\end{subequations}
Notice that in the second equation the potential for $\theta$ is suppressed by the function $f$ that goes as $\exp (\sqrt{2/3} \varphi)$. If inflation occurs in the region $\varphi \gg 1$, the velocity 
$\theta^\prime$ should be small during slow-roll, unless the potential term compensates this exponential suppression. Later, as $\varphi$ moves towards the minimum of the potential, 
the suppression is reduced and the velocity $\theta^\prime$ can increase. 
Let us recall here that inflation occurs for $\varphi \gg 1$ in the single-field case. On the other hand, in the multi-field case, this condition can be relaxed since the inflaton can move in more directions and then the aforementioned suppression can be ineffective. 

\begin{figure}
\centering
\includegraphics[width=.475\textwidth]{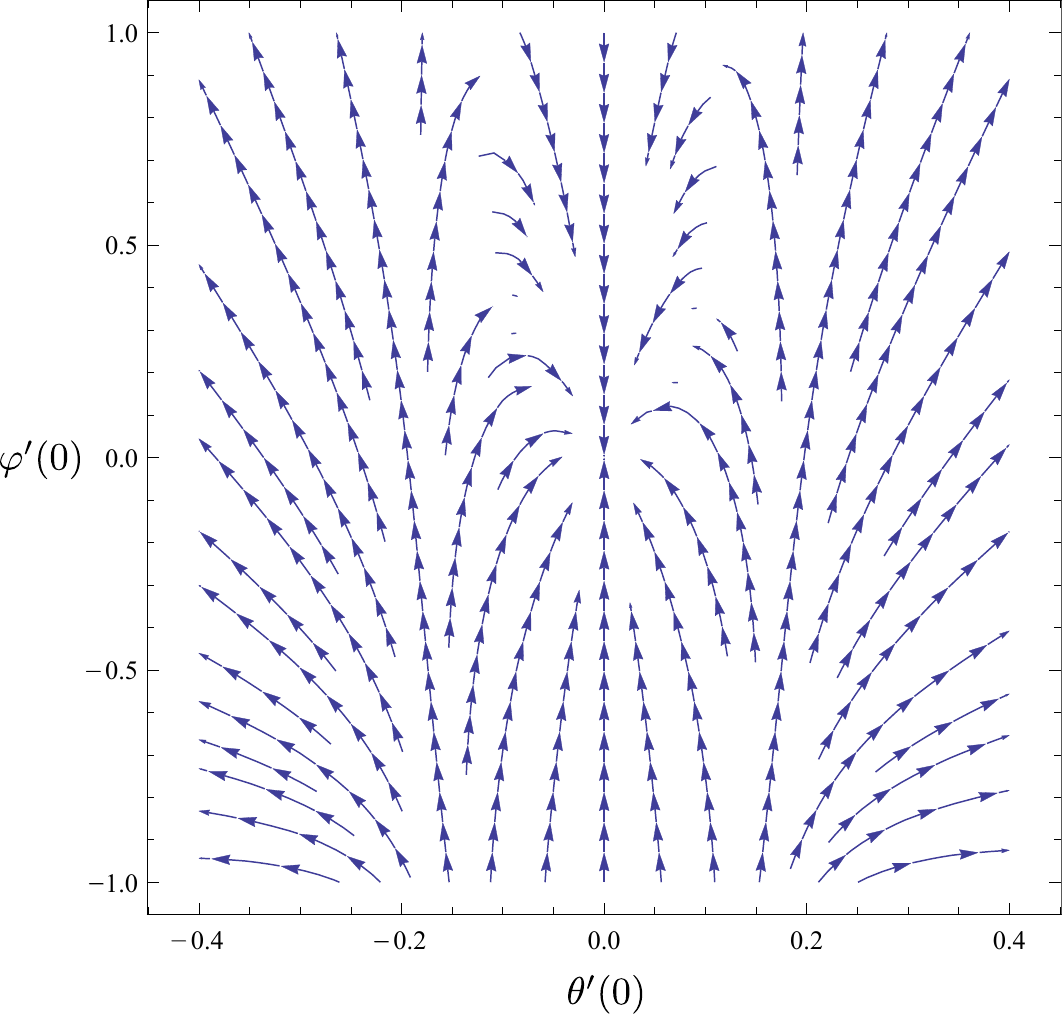}
\includegraphics[width=.475\textwidth]{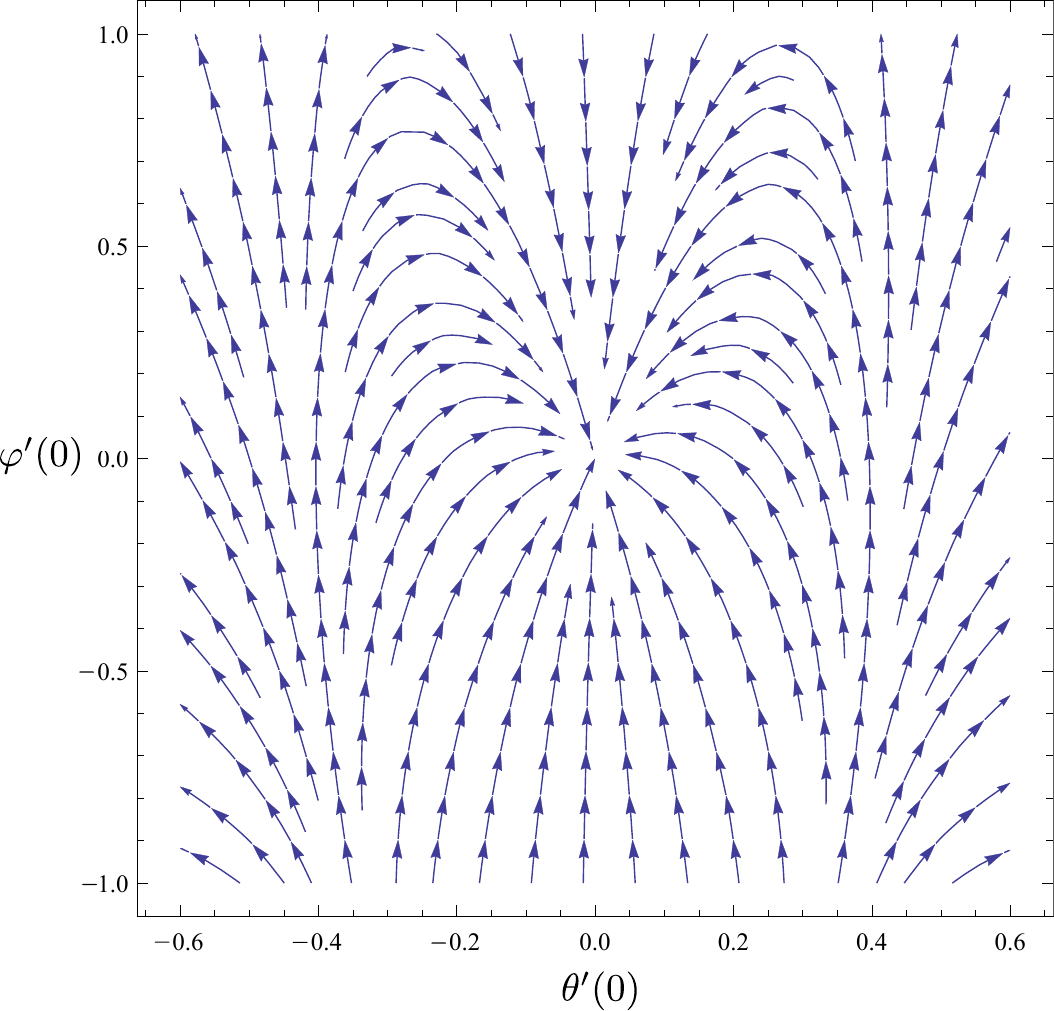}
\caption{Phase-space diagrams for the system described by Eq.~\eqref{eq:flux-no-W}. Points represent initial velocities $\bigl(\theta^\prime(0) , \varphi^\prime(0) \bigl)$ and vectors are 
accelerations $\bigl(\theta^{\prime\prime}(0) , \varphi^{\prime\prime}(0) \bigl)$. These phase-space diagrams are represented at a fixed value of $\varphi(0)$ and hence are reliable in a small neighbourhood of it. The left panel corresponds to $\phi(0)=6$ and $f(\phi(0))=200$, while in the right panel $\phi(0)=4$ and $f(\phi(0))=36$. Inflation only happens for given values of $|\theta^{\prime}| < 0.16$ for the left panel, while it can take place for higher values for the case on the right panel.}
\label{fig:flux-no-W}
\end{figure}

\begin{figure}
\centering
{\includegraphics[width=.475\textwidth]{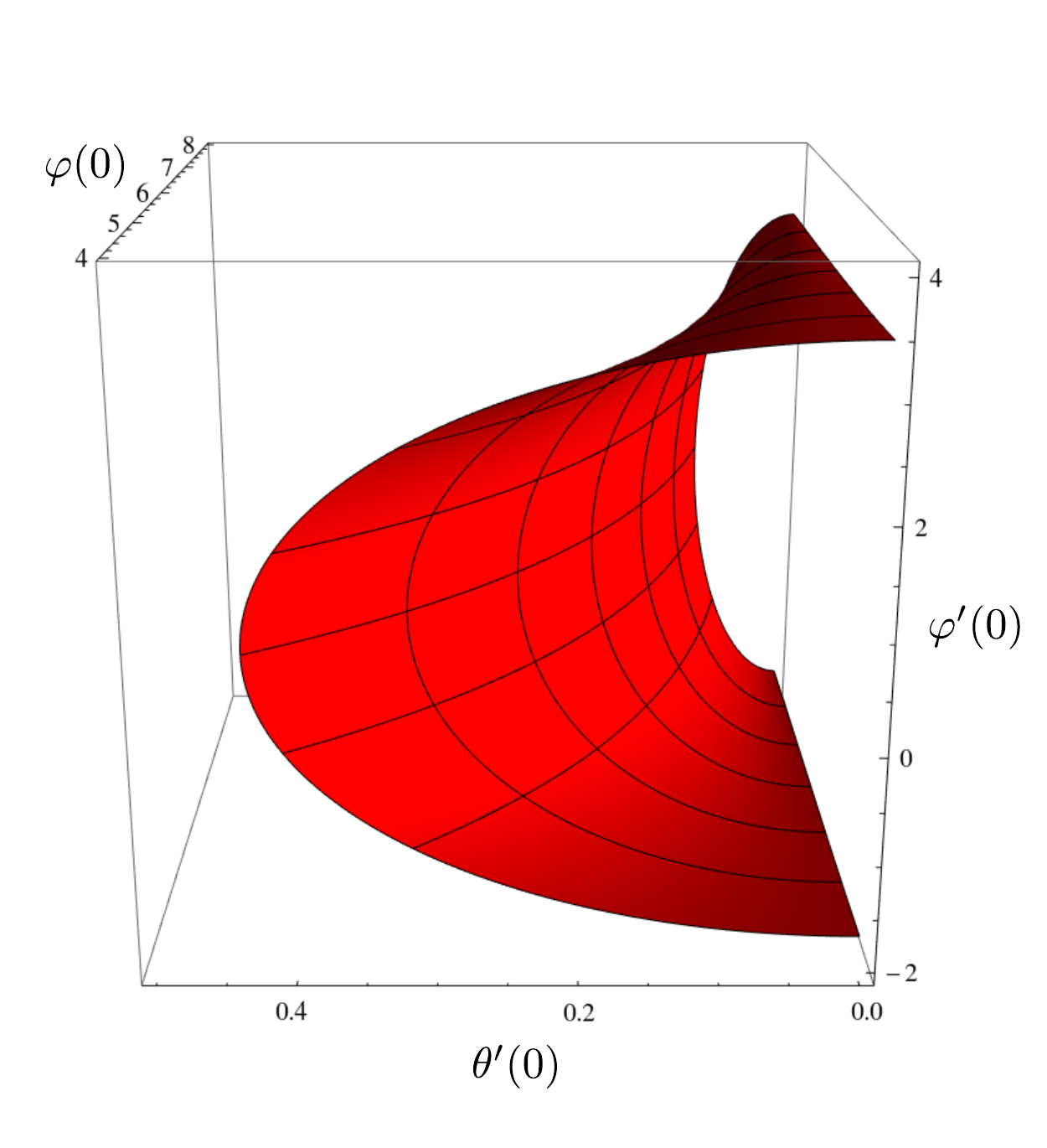}
\caption{Plot of the $J=0$ surface evaluated at $N=0$ (see Eq.~(\ref{eq:function-J})). Inflation occurs if initial conditions lie on the right of the red surface. Notice that $0.4 \gtrsim \theta^\prime(0) \gtrsim -0.4$ (the plot is symmetric in $\theta^\prime(0)$). In addition to this, notice that $4 \gtrsim \varphi^\prime(0) \gtrsim -2$. However, $\lvert \varphi^\prime \lvert < \sqrt{6} $ (due to Eq.~(\ref{eq:H-background-z})) and hence there are practically no restrictions on the initial velocity $\varphi^\prime(0)$ as opposed to $\theta^\prime(0)$.}
\label{fig:sign}}
\end{figure}

\begin{figure}
\centering
{\includegraphics[width=.475\textwidth]{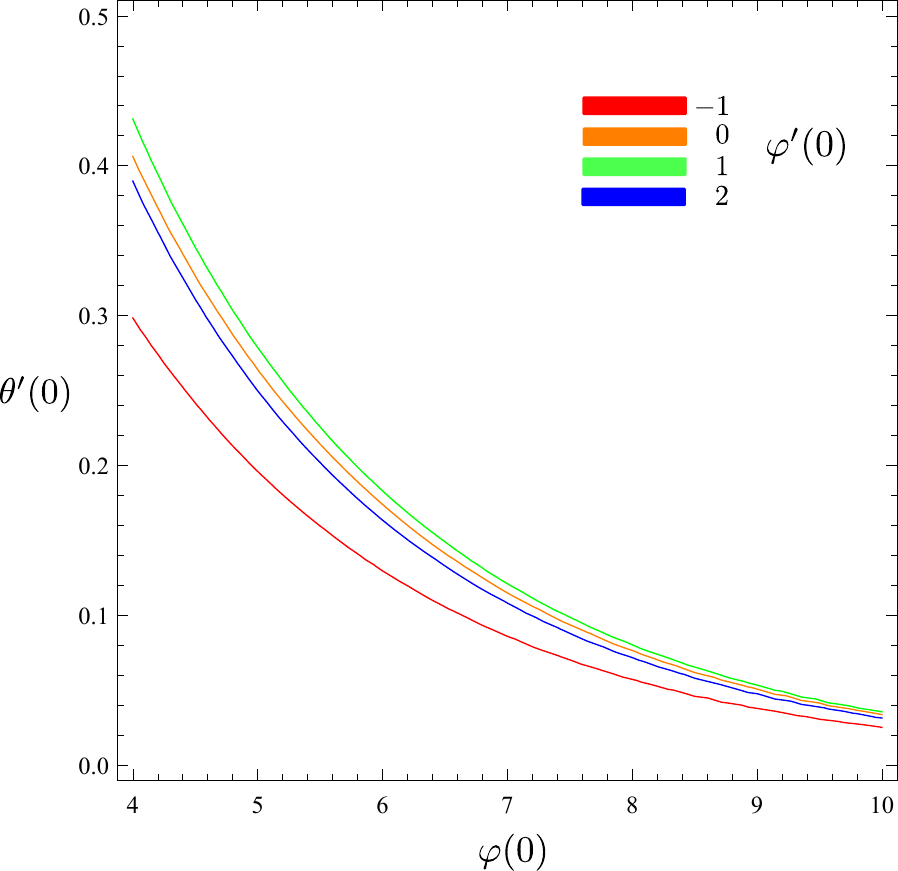}
\label{fig:sign-z}}
\caption{Level curves of the surface in Fig.~\ref{fig:sign} for some values of $\varphi^\prime(0)$. Inflation occurs in the region under the curves. Notice that for large $\varphi(0)$ all the curves converge and hence 
$\varphi^\prime(0)$ has no importance in the dynamics of the inflaton. On the other hand, for $\varphi(0) \gg 1$, $\theta^\prime (0)$ should be very small, otherwise the term 
$\omega_\theta$~\eqref{eq:omega-theta} becomes very large and dominates the dynamics of the system.}
\label{fig:two-init-2}
\end{figure}

Therefore, coming back to the general case of Eqs.~(\ref{eq:phi-background-W-1}-\ref{eqsback4}), let us suppose that $\varphi^\prime \gg V_\varphi / V$ and $\theta^\prime \gg V_\theta / (f V)$. In this limit, the background equations of motion~\eqref{eq:phi-background-W-1} and~\eqref{eq:theta-background-W-1} read
\begin{subequations}
\begin{align}
&\varphi^{\prime\prime} - \frac{f^\prime}{2} \theta^{\prime 2} + (3 - \varepsilon ) \varphi^\prime  = 0 \hspace{1pt}, \\
&\theta^{\prime\prime} + \frac{f^\prime}{f} \theta^\prime \varphi^\prime + (3- \varepsilon)  \theta^\prime  = 0 \hspace{1pt} . \label{eq:flux-no-W-theta}
\end{align}
\label{eq:flux-no-W}
\end{subequations}
Hence, the dynamics of the system is determined only by the non-canonical kinetic term. This approximation is useful when we are dealing with sufficiently high initial velocities (so that the potential terms can be neglected). However, notice that, during slow-roll, the terms related to the potential are not negligible, due to equations~\eqref{eq:SRST-W}.
The phase-space diagram of this system of differential equations is represented in figure~\ref{fig:flux-no-W}. The behaviour of the solutions is quite clear. If the velocities are sufficiently high that the terms $ V_\varphi / V$ and $ V_\theta / (f V)$ can be neglected, then inflation usually does not occur, because the fields accelerate and slow-roll cannot be reached. This acceleration is driven by the term $\omega_\theta$~\eqref{eq:omega-theta} which derives from the non-canonical kinetic term. In order to have inflation, the term $\omega_\theta$ should be sufficiently small and should remain so. Looking at Fig.\ref{fig:flux-no-W},  inflation can happen in regions in the phase space that are ``basin of attraction" for $\varphi'=0$, $\theta'=0$.
Note that, if $f$ is an even function (as the function~\eqref{eq:function-f}), equations~\eqref{eq:flux-no-W} are invariant under the transformations $\varphi \rightarrow - \varphi$ and $\theta \rightarrow - \theta$.

Now, consider equation~\eqref{eq:theta-background-W-1}. 
Let us choose the simplest case: $V_\theta = 0$.
In this case, the equation of motion for $\theta$ coincides with~\eqref{eq:flux-no-W-theta}, i.e. 
\begin{equation}
\theta^{\prime\prime} + J \theta^\prime  = 0 \hspace{1pt}  ,
\end{equation}
where we have defined
\begin{equation}
J \bigl( \varphi,\varphi^\prime,\theta^\prime \bigl) \equiv 3 - \frac{1}{2} {\varphi^\prime}^2  - \frac{f}{2} {\theta^\prime}^2  + \frac{f^\prime}{f} \varphi^\prime .
\label{eq:function-J}
\end{equation}
The sign of the function $J$ (see Fig.~\ref{fig:sign}) is fundamental for the behaviour of the system. If it is negative, $\theta$ accelerates and inflation will never occur, since also $\varphi$ accelerates (due to the term $\omega_\theta$~\eqref{eq:omega-theta} in~\eqref{eq:phi-background-W-1}) and the slow-roll regime cannot be reached. On the other hand, if this term is positive, it acts like a viscous term and $\theta^\prime$ is suppressed. 
Because of the previous considerations, the sign of the function $J$ at $N=0$ determines the initial dynamics of the system. If it is negative, inflation cannot start. However, notice that near $J=0$ some caution must be taken and only numerical analysis of the dynamics can distinguish whether inflation occurs. This function is represented in Fig.~\ref{fig:two-init-2}. During slow-roll the function $J$ is certainly positive and hence $\theta^\prime$ is suppressed (inflation approaches a nearly straight trajectory with small angular velocity $\theta^\prime$).   

Now, let us focus on equation~\eqref{eq:phi-background-W-1} in order to understand what are the effects of the presence of the non-canonical kinetic term of $\theta$. The additional term originating from the non-canonical kinetic term is $\omega_\theta$~\eqref{eq:omega-theta}. Notice that it is always a negative term ($f^\prime$ is positive as well as $\theta^{\prime 2} $, obviously) and hence it induces positive acceleration. Qualitatively, the dynamics of the field $\varphi$ is very simple. If the term $\omega_\theta$ is dominant, the field $\varphi$ accelerates and inflation might be no longer  possible since also the field $\theta$ accelerates for the previous considerations. The remaining term $(3 - \varepsilon ) \varphi^\prime $ is like a viscous force and always reduces the speed ($3- \varepsilon > 0$, due to equation~\eqref{eq:H-background-z}). 

\begin{figure}
\centering
\includegraphics[width=.475\textwidth]{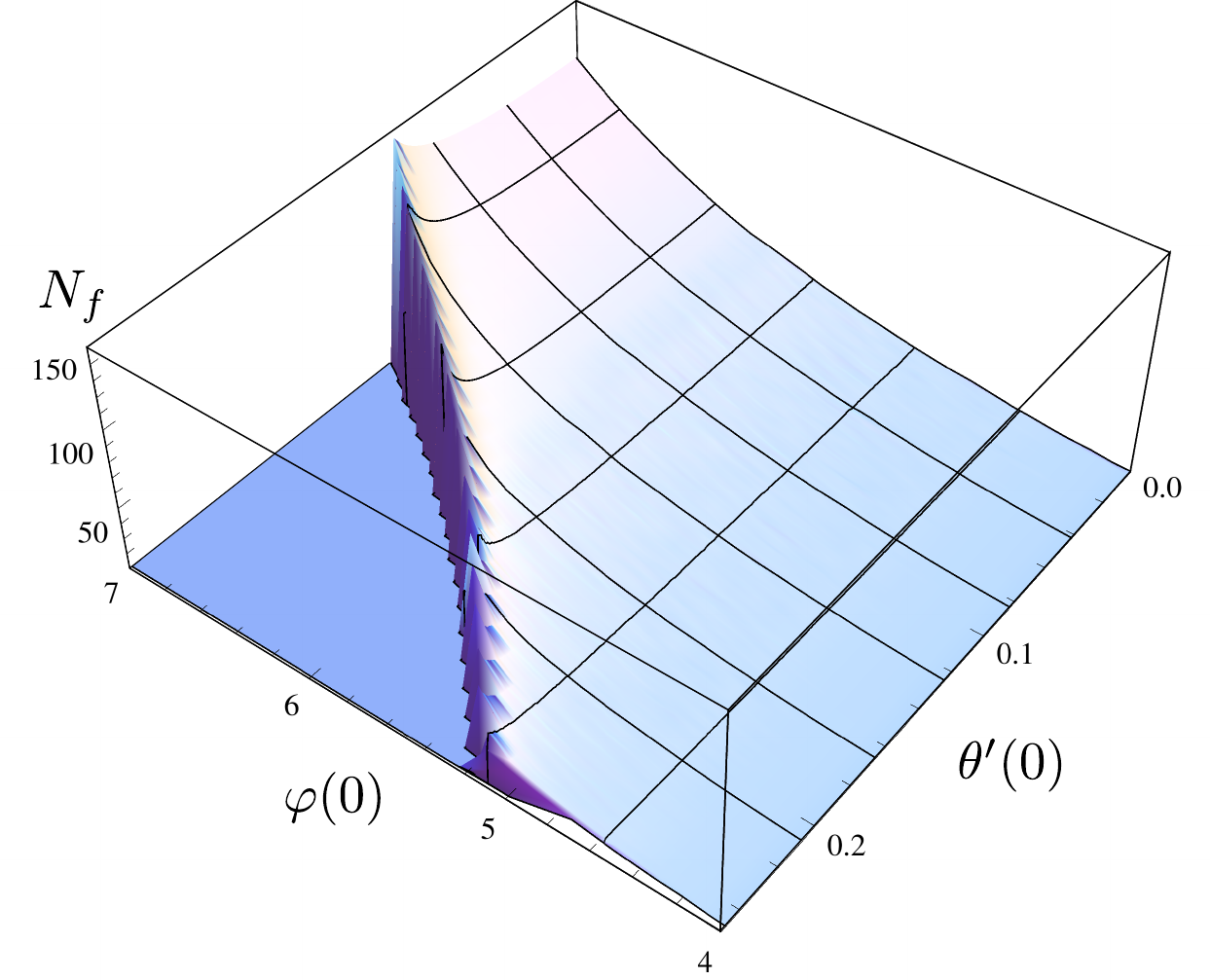}
\includegraphics[width=.475\textwidth]{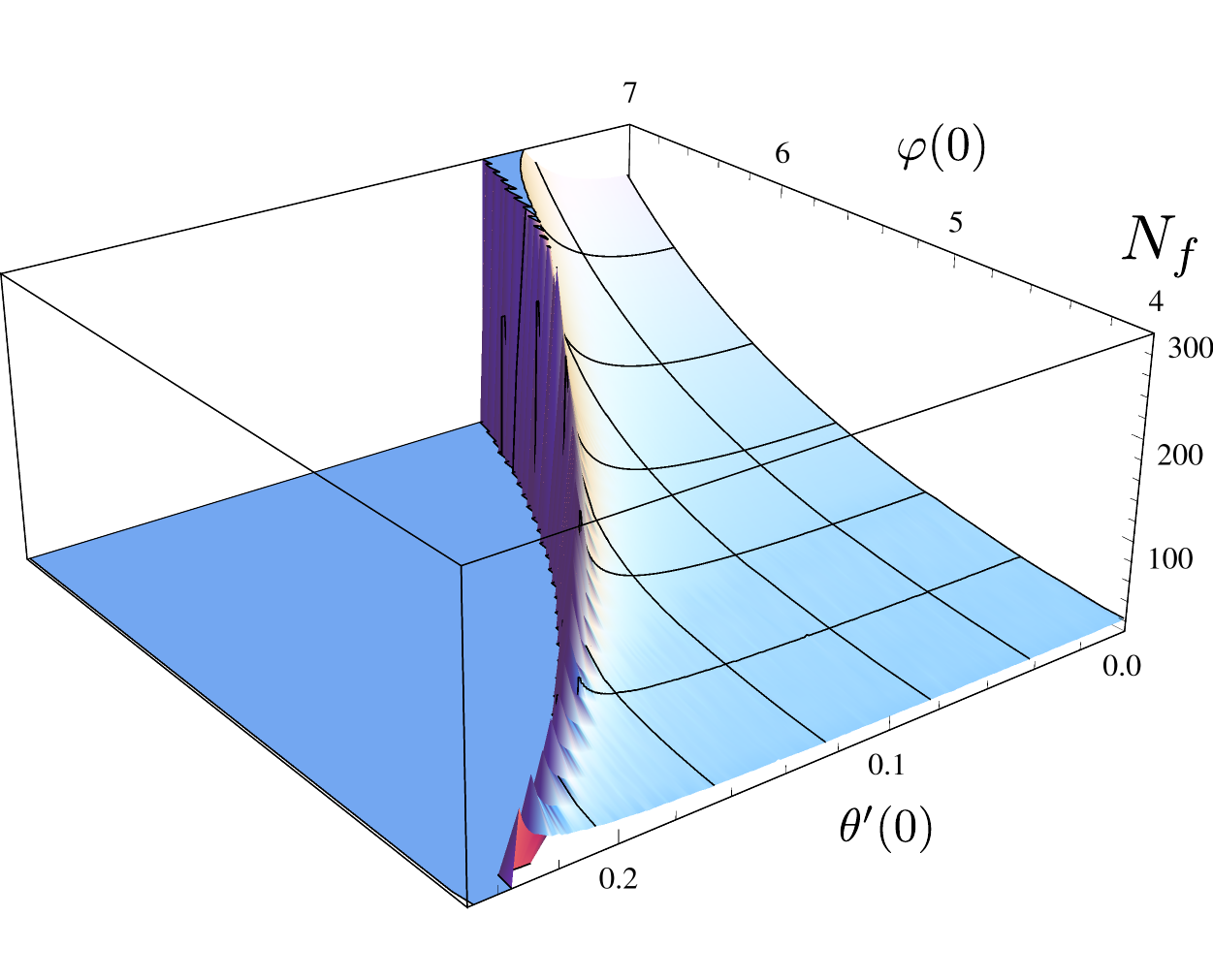}
\label{fig:two-init-20}
\caption{Viability of initial conditions for the model with potential $V(\varphi,\theta) = \tanh^2 \left( \varphi / \sqrt{6} \right)$. Left panel: $\phi^\prime(0) =0$, right panel: $\phi^\prime(0) =2$. The graphs have the same $x$ and $y$ scales, but different $z$ one. Let us focus on a fixed value of $\varphi(0)$. Increasing the initial velocity $\theta^\prime(0)$ increases the number of $e$-folds $N_f$ of inflation until they suddenly drop (inflation does not start). The same occurs increasing the initial position with $\theta^\prime(0)$ fixed (unless $\theta^\prime(0)\rightarrow 0$). In the region 
$\theta^\prime(0)\rightarrow 0$, the fields follow a nearly straight trajectory and the field $\theta$ can be practically ignored. The term $\omega_\theta$ in~\eqref{eq:phi-background-W-1} has the dominant effect on the inflationary region. Notice that increasing the initial velocity $\varphi^\prime$ reduces the region where inflation can occur; moreover, it increases the number of $e$-folds of inflation, keeping fixed the other initial conditions. Finally, note that the boundary of the inflationary region has the shape depicted in figure~\ref{fig:sign-z}.}
\label{fig:two-init}
\end{figure}

 For the simple case of the potential $V(\varphi,\theta) = \tanh^2 \left( \varphi / \sqrt{6} \right)$, numerical calculations were performed in order to study the acceptable initial conditions for inflation. The results are represented in Fig.~\ref{fig:two-init} and they can be interpreted thanks to the previous analysis. Notice that inflation occurs only if $\theta^\prime $ is sufficiently small from the beginning. However, the term $\omega_\theta$~\eqref{eq:omega-theta} could be of order unity at $N=0$, as it can be inferred from figure~\ref{fig:two-init}. Notice that during slow-roll $\omega_\theta$ is certainly small, since it is a slow-roll parameter ($\omega_\varphi$ is also a slow-roll parameter, see section~\ref{sec:non-canonical-kinetic-slow-roll}).

Clearly, introducing a non-trivial potential for $\theta$ opens a Pandora's box. This makes a general study impossible since it can produce a multitude of different and peculiar outcomes. 
If the term $V_\theta/(f V)$ is negligible, we recover one of the previously studied cases. On the other hand, viable initial conditions are mainly determined by the term $\omega_\theta$ in the background equation~\eqref{eq:phi-background-W-1} and by the particular form of the potential. In the next section, we will analyse in detail two relevant representative cases.

As we have seen, the background dynamics and the initial conditions are primarily affected by the term $\omega_\theta$ in~\eqref{eq:phi-background-W-1}. The function $f$~\eqref{eq:function-f} has a first derivative that is  positive for $\varphi > 0$. This constrains the possible initial conditions because the aforementioned term always increases the velocity $\varphi^\prime$ in the region $\varphi > 0$. On the other hand, nothing prevents us from considering a different non-canonical kinetic term with $f^\prime < 0$ for $\varphi > 0$. In this case, the considered term would suppress $\varphi^\prime$ helping the field to reach the attractor solution. However, $f > 0$ (because energy is always positive) and hence the term $\omega_\varphi$ in~\eqref{eq:theta-background-W-1} can accelerate the field $\theta$ breaking the inflationary dynamics. In this case a more detailed study is necessary for each particular  function $f$.

\subsection{Constant turn case}
\label{sec:CT}
\begin{figure}[!t]
  \centering
    \includegraphics[width=0.5\textwidth]{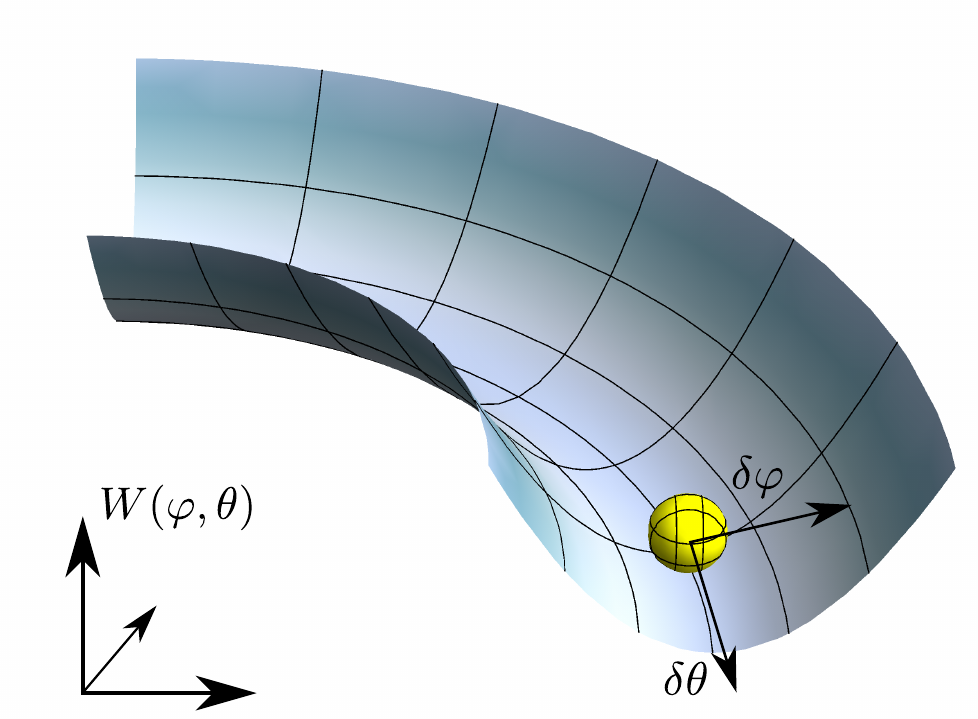}
      \caption{Example of constant-turn trajectory. The field (the yellow ball) moves along a circle, thanks to the particular form of the potential. The fluctuation $\delta\theta$ is related to the curvature perturbation of the inflaton field $\theta$ and $\delta\varphi$ to the isocurvature perturbation of the field $\varphi$. }
      \label{fig:chen-ball}
\end{figure}

Recall that in the previous model the non-canonical kinetic term and the coupling with the potential are the manifestation of  modifications of General Relativity. We wish to study how they generate primordial non-Gaussianities.
Our starting point is the two-field Lagrangian~\eqref{eq:two-alpha-model-2}.
Let us suppose that the background trajectory is a circle with fixed $\varphi$ and constant angular velocity $\dot{\theta}$. We also assume that the potential depends only on $\varphi$ (hence $\theta$ is massless). With all these assumptions, our model is analogous to Chen's quasi-single field~\cite{Chen:2009zp}, except for the presence in our model of a non-trivial coupling in the kinetic term, which is the manifestation of modifications of gravity. Inflation is driven by the field $\theta$ (inflaton), the second field $\varphi$ (isocurvaton) produces the non-Gaussianities that are eventually transferred to the former field~\cite{Bartolo:2001vw,Bartolo:2001cw}. In this approximation we have made the potential $V$  a function only of $\varphi$, neglecting the dependence on $\theta$ (in subsection~\eqref{sec:chen-generalizations} we will consider a potential $V$ that depends on both $\varphi$ and $\theta$). 
This seems too drastic at first sight, since $\theta$ should have a suitable potential in order to be the inflaton. However if we are only interested in evaluating the bispectrum, this approximation allows us to focus on the non-Gaussianities produced by the field $\varphi$ and transferred to $\theta$. Indeed, as we will see, the contributions to the bispectrum due to the dependence of the potential on $\theta$ should be subdominant due to the slow-roll conditions. Hence, for the moment we assume that a suitable potential for $\theta$ was chosen, but that its contribution to the bispectrum can be neglected. 
A pictorial representation of a constant turn trajectory is given in figure~\ref{fig:chen-ball}.

The couplings between the fields depend on the potential and on the non-canonical kinetic term. The most important characteristic of this model is that it can produce large 
non-Gaussianities~\cite{Bartolo:2001vw,Bartolo:2001cw,Chen:2009zp}, because the field $\varphi$ is not subject to the slow-roll conditions, that usually constrain the magnitude of the couplings. 

To perform such an analysis we need the Hamiltonian of the system which allows us to determine both the linear fluctuations of the fields and their interactions (see Appendix~\ref{sec:setup-of-in-in} for details). The  perturbed Hamiltonian  is expanded  in a background term and higher order terms in the field's perturbations.  One can neglect the linear terms in this calculation since they do not contribute to the equations for the fluctuations.  Finally the perturbed Hamiltonian is separated in  a kinetic part and an interaction part and the perturbations in the metric are neglected, since it is well known that the latter give a small contribution~\cite{Acquaviva:2002ud,Maldacena:2002vr}.
With the previous assumptions, the Hamiltonian, Eq.~\eqref{eq:H-0-H-INT}, becomes
\begin{subequations}
\begin{align}
\mathcal{H}_0 &=  \frac{a^3}{2} \dot{\delta\varphi}^2 +  \frac{a^3 f^2 \dot{\theta}^2}{4} \left(\frac{1}{f} \right)^{\hspace{-0.3em} \prime\prime} \delta\varphi^2 + \frac{ a^3 f}{2} \dot{\delta\theta}^2  + \frac{a}{2} \bigl( (\partial_i \delta\varphi)^2 + f (\partial_i \delta\theta)^2 \bigl) \nonumber  \\
&\phantom{=i}+ \frac{a^3}{2} V_{\varphi\varphi} \delta\varphi^2 , \label{eq:H-0-kinetic-CT} \\ 
\mathcal{H}^{(2)}_{INT} &=  -  a^3  f^\prime \dot{\theta}  \delta\varphi \dot{\delta\theta} \hspace{1pt} ,  \label{eq:H-2-INT-CT} \\ 
\mathcal{H}^{(3)}_{INT} &=  \frac{a^3 f^2 \dot{\theta}^2 }{12} \left(\frac{1}{f} \right)^{\hspace{-0.3em} \prime\prime\prime} \delta\varphi^3 + \frac{a^3 f^2 \dot{\theta} }{2} \left(\frac{1}{f} \right)^{\hspace{-0.3em} \prime\prime} \dot{\delta\theta} \delta\varphi^2 + \frac{a^3 f^2 }{2 } \left(\frac{1}{f} \right)^{\hspace{-0.3em} \prime} \dot{\delta\theta}^2 \delta\varphi +  \frac{a}{2} f^\prime \delta\varphi (\partial_i \delta\theta)^2 \nonumber \\
&\phantom{=i}+ \frac{a^3}{6}  V_{\varphi\varphi\varphi} \delta\varphi^3   , \label{eq:H-3-INT-CT} 
\end{align}
\label{eq:H-0-H-INT-CT}
\end{subequations}
where the term ${\cal H}^{(0)}$ describes the
kinetic part, ${\cal H}^{(2)}_{INT}$ and ${\cal H}^{(3)}_{INT}$ are respectively the second and third order interaction terms.
Note that, even if $\dot{\theta}=0$, there are interactions between the fields (but only at third order in perturbations) originating from the potential $V$ and the non-canonical kinetic term. In the following, we will use the same definitions of equations~\eqref{eq:H-0-H-INT} to characterise the different terms of the Hamiltonian. Therefore, in the following  $Z_i,A_i,B_i,C_i,D$ will indicate the various coupling constants as defined in Eq.~(\ref{eq:H-0-H-INT}). For example, $\mathcal{H}^{(3)}_{A_2} = a^3   V_{\varphi\varphi\varphi} \delta\varphi^3 /6 $.

Since $\theta$ is massless, $\tilde{\nu}=3/2$ (see equation~\eqref{eq:nu-theta}). Therefore, the mode functions~\eqref{eq:u-k} and~\eqref{eq:v-k}, respectively for the $\delta\varphi$ and $\delta\theta$ fluctuations, reduce to 
\begin{subequations}
\begin{align}
u_k (\tau)  &= \frac{\sqrt{\pi}}{2} e^{i(\nu + \frac{1}{2}) \frac{\pi}{2}} (-\tau)^{\frac{3}{2}} H \, H^{(1)}_\nu (-k\tau)  \hspace{1pt} , \\
v_k (\tau)  &= \frac{H}{\sqrt{2k^3}\sqrt{f}} (1+ik\tau) e^{-ik\tau} .
\end{align}
\label{eq:CT-mode-functions}
\end{subequations}
Finally, since the trajectory is a circle, the curvature perturbation $\zeta$ can be written~\cite{Chen:2009zp} as~\footnote{\label{f2} The general expression of the curvature perturbation for the Lagrangian~\eqref{eq:two-alpha-model-2} is given by ~\cite{vandeBruck:2014ata}
$\zeta = - \frac{H}{\dot{\sigma}} \delta\sigma$, where $\delta\sigma \equiv \cos\beta \, \delta\varphi + \sin \beta \,\sqrt{f}  \delta\theta$, 
$\dot{\sigma} \equiv \sqrt{\dot{\varphi}^2 + f \dot{\theta}^2}$, $\cos\beta = \frac{\dot{\varphi}}{\dot{\sigma}} $ and 
$\sin \beta = \frac{\sqrt{f} \dot{\theta}}{\dot{\sigma}}$. Since the trajectory is a circle with fixed value of $\varphi$, the velocity $\dot{\varphi}=0$ and then the adiabatic direction is parallel to $\delta\theta$. }

\begin{equation}
\zeta \simeq - \frac{H}{\dot{\theta}} \delta\theta \hspace{1pt} .
\end{equation}
In the following we will use the \textit{in-in} formalism~\cite{Weinberg:2006ac} to compute the correlators of our interest.

\subsubsection{Power-spectrum}
\begin{figure}[t!]
\centering
{\includegraphics[width=.475\textwidth]{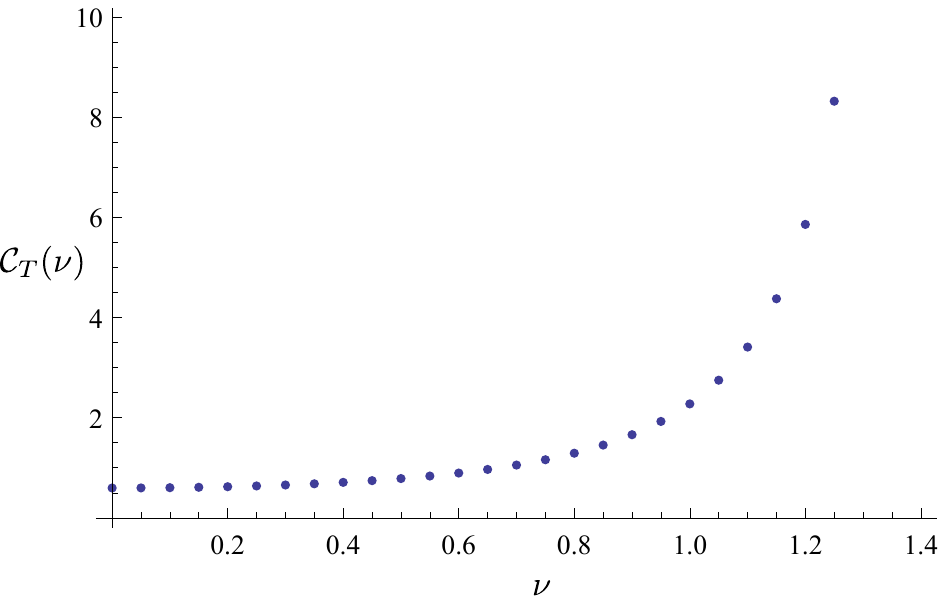}}
{\includegraphics[width=.475\textwidth]{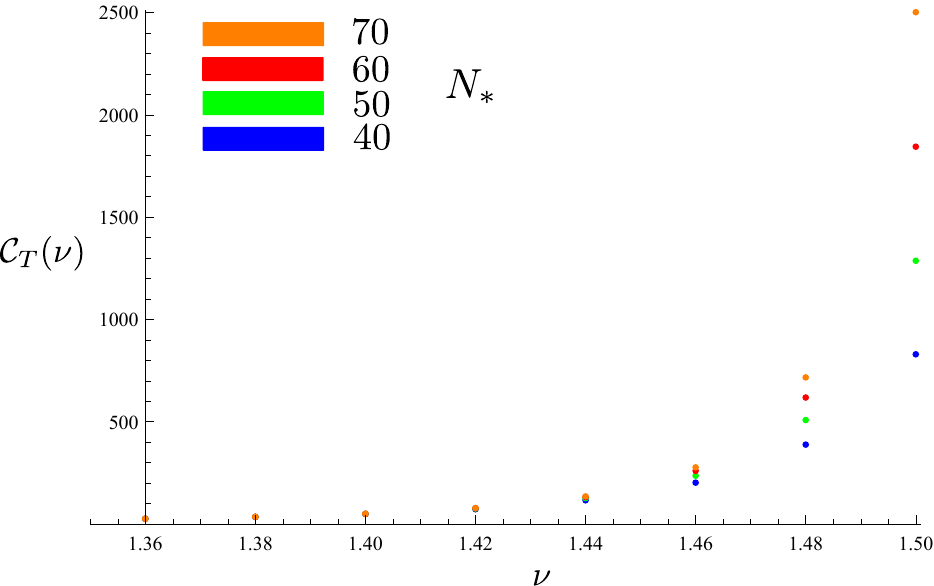}}
\caption{The factor $\mathcal{C}_{T}(\nu)$ defined in~\eqref{eq:CnuT}. For $\nu\ll3/2$ the dependence on the scale is negligible, it becomes significant as $\nu\rightarrow 3/2$. Left panel: $\mathcal{C}_{T}(\nu)$ for a scale that exits the horizon 50 $e$-folds before the end of inflation. Right panel: $\mathcal{C}_{T}(\nu)$ for different scales that exit the horizon $N_*$ $e$-folds before the end of inflation.}
\label{fig:CnuT}
\end{figure}

At lowest order in the \textit{in-in} expansion, the two-point correlation function reads
\begin{align}
\braket{\zeta^2 (\tau)}^{(0)} &= (2\pi)^3 \delta^3 (\textbf{p}_1+\textbf{p}_2) \frac{H^2}{\dot{\theta}^2} \lvert v_{p_1}(\tau) \lvert^2  \nonumber \\
&= (2\pi)^3 \delta^3 (\textbf{p}_1+\textbf{p}_2)   \frac{H^4}{2 f \dot{\theta}^2 \, p_1^3  } (1 + p_1^2 \tau^2) \hspace{1pt}.
\end{align}
On super-horizon scales $-p_1 \tau \ll 1$, the previous expression becomes
\begin{equation}
\braket{\zeta^2 (0)}^{(0)} = (2\pi)^3 \delta^3 (\textbf{p}_1+\textbf{p}_2)   \frac{H^4}{2 f \dot{\theta}^2 \, p_1^3  } \hspace{2pt} .
\end{equation}
The lowest-order correction is given by 
\begin{equation}
\braket{\zeta^2 (t)}^{(2)} = (2\pi)^3 \delta^3 (\textbf{p}_1+\textbf{p}_2) \frac{{f^\prime}^2 H^2}{f^2 p_1^3} \mathcal{C}_{T}(\nu) \hspace{1pt},
\end{equation}
where we have defined
\begin{equation}
\mathcal{C}_{T}(\nu) \equiv \frac{\pi}{4} \text{Re} \Biggl\{ \int_0^\infty dx_1 \int_{x_1}^\infty dx_2 \Biggl( \frac{e^{i(x_1-x_2)}}{\sqrt{x_1 x_2}} H_\nu^{(1)} (x_1) H_\nu^{(2)} (x_2) - \frac{e^{-i(x_1+x_2)}}{\sqrt{x_1 x_2}} H_\nu^{(1)} (x_1)H_\nu^{(2)} (x_2)  \Biggl) \Biggl\},
\label{eq:CnuT}
\end{equation}
and $x\equiv - p_1 \tau$. In the previous expression, we have considered the superhorizon limit. The factor $\mathcal{C}_{T}(\nu)$ is plotted in figure~\ref{fig:CnuT}. 
The lowest integration limit is not zero, but practically it should be taken at the time inflation ends, thus imposing a cutoff at 
\begin{equation}
e^{N_*} = - \frac{1}{p_1 \tau_f} \hspace{2pt},
\end{equation}
where $N_*$ is the number of $e$-folds before the end of inflation at which the considered scale $p_1$ leaves the horizon and $\tau_f$ is the end of inflation. However, in the following, we will suppose that inflation ends at $\tau_f = 0$. 

In summary, the power-spectrum (up to second order in the \textit{in-in} expansion) reads
\begin{align}
\braket{\zeta^2 (t)} &= \braket{\zeta^2 (t)}^{(0)} + \braket{\zeta^2 (t)}^{(2)} \nonumber \\
&= (2\pi)^3 \delta^3 (\textbf{p}_1+\textbf{p}_2)\frac{(2\pi)^2}{2p_1^3}  \biggl( \frac{H^4}{ 4\pi^2 f \dot{\theta}^2  }  + \frac{{f^\prime}^2 H^2}{2 \pi^2 f^2 } \mathcal{C}_T(\nu)   \biggl) ,
\end{align}
and the dimensionless power-spectrum is 
\begin{equation}
\mathcal{P}_{\zeta} = \frac{H^4}{ (2 \pi  \dot{\theta} )^2 f} \Biggl[1 + \frac{2 {f^\prime}^2}{f} \, \mathcal{C}_T(\nu) \, \biggl( \frac{\dot{\theta}^2}{H^2} \biggl)   \Biggl] .
\label{eq:power-spectrum-CT}
\end{equation}
Note that
\begin{equation}
\frac{2 {f^\prime}^2}{f}  \left( \frac{\dot{\theta}^2}{H^2} \right) = 2 \xi_\theta \hspace{1pt},
\end{equation}
where $\xi_\theta$ is the slow-roll parameter defined in~\eqref{eq:xi-theta}.
Therefore, the correction to the dimensionless power-spectrum is of first order in the slow-roll parameters. This is fundamental in order for the theory to remain perturbative. Another feature to notice is the dependence of $\mathcal{P}_{\zeta}$ on $f^{-1}$. In the constant-turn case, this is like a normalization factor since $\varphi$ is assumed constant. This discussion for the power spectrum is similar to the original single-field models of inflation~\cite{Chen:2009zp}, however in our case the non-trivial coupling in the kinetic term comes into play. 

\subsubsection{Bispectrum}

The definition of the non-linearity parameter $f_{\rm NL}$ is obtained considering the amplitude of the bispectrum in the equilateral configuration, i.e. 
\begin{equation}
\braket{\zeta(\textbf{p}_1) \zeta(\textbf{p}_2) \zeta(\textbf{p}_3)}  \xrightarrow{p_1=p_2=p_3} (2\pi)^7 \delta^3 (\textbf{p}_1 +\textbf{p}_2 +\textbf{p}_3 ) \mathcal{P}_\zeta^2 \biggl( \frac{9}{10} f_{\rm NL} \biggl) \frac{1}{p_1^6} \hspace{2pt} .
\label{eq:fNL-def-2}
\end{equation}
The magnitude of $f_{\rm NL}$ can be estimated using Feynman rules. At each vertex we associate the corresponding dimensionless coupling constant (it is convenient to work with dimensionless constants, since $f_{\rm NL}$ is dimensionless).
The $n$-th order term of the \textit{in-in} expansion can be estimated as~\cite{Chen:2009zp}
\begin{equation}
\braket{\zeta^3}^{(n)} \sim I \prod_i^n C_i \frac{\sqrt{f} \dot{\theta}}{H^2} \mathcal{P}_\zeta^2 \frac{1}{p_1^6} \hspace{2pt} ,
\end{equation}
where $I$ denotes an integral term and $C_i$ are the coupling constants. The integral term can be considered $I\sim\mathcal{O}(1)$, even in the limit $\nu \rightarrow 3/2$~\cite{Chen:2009zp}.
Let us notice that the power-spectrum~\eqref{eq:power-spectrum-CT} (at zeroth order) can be rewritten as
\begin{equation}
\mathcal{P}_\zeta = \frac{H^2}{8 \pi^2 \varepsilon_\theta} \hspace{2pt} \, ,
\end{equation}
where $\varepsilon_\theta$ in this case corresponds to the usual slow-roll parameter $\epsilon$ according to Eq.~(\ref{eq:epsilon-phi-theta}). 
Finally, the amplitude of $f_{\rm NL}$ can be approximated as
\begin{equation}
f_{\rm NL} \sim \prod_i^n C_i \frac{\sqrt{f} \dot{\theta}}{H^2}  \hspace{2pt} .
\end{equation}

The lowest order non-vanishing contribution to the bispectrum reads 
\begin{equation}
\braket{\delta\theta^3 }^{(2)} = \braket{\delta\theta^3 H^{(2)} H^{(3)}_{C_1} } .
\end{equation}
The dimensionless coupling constant of $H^{(2)}$ is $f^\prime f^{-1/2} \dot{\theta} H^{-1} $, and that of $H^{(3)}_{C_1}$ is given by $f^\prime f^{-1} H$.
Then, it is simple to show that
\begin{equation}
\braket{\zeta^3 }^{(2)}_{C_1}  \sim \frac{{f^\prime}^2}{f^2} \varepsilon_\theta  \mathcal{P}_\zeta^{2} \frac{1}{p_1^6} \hspace{2pt} .
\end{equation}
We recall that with $Z_i,A_i,B_i,C_i,D$ we indicate the various coupling constants defined in Eq.~(\ref{eq:H-0-H-INT}). We have introduced the notation $\braket{\zeta^3 }^{(2)}_{C_1}$, that denotes the second-order term in the expansion of the three-point correlator $\braket{\zeta^3 }$ that contains the free vacuum expectation value $\bra{0} \zeta^3 H^{(2)} H^{(3)}_{C_1} \ket{0}$. In what follows we will use this short-hand notation to characterise the terms of the \textit{in-in} expansion. We will consider only the connected diagrams and write down only the relevant couplings to identify the different terms (e.g., in the previous example we did not write the coupling $Z_1$ for $H^{(2)}$ since it was uniquely determined).

At third order,
\begin{equation}
\braket{\delta\theta^3 }^{(3)} \supseteq \braket{\delta\theta^3 H^{(2)} H^{(2)} H^{(3)}_{B_1} } ,
\end{equation}
since $H^{(3)}$~\eqref{eq:H-3-INT-CT} does not contain a $\delta\theta^3$ term (in the following subsection we will consider a massive inflaton field $\theta$). Its contribution to the 
bispectrum is given by
\begin{equation}
\braket{\zeta^3}^{(3)}_{B_1}  \sim \varepsilon_\theta^2 \frac{f^\prime}{f} \frac{2 {f^\prime}^3 - f f^\prime f^{\prime\prime}}{f^3} \mathcal{P}_\zeta^{2} \frac{1}{p_1^6} \hspace{2pt} .
\end{equation}
Finally, at fourth order we need to calculate terms like
\begin{equation} 
\braket{\delta\theta^3 }^{(4)} \supseteq \braket{\delta\theta^3 H^{(2)} H^{(2)} H^{(2)}  H^{(3)} },
\end{equation}
where $H^{(3)}=H^{(3)}_A$ or $H^{(3)}=H^{(3)}_{C_1}$.
The naive evaluations of these contributions are
\begin{subequations}
\begin{align}
\braket{\zeta^3}^{(4)}_{A_1} &  \sim  \varepsilon^3_\theta \frac{{f^\prime}^3}{f^3}  \biggl( \frac{6 f f^\prime f^{\prime\prime} - 6 {f^\prime}^3 - f^2 f^{\prime\prime\prime} }{f^3} \biggl)   \mathcal{P}_\zeta^{2} \frac{1}{p_1^6} \hspace{2pt} , \label{eq:zeta-4-a1} \\
\braket{\zeta^3}^{(4)}_{A_2} & \sim \frac{V_{\varphi\varphi\varphi}}{H} \varepsilon^2_\theta \frac{{f^\prime}^3 }{f^3}  \mathcal{P}_\zeta^{2} \frac{1}{p_1^6} \hspace{2pt}, \label{eq:CT-chens}\\
\braket{\zeta^3}^{(4)}_{C_1} &\sim \varepsilon_\theta^2 \frac{{f^\prime}^4}{f^4} \mathcal{P}_\zeta^{2} \frac{1}{p_1^6} \hspace{2pt}.
\end{align}
\end{subequations}
All the previously estimated contributions are suppressed by the slow-roll parameters. In addition to this, notice that $f \sim f^\prime \sim f^{\prime\prime}$ for the function~\eqref{eq:function-f} and hence the non-canonical kinetic term cannot affect the order of magnitude of $f_{\rm NL}$.
Chen~\cite{Chen:2009zp} computes a term similar to $\braket{\zeta^3}^{(4)}_{A2}$, since it is the only one that can be enhanced thanks to $V_{\varphi\varphi\varphi}$, which is not suppressed 
by the slow-roll conditions (the field $\varphi$ is not the inflaton).
Since our model is very similar to that of Chen\footnote{The main difference is the presence in our model of a non-canonical kinetic term, but this does not alter significantly the results, since, for the function $f$~\eqref{eq:function-f}, we have $f\sim f^\prime \sim f^{\prime\prime}$. However, our treatment is general and can be generalized considering other possibilities for $f$.}, the aforementioned term is the only one that can be enhanced, for the same reason; all other contributions are suppressed by the slow-roll parameters. Therefore we provide some more details about the term $\braket{\zeta^3}^{(4)}_{A2}$.

\begin{figure}[!t]
  \centering
    \includegraphics[width=1\textwidth]{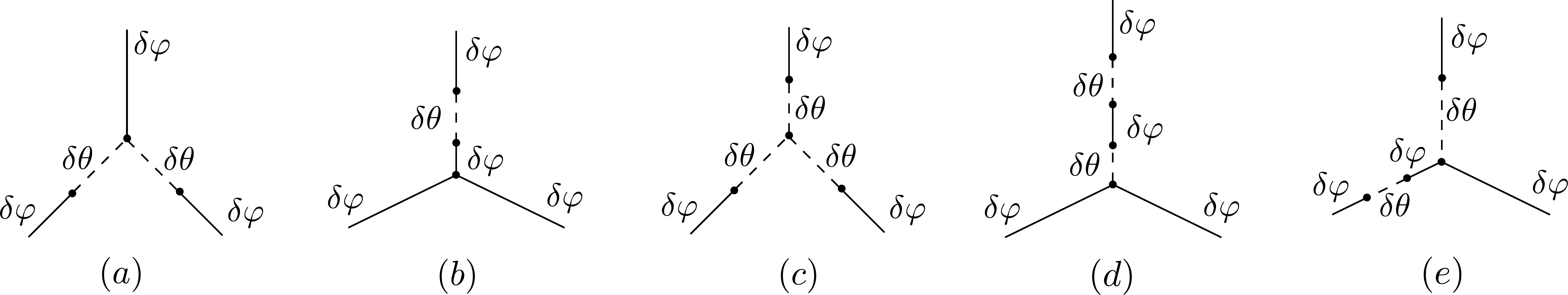}
      \caption{Diagrams associated to third and fourth-order contributions to bispectrum in the \textit{in-in} expansion. }
      \label{fig:diagrams-2}
\end{figure}

The diagrammatic representation of the term~\eqref{eq:CT-chens} is given by the diagram (c) in figure~\ref{fig:diagrams-2} with the fields $\delta\varphi$ and $\delta\theta$ swapped.
The full expression of the term~\eqref{eq:CT-chens} in the commutator form reads
\begin{align}
\braket{\zeta^3}^{(4)}_{A_2} =& - 12 (2\pi)^3 \delta^3(\textbf{p}_1 +\textbf{p}_2 +\textbf{p}_3) \biggl( \frac{H f^\prime}{\dot{\theta}} \biggl)^3 v_{p_1} (0) v_{p_2} (0) v_{p_3} (0) \nonumber \\
&\text{Re} \Biggl\{ \int^0_{-\infty} d\tau_1 \int^{\tau_1}_{-\infty} d\tau_2 \int^{\tau_2}_{-\infty} d\tau_3 \int^{\tau_3}_{-\infty} d\tau_4  \prod_{i=1}^4 \Bigl( a^3(\tau_i)  \Bigl) \dot{\theta}^3 \bigl( v^\prime_{p_1} (\tau_1) - \text{c.c.} \bigl) \nonumber \\
& \biggl(  \frac{a(\tau_2) V_{\varphi\varphi\varphi} (\tau_2) }{6 }  A  +  \frac{a(\tau_3) V_{\varphi\varphi\varphi} (\tau_3) }{6 } B +  \frac{a(\tau_4) V_{\varphi\varphi\varphi} (\tau_4) }{6 } C \biggl)  \Biggl\} + 5\,\text{perm} \hspace{1pt} ,
\label{eq:CT-chens-comm-form}
\end{align}
where
\begin{align}
A &=  \bigl(  u_{p_1}(\tau_1) u^*_{p_1}(\tau_2)  - \text{c.c.} \bigl) \bigl( u_{p_3} (\tau_2) u^*_{p_3} (\tau_4) v^{\prime *}_{p_3} (\tau_4) - \text{c.c.} \bigl) u_{p_2} (\tau_2) u^*_{p_2} (\tau_3) v^{\prime *}_{p_2} (\tau_3) \hspace{1pt}, \nonumber \\
B &= ( v^\prime_{p_2} (\tau_2) - \text{c.c.} \bigl) \bigl( u^*_{p_1} (\tau_1) u^*_{p_2} (\tau_2) u_{p_1} (\tau_3) u_{p_2} (\tau_3) - \text{c.c.} \bigl)  u_{p_3} (\tau_3)  u^*_{p_3} (\tau_4)  v^{\prime *}_{p_3} (\tau_4) \hspace{1pt}, \nonumber\\
C &= - ( v^\prime_{p_2} (\tau_2) - \text{c.c.} \bigl) ( v^\prime_{p_3} (\tau_3) - \text{c.c.} \bigl) u^*_{p_1} (\tau_1) u^*_{p_2} (\tau_2) u^*_{p_3} (\tau_3) u_{p_1} (\tau_4) u_{p_2} (\tau_4) u_{p_3} (\tau_4) \hspace{1pt}. \nonumber
\end{align}
In order to calculate $f_{\rm NL}$, we consider the equilateral limit $p_1=p_2=p_3=p$ (see equation~\eqref{eq:fNL-def-2}).
Using the mode functions~\ref{eq:CT-mode-functions}, equation~\eqref{eq:CT-chens-comm-form} becomes
\begin{align}
\braket{\zeta^3}^{(4)}_{A_2} &=   (2\pi)^3 \delta^3(\textbf{p}_1 +\textbf{p}_2 +\textbf{p}_3) \frac{3 \pi^3}{2^7} \frac{{f^\prime}^3 H^2}{f^3 p^6 } V_{\varphi\varphi\varphi}  \, \text{Re} \Biggl\{  \int_0^{+\infty} dx_1 \int_{x_1}^{+\infty} dx_2 \int_{x_2}^{+\infty} dx_3 \int_{x_3}^{+\infty} dx_4  \nonumber \\
& \phantom{=i} \prod_{i=1}^4 \biggl(  \frac{ 1 }{x_i^4}  \biggl)  \bigl( \tilde{v}^\prime (x_1) - \text{c.c.} \bigl)  \biggl(  \frac{  \tilde{A}  }{  x_2}   +  \frac{\tilde{B}  }{ x_3 }  +  \frac{ \tilde{C}   }{  x_4 }\biggl)  \Biggl\}  \, ,
\label{eq:CT-chens-equilateral}
\end{align}
where we have defined $x_i \equiv -p\tau_i$ and
\begin{subequations}
\begin{align}
&\tilde{v} (x) \equiv (1- ix ) e^{i x} , \\
&\tilde{u} (x) \equiv x^{\frac{3}{2}} H_\nu^{(1)} (x) \hspace{1pt},  \\
&\tilde{A} =  \bigl(  \tilde{u} (x_1) \tilde{u}^* (x_2)  - \text{c.c.} \bigl) \bigl( \tilde{u} (x_2) \tilde{u}^* (x_4) \tilde{v}^{\prime *} (x_4) - \text{c.c.} \bigl) \tilde{u} (x_2) \tilde{u}^* (x_3) \tilde{v}^{\prime *} (x_3) \hspace{1pt},  \\
&\tilde{B} = ( \tilde{v}^\prime (x_2) - \text{c.c.} \bigl) \bigl( \tilde{u}^* (x_1) \tilde{u}^* (x_2) \tilde{u} (x_3) \tilde{u} (x_3) - \text{c.c.} \bigl)  \tilde{u} (x_3)  \tilde{u}^* (x_4)  \tilde{v}^{\prime *} (x_4) \hspace{1pt}, \\
&\tilde{C} = - ( \tilde{v}^\prime (x_2) - \text{c.c.} \bigl) ( \tilde{v}^\prime (x_3) - \text{c.c.} \bigl) \tilde{u}^* (x_1) \tilde{u}^* (x_2) \tilde{u}^* (x_3) \tilde{u} (x_4) \tilde{u} (x_4) \tilde{u} (x_4) \hspace{1pt}. 
\end{align}
\end{subequations}
Finally, using the definition~\eqref{eq:fNL-def-2} and the expression of the power-spectrum~\eqref{eq:power-spectrum-CT}, the contribution of the previous correlator to the parameter $f_{\rm NL}$ 
reads
\begin{align}
f_{\rm NL}^{(4)} &=  \frac{5 \pi^3}{3\cdot2^6} \, \frac{{f^\prime}^3 }{f^3 } \, \biggl(f \frac{ \dot{\theta}^2}{H^2} \biggl)^{\hspace{-2pt} 2} \, \frac{V_{\varphi\varphi\varphi}}{H^2} \,  \text{Re} \Biggl\{ \int_0^{+\infty} dx_1 \int_{x_1}^{+\infty} dx_2 \int_{x_2}^{+\infty} dx_3 \int_{x_3}^{+\infty} dx_4 \nonumber \\
&\phantom{=i}\prod_{i=1}^4 \biggl(  \frac{ 1 }{x_i^4}  \biggl)  \bigl( \tilde{v}^\prime (x_1) - \text{c.c.} \bigl) \biggl(  \frac{ \tilde{A} }{  x_2}   +  \frac{ \tilde{B} }{ x_3 }  +  \frac{ \tilde{C} }{  x_4 }  \biggl)  \Biggl\}  \, .
\label{eq:CT-chens-fNL}
\end{align}
Notice that $f_{\rm NL}^{(4)}$ is proportional to the square of the slow-roll parameter $\varepsilon_\theta$~\eqref{eq:epsilon-phi-theta}. 
As we have already said, the amplitude can be increased, thanks to the third derivative $V_{\varphi\varphi\varphi}$. Obviously, this enhancement depends on the particular form of the potential $V$. We postpone this discussion to section~\ref{sec:chen-alpha-introduction}, where we are going to study the effects of the presence of the parameter $\alpha$ which characterizes the specific models we are analyzing.

\subsubsection{Generalizations}
\label{sec:chen-generalizations}

In this section, we will generalize the previous model, by introducing a potential for the inflaton $\theta$. We will see that the new contributions to the power-spectrum and the bispectrum are small, since they are suppressed by the slow-roll conditions.
Let us suppose that the potential is separable, $V(\varphi,\theta) = {\cal V}(\varphi) + U(\theta)$.
If the field $\theta$ is massive, the expression of the power-spectrum~\eqref{eq:power-spectrum-CT} is slightly modified by the mass term $\frac{a^3}{2} V_{\theta\theta} \delta\theta^2$ in the Hamiltonian~\eqref{eq:H-0-H-INT}. The mode functions are given by~\eqref{eq:u-k} and~\eqref{eq:v-k}. Now the power-spectrum reads
\begin{equation}
\mathcal{P}_{\zeta} = \frac{H^4}{ (2 \pi  \dot{\theta} )^2 f} \Biggl[1 + \frac{2 {f^\prime}^2}{f} \, \mathcal{C}_T(\nu,\tilde{\nu}) \, \biggl( \frac{\dot{\theta}^2}{H^2} \biggl)   \Biggl] ,
\label{eq:CT-power-spectrum-massive}
\end{equation}
where
\begin{align}
\mathcal{C}_T(\nu,\tilde{\nu})  & \equiv  \frac{\pi^3}{4^3} \text{Re} \Biggl\{  \int_0^\infty dx_1 \int_{x_1}^\infty dx_2 H_\nu^{(1)} (x_1) x_1^{-\frac{3}{2}} \frac{d}{d x_2} \biggl( x_2^{\frac{3}{2}} H_{\tilde{\nu}}^{(2)} (x_2) \biggl) H_\nu^{(2)} (x_2) x_2^{-\frac{3}{2}} \nonumber \\
& \phantom{= i} \Biggl[  H_{\tilde{\nu}}^{(1)} (0) H_{\tilde{\nu}}^{(2)} (0) \frac{d}{d x_1} \biggl( x_1^{\frac{3}{2}} H_{\tilde{\nu}}^{(1)} (x_1) \biggl) -  H_{\tilde{\nu}}^{(1)} (0) H_{\tilde{\nu}}^{(1)} (0) \frac{d}{d x_1} \biggl( x_1^{\frac{3}{2}} H_{\tilde{\nu}}^{(2)} (x_1) \biggl)  \Biggl] \Biggl\}. 
\end{align}

The cubic Hamiltonian~\eqref{eq:H-3-INT-CT} has the additional term
\begin{equation}
\mathcal{H}^{(3)}_D = \frac{a^3}{6} V_{\theta\theta\theta} \delta\theta^3\, ,
\end{equation}
and hence new corrections arise for the bispectrum. At first order, the term
\begin{equation}
\braket{\delta\theta^3 }^{(1)} \supseteq \braket{\delta\theta^3 H^{(3)}_D}
\end{equation}
constitutes a self-interaction of the field $\theta$. As previously done, its contribution to the curvature perturbation can be approximated as
\begin{equation}
\braket{\zeta^3}^{(1)}_{D}   \sim \frac{V_{\theta\theta\theta}}{H} \sqrt{\varepsilon_\theta} f^{-\frac{3}{2}} \mathcal{P}_\zeta^2 \frac{1}{p_1^6} \hspace{2pt}.
\label{eq:chens-zeta-1-D}
\end{equation}
This contribution is negligible since $V_{\theta\theta\theta} H^{-1}$ should be small during slow-roll\footnote{The third derivative of the potential is related to a second-order slow-roll parameter. In the notations of~\cite{Ade:2015lrj} the slow-roll parameter $\xi_V^2 \equiv V_{\varphi} V_{\varphi\varphi\varphi} / V^2$ is defined, where $V$ is the potential of the inflaton. \label{ft:third-derivative} }.
At second order, there are no new terms. Instead, at third order, the term
\begin{equation}
\braket{\delta\theta^3 }^{(3)} \supseteq \braket{\delta\theta^3 H^{(3)}_D H^{(2)} H^{(2)} }
\end{equation}
appears, and its contribution to the bispectrum is given by
\begin{equation}
\braket{\zeta^3}^{(3)}_{D}   \sim   \frac{V_{\theta\theta\theta}}{H} \frac{{f^\prime}^2}{f^{\frac{7}{2}}} \varepsilon_\theta^{\frac{3}{2}}  \mathcal{P}_\zeta^2 \frac{1}{p_1^6} \hspace{2pt} .
\label{eq:chens-zeta-3-D}
\end{equation}
However, this term cannot contribute significantly to the bispectrum for the same reason as the term $\braket{\zeta^3}^{(1)}_{D}$. Therefore, the field $\theta$ can be considered massless with good approximation, as long as the slow-roll conditions hold and as far as the bispectrum is concerned.

If $f$ is a generic function, one can increase the ratio $f^\prime / f$, in order to produce large non-Gaussianities, since the previously estimated corrections usually depend on this ratio. However, some caution must be used. Indeed, this ratio affects the first non-vanishing correction to the power-spectrum~\eqref{eq:power-spectrum-CT} (or~\eqref{eq:CT-power-spectrum-massive}), and hence we have to impose a constraint like 
\begin{equation}
\frac{f^{\prime 2}}{f^2} \varepsilon_\theta^2 \ll 1 \hspace{1pt}\, ,
\end{equation}
in order not to loose the perturbativity of the computation. 
In addition to this, since $\theta$ is the inflaton we should require that the slow-roll parameters in the definition of $\tilde{\nu}$~\eqref{eq:nu-theta} are small. The latter condition implies that
\begin{subequations}
\begin{align}
\varepsilon_\theta &\ll 1 \hspace{1pt}, \\
\frac{f^{\prime\prime}}{f} \varepsilon_\theta &\ll 1 \hspace{1pt} \, .
\end{align}
\end{subequations}
These conditions prevent an enhancement of all the estimated contributions except for~\eqref{eq:zeta-4-a1} and the self-interaction terms~\eqref{eq:chens-zeta-1-D} and~\eqref{eq:chens-zeta-3-D}. The latter terms can be enhanced if $f$ is sufficiently small, because they are proportional to inverse powers of $f$. The actual enhancement depends on the specific expression of $f$ and thus we do not make a deeper analysis. However, notice that these two terms are suppressed by the slow-roll parameter $\varepsilon_\theta$ and also by the factor $V_{\theta\theta\theta} / H$ that should be small during inflation if $\theta$ drives it, as we mentioned before. The term~\eqref{eq:zeta-4-a1} can be enhanced if $f^{\prime\prime\prime} f^2 f^{\prime -3} \gg 1$ thanks to the third derivative $f^{\prime\prime\prime}$ which is not constrained by the slow-roll conditions. Again, the magnitude of this enhancement depends on the particular form of $f$.

\subsubsection{The \texorpdfstring{$\alpha$}{a} parameter}
\label{sec:chen-alpha-introduction}

Similarly to what is done in the original $\alpha-$attractor models~\cite{Kallosh:2013xya,Kallosh:2013hoa,Kallosh:2013yoa,Kallosh:2014rga}, let us now introduce the dimensionless parameter $\alpha$ into the Lagrangian~\eqref{eq:two-alpha-model-2}. This parameter  interpolates between the standard predictions of the chaotic inflation scenario and the universal attractor regime while describing a  broad class of inflationary models that arise naturally in supergravity. In fact, the parameter $\alpha$, besides allowing one to parametrize a broad class of models, arises naturally in supergravity theories and is related to the K\"{a}hler curvature of the manifold. For our purposes, it is sufficient to think of $\alpha$ as a parameter influencing the UV cutoff (see Refs.~\cite{Kallosh:2013yoa,Kallosh:2014rga}).  In practice, it can be introduced simply by replacing $\varphi / M_{Pl} \rightarrow\varphi / (\sqrt{\alpha} M_{Pl} ) $ into the function $f$ and the potential $V$ of the Lagrangian~\eqref{eq:two-alpha-model-2}.

What is the effect  of the  the parameter $\alpha$ on the generation primordial  non-Gaussianities? Can certain values of this parameter generate large non-Gaussianities? This parameter can be introduced into the previous estimates, noticing that each derivative w.r.t. $\varphi$ corresponds to a factor $1/ \sqrt{\alpha}$. This is clear after the introduction of a rescaled field $\phi \equiv \sqrt{\alpha} \varphi$. Obviously, $\alpha$ should be small, if we want to enhance through it some of the previous contributions. However, there is a tight constraint deriving from the correction to the 
power-spectrum~\eqref{eq:CT-power-spectrum-massive}.
Since the correction should be small, in order to preserve almost scale-invariance, we require 
\begin{equation}
\frac{\varepsilon_\theta }{\alpha} \ll 1 \hspace{1pt} .
\label{eq:CT-alpha-constraint}
\end{equation}
One can show that this condition prevents an enhancement of all the terms estimated in the previous two subsections, except for~\eqref{eq:CT-chens}. Indeed, this term can be estimated as
\begin{equation}
\braket{\zeta^3}^{(4)}_{A_2} \sim \frac{1}{\alpha} \frac{V_{\phi\phi\phi}}{H^2} \frac{\varepsilon^2_\theta}{\alpha^2} \frac{{f^\prime}^3 }{f^3}  \mathcal{P}_\zeta^{2} \frac{1}{p_1^6} \hspace{2pt} ,
\label{eq:chen-term-enhancement-alpha}
\end{equation}
where we have supposed that the potential $V$ has the same dependence as~\eqref{eq:two-alpha-model-2}.
This is rather interesting. We have obtained a term that can be enhanced through the parameter $\alpha$ as $\alpha\rightarrow 0$,  since it scales as $\alpha^{-1}$.  Furthermore, the main contribution to the bispectrum comes from a single term. Since the model we have investigated is a specific realization of the quasi-single field mechanism~\cite{Chen:2009zp} the shape of the primordial non-Gaussianity generated can span from equilateral to local according to the value of the $\nu$ parameter for the field $\varphi$~\cite{Chen:2009zp}.

\begin{figure}[!t]
  \centering
    \includegraphics[width=0.6\textwidth]{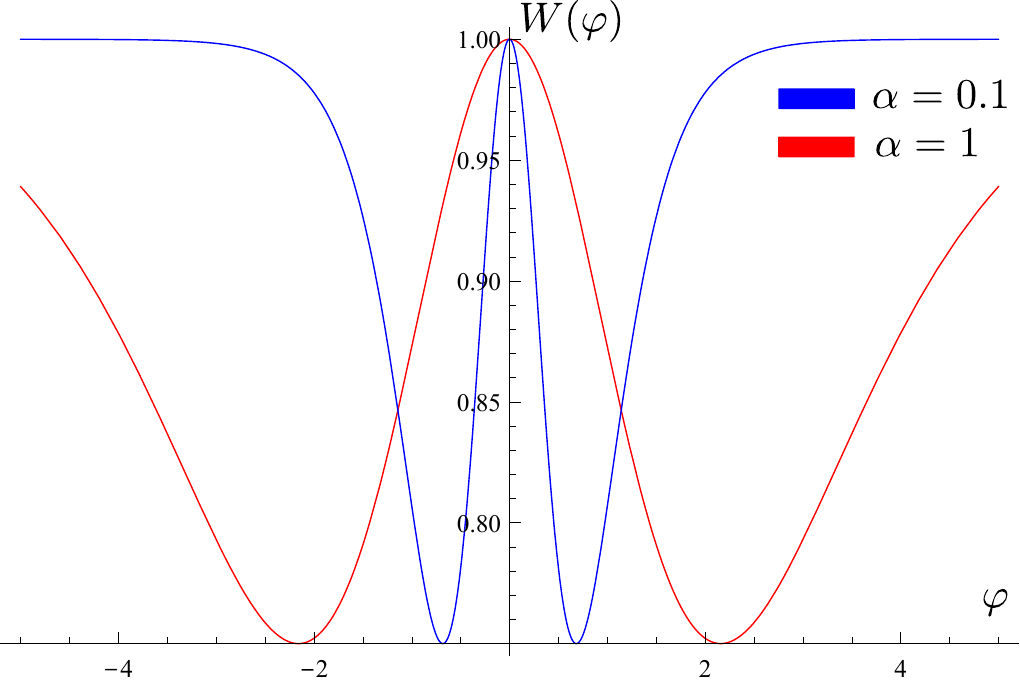}
      \caption{The double-well potential~\eqref{eq:chen-potential}. Notice that a little variation of $\alpha$ could make the potential more confining. For comparison, the third derivative of the potential with $\alpha=1$ evaluated at the minima is $ -0.15 \, M_{Pl} $, that of the potential with $\alpha=0.1$ is $ -4.56 \, M_{Pl} $.  }
      \label{fig:chen-potential}
\end{figure}

Chen, in his paper~\cite{Chen:2009zp}, does not provide an explicit potential that could produce large non-Gaussianities since his main concern is to discuss a mechanism to produce 
non-Gaussianities, rather than a specific model. In the context of $\alpha$-attractors, we have seen (equation~\eqref{eq:F-full-dependence}) that the potential depends on the hyperbolic tangent of 
$\varphi$. Remember that we have supposed that the trajectory is a circle in field-space and hence we need a confining potential along the radial direction $\varphi$, in order to satisfy the initial assumptions. One possibility is the potential
\begin{equation}
V(\varphi) = 1 - \tanh^2 \frac{\varphi}{\sqrt{6 \alpha}} +  \tanh^4 \frac{\varphi}{\sqrt{6 \alpha}} \hspace{2pt} ,
\label{eq:chen-potential}
\end{equation}
where we have only given the dependence on $\varphi$, with the assumption that a suitable potential for $\theta$ should be chosen in order to have inflation\footnote{Given the full dependence~\eqref{eq:F-full-dependence} of the potential term, one possibility for the potential of the inflaton $\theta$ could be $U(\theta) = (1+\cos(\theta / \mu) ) $, where $\mu$ is a scale that determines the curvature of the potential. This potential is typical of natural inflation models~\cite{Freese:1990rb,Adams:1992bn}.}. The previous example is a double-well potential with minima at 
$\varphi_\pm \simeq \pm 2.2 \sqrt{\alpha}$ and it is represented in figure~\ref{fig:chen-potential}. The third derivative of the potential, evaluated at the minima, is $V_{\varphi\varphi\varphi}(\varphi_\pm) \simeq -0.15 \alpha^{-\frac{3}{2}}$. Notice that, decreasing $\alpha$ makes the potential more confining and it increases the magnitude of non-Gaussianities, enhancing the third derivative of the potential.

Eq.~\eqref{eq:chen-term-enhancement-alpha} is our main result and shows that indeed large non-Gaussianities can be obtained by the dependence on $\alpha^{-1}$. Moreover it provides 
a well motivated realization of the quasi-single field scenario~\cite{Chen:2009zp}.  The analysis we perform in this section can be also applicable to realizations of natural inflation models within $\alpha$-attractor supergravity, as studied in~\cite{Linde:2018hmx} (see also~\cite{Yamada:2018nsk}). For example, the findings of this section might be applied to natural inflation after the alpha-attractor regime ends, or at an intermediate stage between two regimes driven by the radial component of the inflaton field~\cite{Linde:2018hmx}.

Note, though, that these non-Gaussianities cannot grow as large as $\alpha^{-1}$ because as $\alpha \rightarrow 0$, the slow-roll parameter $\varepsilon_\theta \rightarrow 0$ (due to the constraint~\eqref{eq:CT-alpha-constraint}) and this can become a very stringent requirement on the potential of $\theta$, which should be accurately fine-tuned to be sufficiently flat, in order to have $\varepsilon_\theta \rightarrow 0$. However, it is easy to see that values of $f_{\rm NL}$ of order one can be obtained for \footnote{The current observational limit by Planck ($\alpha < 14$ at $95 \%$C.L., see~\cite{Ade:2015lrj}) applies to single-field models of $\alpha$-attractors, where the inflaton field is $\phi$, so that our result on the level of primordial non-Gaussianity can still hold even for improved constraints of this type.} $\alpha <  1$.  


\subsection{Nearly single-field trajectory}
\label{sec:NSF}

We wish to study now the case in which the trajectory of the fields is nearly straight. This represents the minimum amount of non-Gaussianity that can be generated, as it will mimic a near single-field trajectory.

\begin{figure}
\centering
{\includegraphics[width=.475\textwidth]{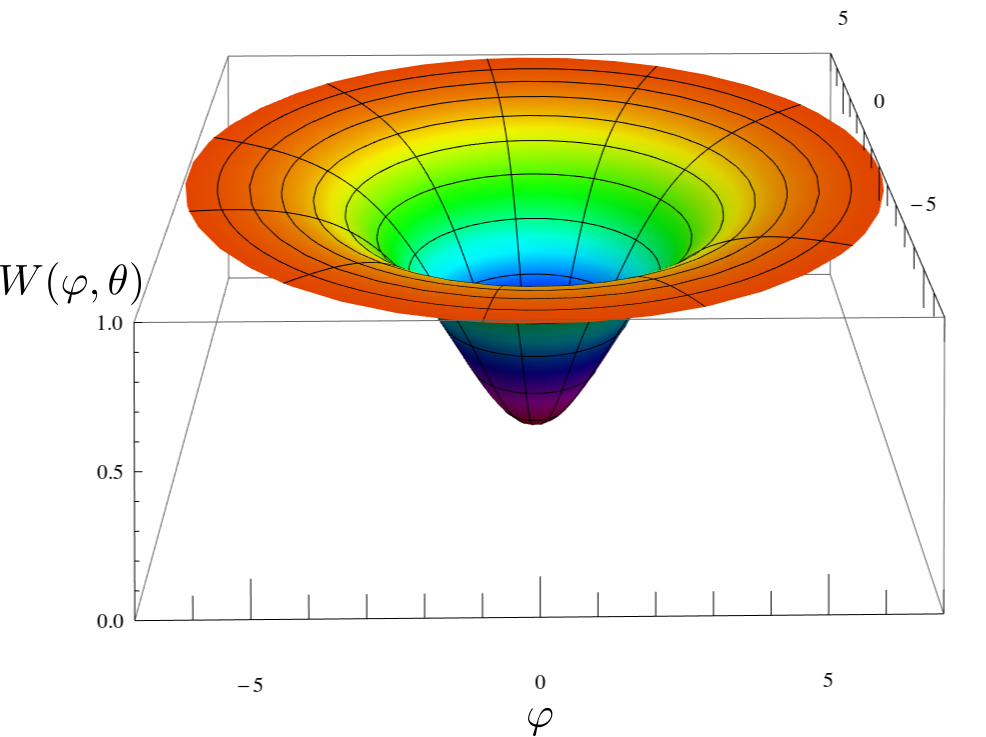}}
{\includegraphics[width=.475\textwidth]{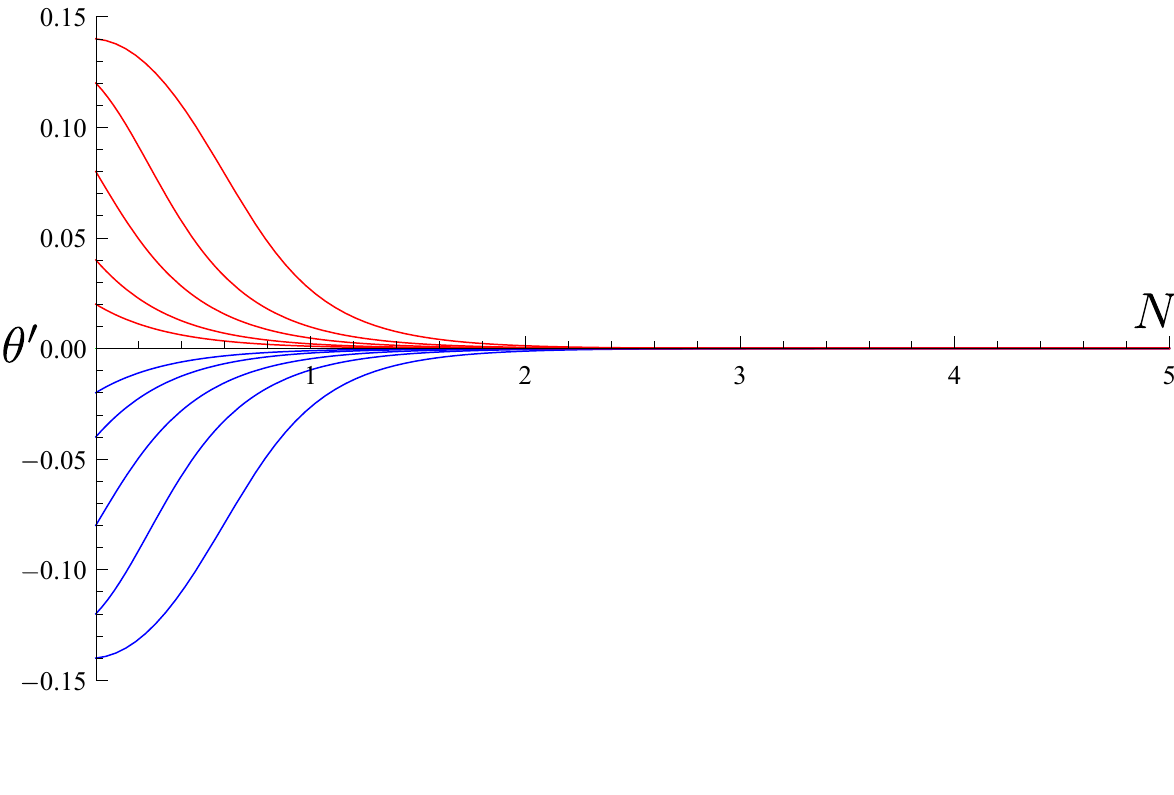} }
\caption{Left panel: The potential~\eqref{eq:NSF-potential-phi} $V(\varphi,\theta) =  \tanh^2 \left( \varphi / \sqrt{6} \right)$ represented in cylindrical coordinates. Right panel: The velocity 
$\theta^\prime$ as a function of the number of $e$-folds $N$. In this plot, the same notation and conventions of section~\ref{sec:W-model-initials} are used. 
The initial conditions are $\varphi (0) = 6.5$ and $\varphi^\prime(0) = 0$. Notice the suppression of the velocity $\theta^\prime$.}
\label{fig:TANH2-potential-dynamics}
\end{figure}

Let us consider, as we have done before for pedagogical reasons, one of the simplest models that the Lagrangian~\eqref{eq:two-alpha-model-2} can describe, namely the one with potential~\cite{Kallosh:2013daa}
 
\begin{equation}
V(\varphi,\theta) =  \tanh^2 \frac{\varphi}{\sqrt{6}} \hspace{2pt} .
\label{eq:NSF-potential-phi}
\end{equation}
This potential is shown in  the left panel of Fig.~\ref{fig:TANH2-potential-dynamics}. We have seen that the angular velocity must obey $\dot{\theta} / H \ll 1$, in order to have inflation. Hence we suppose that inflation is driven by the field $\varphi$, which slowly rolls down the potential on a nearly straight trajectory\footnote{We call ``nearly straight" a trajectory with $\dot{\theta} / H \ll 1$ during slow-roll, which corresponds to 
$\theta^\prime \ll 1$, using the notation of section~\ref{sec:W-model-initials}.}. We assume that the angular velocity $\dot{\theta} $ is small (with respect to $H$) and constant. 
We want to study whether it is possible to produce large non-Gaussianities, even if the background trajectory reduces to that of the single-field case, thanks to the non-canonical kinetic term of 
$\theta$ or to the introduction of the parameter $\alpha$. Indeed, this is an interesting question, since we know that for single-field inflation non-Gaussianities are limited by the smallness of the slow-roll parameters~\cite{Gangui,Acquaviva:2002ud,Maldacena:2002vr}.
In fact, our (background) trajectory is the same as in single-field case but it lives in a two dimensional field space. Inflation is driven by the field $\varphi$ and $\theta$ does not contribute significantly to it since it has no potential energy and small kinetic energy.

Assuming that the trajectory is a straight line seems, at first sight, to be a restrictive assumption. However, for the type of models we are studying this is not the case. Indeed, the term 
$\omega_\theta$ in the background equation~\eqref{eq:phi-background-W-1} forces the velocity $\theta^\prime$ to be small, at least during slow-roll (see section~\ref{sec:W-model-initials}). Furthermore, we have seen that in the limit $\varphi \gg 1$ the trajectories are nearly straight, thanks to the suppression of the potential term in equation~\eqref{eq:theta-background-W-1}, due to the factor $f^{-1}$ ($f$ is defined in equation~\eqref{eq:function-f}). Therefore, this simple trajectory can be a good starting point to study the more general models described by the Lagrangian~\eqref{eq:two-alpha-model-2}, as we will see in the following section. 

The Hamiltonian~\eqref{eq:H-0-H-INT} for this model coincides with~\eqref{eq:H-0-H-INT-CT}.
The first, second and third derivatives of the potential are respectively given by
\begin{subequations}
\begin{align}
V_\varphi &= \sqrt{\frac{2}{3}} \frac{\sinh (\varphi/\sqrt{6})}{\cosh^3(\varphi/\sqrt{6})} \hspace{2pt} ,\\
V_{\varphi\varphi} &= \frac{1-2 \sinh^2(\varphi/\sqrt{6})}{3 \cosh^4(\varphi/\sqrt{6})} \hspace{2pt} ,\\
V_{\varphi\varphi\varphi} &= \frac{2}{3} \sqrt{\frac{2}{3}} \frac{\sinh(\varphi/\sqrt{6})(\sinh^2(\varphi/\sqrt{6})-2)}{\cosh^5(\varphi/\sqrt{6})} \hspace{1pt} , \label{eq:NSF-potential-ppp}
\end{align}
\end{subequations}
and hence they behave as $\exp\bigl(-\sqrt{2/3} \hspace{2pt} \varphi \bigl)$ in the limit $\varphi \gg 1$, i.e., the region where inflation occurs. Thanks to this, we can consider both $\delta\varphi$ and $\delta\theta$ as almost massless fields ($\nu=\tilde{\nu}=3/2$). Therefore, the mode functions~\eqref{eq:u-k} and~\eqref{eq:v-k} reduce to 
\begin{subequations}
\begin{align}
u_k (\tau)  &= \frac{H}{\sqrt{2k^3}} (1+ik\tau) e^{-ik\tau} , \\
v_k (\tau)  &= \frac{H}{\sqrt{2k^3}\sqrt{f}} (1+ik\tau) e^{-ik\tau} .
\end{align}
\label{eq:NSF-mode-functions}
\end{subequations}
Finally, since the trajectory is nearly straight, it is characterized by $\dot{\theta} / H \ll 1$, and the adiabatic direction is approximately parallel to $\delta\varphi$, see footonote~\ref{f2}. Therefore we can approximate the curvature perturbation $\zeta$ as 
\begin{equation}
\zeta \simeq - \frac{H}{\dot{\varphi}} \delta\varphi \hspace{1pt} .
\end{equation}

\subsubsection{Power-spectrum}

At lowest order in the \textit{in-in} expansion, the correlator $\braket{\zeta^2}$ is given by
\begin{align}
\braket{\zeta^2 (\tau)}^{(0)} &= (2\pi)^3 \delta^3 (\textbf{p}_1+\textbf{p}_2) \frac{H^2}{\dot{\varphi}^2} \lvert u_{p_1}(\tau) \lvert^2  \nonumber \\
&= (2\pi)^3 \delta^3 (\textbf{p}_1+\textbf{p}_2)   \frac{H^4}{2 p_1^3 \dot{\varphi}^2 } (1+p_1^2 \tau^2) \hspace{1pt} .
\end{align}
On super-horizon scales $-p_1 \tau \ll 1$, the previous expression becomes
\begin{equation}
\braket{\zeta^2 (0)}^{(0)} = (2\pi)^3 \delta^3 (\textbf{p}_1+\textbf{p}_2)   \frac{H^4}{2 p_1^3 \dot{\varphi}^2 } \hspace{2pt} .
\end{equation}

The next correction is given by
\begin{align}
\braket{\zeta^2 (\tau) }^{(2)} &= 4 \frac{H^2}{\dot{\varphi}^2} \text{Re} \Biggl\{  \int_{-\infty}^\tau d\tau_1 \int_{-\infty}^{\tau_1} d\tau_2  a^3(\tau_1) a^3(\tau_2) f^\prime(\tau_1) f^\prime(\tau_2) \dot{\theta}^2 \nonumber \\
&\phantom{=i}\Biggl[  \lvert u_{p_1} (\tau) \lvert^2     u_{p_1}(\tau_1) v^\prime_{p_1}(\tau_1)    u_{p_1}^* (\tau_2) {v^\prime}^*_{p_1}(\tau_2) -  u^2_{p_1}(\tau)    u_{p_1}^* (\tau_1)  v^\prime_{p_1}(\tau_1)   u_{p_1}^* (\tau_2) {v^\prime}^*_{p_1} (\tau_2) \Biggl] \Biggl\} ,
\label{eq:NSF-two-point-correlator-zeta}
\end{align}
where we have omitted the factor $(2\pi)^3 \delta^3 (\textbf{p}_1+\textbf{p}_2)$. Therefore dimensionless power-spectrum turns out to be
\begin{equation}
\mathcal{P}_\zeta = \frac{H^4}{(2\pi \dot{\varphi})^2} \biggl( 1 + \mathcal{C} f \frac{\dot{\theta}^2}{H^2}  \biggl) ,
\label{eq:NSF-power-spectrum}
\end{equation}
where
\begin{equation}
\mathcal{C} \simeq \frac{1}{2} \hspace{1pt} ,
\end{equation}
and where we have assumed $\dot{\theta}$, $f^\prime$ and $f$ nearly constant for simplicity.

\subsubsection{Bispectrum}

Let us first estimate naively the amplitude of the different contributions of the terms in the expansion of $\braket{\zeta^3}$, following the strategy of the previous section. Now, the function $f(\varphi)$ is integrated over time, but we will neglect this fact and consider $f$ and its derivatives constant during inflation. 

The lowest-order non-vanishing term is given by
\begin{equation}
\braket{\delta\varphi^3}^{(1)} \supseteq \braket{\delta\varphi^3 H_{A}^{(3)} } ,
\end{equation}
and its contributions to the bispectrum are
\begin{align}
\braket{\zeta^3}^{(1)}_{A_1} &\sim \frac{6 f f^\prime f^{\prime\prime}  - 6{f^{\prime}}^3 -f^2 f^{\prime\prime\prime}}{f^3}  \varepsilon_\theta \sqrt{\varepsilon_\varphi}  \mathcal{P}_\zeta^2 \frac{1}{p^6} \hspace{2pt} , \label{eq:NSF-bispectrum-1-A1-f} \\
\braket{\zeta^3}^{(1)}_{A_2} &\sim \frac{V_{\varphi\varphi\varphi}}{H} \sqrt{\varepsilon_\varphi} \mathcal{P}_\zeta^2 \frac{1}{p^6} \hspace{2pt} . \label{eq:NSF-bispectrum-self-interaction}
\end{align}
At second order, the correlator contains the term
\begin{equation}
\braket{\delta\varphi^3}^{(2)} \supseteq \braket{\delta\varphi^3 H_{B_1}^{(3)} H^{(2)} } ,
\end{equation}
and its contribution can be estimated as
\begin{equation}
\braket{\zeta^3}^{(2)}_{B_1} \sim \frac{2 {f^\prime}^3 - f f^\prime f^{\prime\prime} }{f^3} \sqrt{\varepsilon_\varphi} \varepsilon_\theta  \mathcal{P}_\zeta^2  \frac{1}{p^6} \hspace{2pt} .
\label{eq:NSF-bispectrum-2}
\end{equation}
At third order, the correlator can be estimated as
\begin{equation}
\braket{\delta\varphi^3}^{(3)} \supseteq \braket{\delta\varphi^3 H^{(3)} H^{(2)} H^{(2)} },
\end{equation}
where $H^{(3)}=H^{(3)}_A$ or $H^{(3)}=H^{(3)}_C$.
The contributions of these terms are given by
\begin{subequations}
\begin{align}
\braket{\zeta^3}^{(3)}_{A_1} & \sim \frac{6 f f^\prime f^{\prime\prime}  - 6{f^{\prime}}^3 -f^2 f^{\prime\prime\prime}}{f^3} \frac{{f^\prime}^2}{f^2} \sqrt{\varepsilon_\varphi} \varepsilon_\theta^{2}   \mathcal{P}_\zeta^2 \frac{1}{p^6} \hspace{2pt}, \label{eq:NSF-bispectrum-3-A1-f}  \\
\braket{\zeta^3}^{(3)}_{A_2} & \sim \frac{V_{\varphi\varphi\varphi}}{H} \frac{{f^\prime}^2}{f^{2}} \sqrt{\varepsilon_\varphi} \varepsilon_\theta   \mathcal{P}_\zeta^2 \frac{1}{p^6} \hspace{2pt}, \label{eq:NSF-bispectrum-3-A2} \\
\braket{\zeta^3}^{(3)}_{C_1} & \sim \frac{{f^\prime}^3}{f^3} \sqrt{\varepsilon_\varphi} \varepsilon_\theta  \mathcal{P}_\zeta^2 \frac{1}{p^6} \hspace{2pt}. \label{eq:NSF-bispectrum-3-C1} 
\end{align}
\end{subequations}
Finally, at fourth order,
\begin{equation}
\braket{\delta\varphi^3}^{(4)} \supseteq \braket{\delta\varphi^3 H_{B_1}^{(3)} H^{(2)} H^{(2)} H^{(2)} } ,
\end{equation}
and we obtain the contribution
\begin{equation}
\braket{\zeta^3}^{(4)}_{B_1} \sim \frac{2 {f^\prime}^2 - f f^{\prime\prime} }{f^2} \frac{{f^\prime}^3}{f^3} \sqrt{\varepsilon_\varphi} \varepsilon_\theta^2  \mathcal{P}_\zeta^2 \frac{1}{p^6} \hspace{2pt}.
\end{equation}

Unfortunately, all these terms are suppressed by the slow-roll parameters. If $V_\theta = 0$ (e.g., the potential~\eqref{eq:NSF-potential-phi}), one can show that the main suppression comes from 
$\varepsilon_\theta$ during slow-roll, due to previous considerations of section~\ref{sec:W-model-initials} (typically, $\varepsilon_\theta \lesssim 10^{-4}$ few $e$-folds after the beginning of inflation, see figure~\ref{fig:TANH2-potential-dynamics}).
Since we are considering a nearly single-field trajectory, this result is not surprising. Notice that there is no enhancement thanks to the non-canonical kinetic term, since $f\sim f^\prime \sim f^{\prime\prime}$ for the function~\eqref{eq:function-f}.  Hence, in this model it is impossible to obtain large non-Gaussianities. In the following sections, we will see that a more refined model produces more interesting results.
In particular we will discuss in section~\ref{2.4.3} and~\ref{sec:NSF-generalizations} the possibility of obtaining large non-Gaussianities, changing the function $f$.

\subsubsection{The \texorpdfstring{$\alpha$}{a} parameter}
\label{2.4.3}
Reintroducing the parameter $\alpha$ into the previously estimated contributions, one can show that some terms are enhanced, if $\alpha$ is sufficiently small. Remember that, in order to introduce the $\alpha$ parameter one should substitute $\varphi / M_{Pl} \rightarrow\varphi / (\sqrt{\alpha} M_{Pl} )$ into the non-canonical kinetic term of 
$\theta$ and into the potential $V$ of the Lagrangian~\eqref{eq:two-alpha-model-2}. Again, the parameter $\alpha$ enters in the correction to the 
power-spectrum~\eqref{eq:NSF-power-spectrum}, leading to the constraint~\eqref{eq:CT-alpha-constraint}. As $\alpha \rightarrow 0$, this implies a very tight constraint on the initial velocity $\dot{\theta}$. This constraint can be satisfied reducing the initial angular velocity $\dot{\theta}$, but practically we reduce to the single-field case and so we would not expect large non-Gaussianities. 

With the introduction of $\alpha$, the function $f$~\eqref{eq:function-f} becomes
\begin{equation}
f(\varphi) \equiv 6 \sinh^2 \frac{\varphi}{\sqrt{6 \alpha}} \hspace{2pt} ,
\label{eq:NSF-f-dependence}
\end{equation}
and this affects the background dynamics through the term $\omega_\theta = - \frac{f^\prime}{2} \theta^{\prime 2}$ in the background equation~\eqref{eq:phi-background-W-1}. Defining $\phi = \sqrt{\alpha} \varphi$, the term $\omega_\theta$ scales as $\alpha^{-1/2}$. Remember that this term should be small in order to have inflation, hence as $\alpha \rightarrow 0$ the inflationary region shrinks and the initial velocity $\theta^\prime (0) \rightarrow 0$ (see section~\ref{sec:W-model-initials} and figure~\ref{fig:two-init}). At the end, reducing $\alpha$ makes the trajectory a straight line in field space. Notice that this behaviour is insensitive to the the form of the potential, it depends only on the non-trivial field-space metric.

Let us consider the general potential 
\begin{equation}
V(\varphi) = \tanh^{2l} \frac{\varphi}{\sqrt{6 \alpha }} \hspace{2pt} \, ,
\label{eq:NSF-general-potential}
\end{equation}
where $l$ is a positive power. This potential is a typical one of the so called ``T-models" for single-field $\alpha$-attractors~\cite{Kallosh:2013hoa,Kallosh:2013yoa}.  
Let us see that, in order to produce large non-Gaussianities, $\alpha$ should be small. In the limit $\alpha \ll 1$ and during slow-roll, one can show from equation~\eqref{eq:SRST-W-phi} that the velocity $\varphi^\prime $ scales roughly as
\begin{equation}
\varphi^\prime \, \propto \, \sqrt{\alpha} \hspace{1pt} .
\label{eq:NSF-velocity-scale}
\end{equation}
Thanks to the previous scaling property~\eqref{eq:NSF-velocity-scale} and to~\eqref{eq:CT-alpha-constraint}, it is simple to show that, for the potential~\eqref{eq:NSF-general-potential}, the only terms that might be  enhanced through $\alpha$ are the self-interaction term $\braket{\zeta^3}^{(1)}_{A_2}$~\eqref{eq:NSF-bispectrum-self-interaction} and the third-order correction $\braket{\zeta^3}^{(3)}_{A_2}$~\eqref{eq:NSF-bispectrum-3-A2}. Let us first consider the term ~\eqref{eq:NSF-bispectrum-self-interaction}. In fact this self-interaction term is the same that one would get in single-field slow-roll models, since the second field does not participate to inflation and it cannot modify the self-interaction at first order. It  is well known that this term gives a negligible contribution w.r.t. to the leading-order result one obtains in single-field models of slow-roll inflation~\cite{Maldacena:2002vr,Acquaviva:2002ud}. 

Therefore, now let us focus on the third-order term~\eqref{eq:NSF-bispectrum-3-A2}. This term constitutes a correction of the self-interaction term, due to the presence of the second field (see diagram (b) of figure~\ref{fig:diagrams-2}). Notice that the coupling constant of this interaction between $\varphi$ and 
$\theta$ is $f^\prime /f \sqrt{\varepsilon_\theta} $. 
The term~\eqref{eq:NSF-bispectrum-3-A2} is explicitly given by
\begin{align}
\braket{\zeta^3}^{(3)}_{A_2} =& - (2\pi)^3 \delta^3 (\textbf{p}_1 + \textbf{p}_2 +\textbf{p}_3)  \frac{H^3}{\dot{\varphi}^3}  2 u_{p_1} (0) u_{p_2} (0) u_{p_3} (0)   \nonumber \\
&\text{Im} \biggl\{ \int^{0}_{-\infty} d\tau_1 \int^0_{-\infty} d\tau_2 \int^{\tau_2}_{-\infty} d\tau_3 \dot{\theta}^2  \prod_{i=1}^3 \bigl( a^3(\tau_i) f^\prime(\tau_i) \bigl) \sum_{j=1}^3 \biggl(  a(\tau_j) \frac{V_{\varphi\varphi\varphi}  (\tau_j) }{f^\prime (\tau_j)} \mathcal{A}_j \biggl) - \nonumber \\
&\int^{0}_{-\infty} d\tau_1 \int^{\tau_1}_{-\infty} d\tau_2 \int^{\tau_2}_{-\infty} d\tau_3 \dot{\theta}^2  \prod_{i=1}^3 \bigl( a^3(\tau_i) f^\prime(\tau_i) \bigl) \sum_{j=1}^3 \biggl(  a(\tau_j) \frac{V_{\varphi\varphi\varphi}  (\tau_j) }{f^\prime (\tau_j)} \mathcal{B}_j \biggl) + \, 5 \, \text{perm} \biggl\},
\end{align}
where
\begin{subequations}
\begin{align}
\mathcal{A}_1 &=  2 u_{p_1} (\tau_1)  u_{p_2} (\tau_1)  u_{p_3} (\tau_1) u^*_{p_3} (\tau_2) v^\prime_{p_3}(\tau_2) u^*_{p_3} (\tau_3) v^{\prime *}_{p_3} (\tau_3) \hspace{1pt},  \\
\mathcal{A}_2 &= u_{p_3}(\tau_1) v_{p_3}^\prime(\tau_1) u_{p_1}^* (\tau_2) u_{p_2}^* (\tau_2) u^*_{p_3} (\tau_3) v^{\prime *}_{p_3} (\tau_3) \bigl(u_{p_3} (\tau_2) + \text{c.c.}  \bigl)  ,  \\
\mathcal{A}_3 &=  u_{p_3} (\tau_1) v^\prime_{p_3} (\tau_1) v^{\prime *}_{p_3} (\tau_2) u^*_{p_1} (\tau_3) u^*_{p_2} (\tau_3) u^*_{p_3} (\tau_3) \bigl( u_{p_3}(\tau_2) + \text{c.c.}  \bigl)  ,  \\
\mathcal{B}_1 &= 2 u^*_{p_1}(\tau_1) u^*_{p_2}(\tau_1) u_{p_3}(\tau_1) u^*_{p_3}(\tau_2) v^\prime_{p_3}(\tau_2) u^*_{p_3}(\tau_3) v^\prime_{p_3}(\tau_3)  \hspace{1pt} ,  \\
\mathcal{B}_2 &=  v^\prime_{p_3}(\tau_1) u^*_{p_1}(\tau_2) u^*_{p_2}(\tau_2) u^*_{p_3}(\tau_3) v^{\prime *}_{p_3}(\tau_3) \bigl(u^*_{p_3}(\tau_1) u_{p_3}(\tau_2) +\text{c.c.}  \bigl)   ,  \\
\mathcal{B}_3 &= v^\prime_{p_3}(\tau_1) v^{\prime *}_{p_3}(\tau_2) u^*_{p_1}(\tau_3) u^*_{p_2}(\tau_3) u^*_{p_3}(\tau_3) \bigl( u^*_{p_3}(\tau_1) u_{p_3}(\tau_2) + \text{c.c.}  \bigl)  .  
\end{align}
\end{subequations}
In the equilateral limit, the previous expression becomes
\begin{equation}
\braket{\zeta^3}^{(3)}_{A_2} = \frac{3}{2^3} (2\pi)^3 \delta^3 (\textbf{p}_1 + \textbf{p}_2 +\textbf{p}_3)  \frac{H^3}{\dot{\varphi}^3}    \frac{H^2}{p^6}  \mathcal{I} \hspace{1pt} ,
\end{equation}
where we have defined $x_i \equiv - p \tau_i $ and
\begin{subequations}
\begin{align}
\mathcal{I} &= \text{Im} \biggl\{ \int_{0}^{+\infty} dx_1 \int_0^{+\infty} dx_2 \int_{x_2}^{+\infty} dx_3 \frac{\dot{\theta}^2}{H^2}  \prod_{i=1}^3 \left( \frac{f^\prime(x_i)}{x_i^3}  \right) \sum_{j=1}^3 \biggl(  \frac{1}{x_j} \frac{V_{\varphi\varphi\varphi}  (x_j) }{H f^\prime (x_j)} \mathcal{A}_j \biggl)  \nonumber \\
&\phantom{=i}-\int_{0}^{+\infty} dx_1 \int_{x_1}^{+\infty} dx_2 \int_{x_2}^{+\infty} dx_3 \frac{\dot{\theta}^2}{H^2}  \prod_{i=1}^3 \left( \frac{f^\prime(x_i)}{x_i^3}  \right) \sum_{j=1}^3 \biggl(  \frac{1}{x_j} \frac{V_{\varphi\varphi\varphi}  (x_j) }{H f^\prime (x_j)} \mathcal{B}_j \biggl) \biggl\}, \\
\tilde{u} (x) &\equiv (1-i x) e^{i x} , \\
\tilde{v} (x) &\equiv \frac{1}{\sqrt{f(x)}} (1-i x) e^{i x} , \\
\tilde{\mathcal{A}}_1 &=  2 \tilde{u}_{p} (x_1)  \tilde{u}_{p} (x_1)  \tilde{u}_{p} (x_1) \tilde{u}^*_{p} (x_2) \tilde{v}^\prime_{p}(x_2) \tilde{u}^*_{p} (x_3) \tilde{v}^{\prime *}_{p} (x_3) \hspace{1pt} ,  \\
\tilde{\mathcal{A}}_2 &= \tilde{u}_{p}(x_1) \tilde{v}_{p}^\prime(x_1) \tilde{u}_{p}^* (x_2) \tilde{u}_{p}^* (x_2) \tilde{u}^*_{p} (x_3) \tilde{v}^{\prime *}_{p} (x_3) \bigl( \tilde{u}_{p} (x_2) + \text{c.c.}  \bigl)  ,  \\
\tilde{\mathcal{A}}_3 &=  \tilde{u}_{p} (x_1) \tilde{v}^\prime_{p} (x_1) \tilde{v}^{\prime *}_{p} (x_2) \tilde{u}^*_{p} (x_3) \tilde{u}^*_{p} (x_3) \tilde{u}^*_{p} (x_3) \bigl( \tilde{u}_{p}(x_2) + \text{c.c.}  \bigl)  ,  \\
\tilde{\mathcal{B}}_1 &= 2 \tilde{u}^*_{p}(x_1) \tilde{u}^*_{p}(x_1) \tilde{u}_{p}(x_1) \tilde{u}^*_{p}(x_2) \tilde{v}^\prime_{p}(x_2) \tilde{u}^*_{p}(x_3) \tilde{v}^\prime_{p}(x_3) \hspace{1pt} ,  \\
\tilde{\mathcal{B}}_2 &=  \tilde{v}^\prime_{p}(x_1) \tilde{u}^*_{p}(x_2) \tilde{u}^*_{p}(x_2) \tilde{u}^*_{p}(x_3) \tilde{v}^{\prime *}_{p}(x_3) \bigl( \tilde{u}^*_{p}(x_1) \tilde{u}_{p}(x_2) +\text{c.c.}  \bigl)   ,  \\
\tilde{\mathcal{B}}_3 &= \tilde{v}^\prime_{p}(x_1) \tilde{v}^{\prime *}_{p}(x_2) \tilde{u}^*_{p}(x_3) \tilde{u}^*_{p}(x_3) \tilde{u}^*_{p}(x_3) \bigl( \tilde{u}^*_{p}(x_1) \tilde{u}_{p}(x_2) + \text{c.c.}  \bigl)  .  
\end{align}
\end{subequations}
Finally, the contribution of~\eqref{eq:NSF-bispectrum-3-A2} to $f_{\rm NL}$~\eqref{eq:fNL-def-2} is given by
\begin{equation}
\label{fNL3}
f_{\rm NL}^{(3)} =  \frac{5}{48}  \frac{\dot{\varphi}}{H^2}     \mathcal{I} \hspace{1pt} .
\end{equation}
If the potential $V$ is given by~\eqref{eq:NSF-general-potential}, we would expect an enhancement of order $\alpha^{-1}$. The details of this enhancement depends on the precise background dynamics. 
Nonetheless, this term constitutes an interaction between the fields and our estimate suggests that it can be enhanced through the parameter $\alpha$. In this case, the non-Gaussianities can become large, because the field-space is not one-dimensional and the existence of the second field is necessary. As we explained in this model both fields can be considered as massless, therefore we expect the resultant non-Gaussianity to be of the local type. In addition to this, notice that the enhancement is possible only if the function $f$ has the dependence~\eqref{eq:NSF-f-dependence} on $\alpha$. If $f$ had been independent of $\alpha$, we would have not been able to enhance the term 
$\braket{\zeta^3}^{(3)}_{A_2}$. This is an interesting result. 

In this case the second-field $\theta$ does not contribute to the background dynamics, and produces a negligible correction to the power spectrum, so up to  this order it would correspond to a single-field ``T-model". However the second scalar field $\theta$ has a  non-negligible impact on the level of primordial non-Gaussianities thanks to a non trivial interaction with the inflaton field. We  elaborate more on this in the following section.

\subsubsection{Generalizations}
\label{sec:NSF-generalizations}

Let us now study under which conditions $f$ could produce large non-Gaussianities even if the background trajectory is nearly indistinguishable from that of the single-field case. Notice that many of the previously estimated contributions contain the ratio $f^\prime / f$, and then a possibility to enhance the previous terms is to increase this ratio. 
Since we suppose that $\varphi$ is the inflaton, we should verify that at least the slow-roll parameters in the definition of $\nu$~\eqref{eq:nu-phi} are small.
This fact leads to the conditions
\begin{subequations}
\begin{align}
\varepsilon_\theta &\ll 1 \hspace{1pt}, \\
\frac{ {f^\prime}^2 }{f^2} \varepsilon_\theta &\ll 1 \hspace{1pt} , \\
\frac{f^{\prime\prime}}{f} \varepsilon_\theta &\ll 1 \hspace{1pt} .
\end{align}
\label{eq:NSF-f-slow-roll-generic}
\end{subequations}
The first of these conditions implies that the trajectory should be nearly a straight line, since $\theta^\prime$ should be small. Notice that if $\theta^\prime \equiv 0$ many of the estimated contributions vanish, since some of the couplings of the interactions in the Hamiltonian~\eqref{eq:H-0-H-INT} depend on $\theta^\prime$.

Consider the terms~\eqref{eq:NSF-bispectrum-2} and~\eqref{eq:NSF-bispectrum-3-C1}. Their contributions have the same magnitude and they can be approximated as
\begin{equation}
\braket{\zeta^3}^{(3)}  \sim  \frac{{f^\prime}^3}{f^3} \sqrt{\varepsilon_\varphi} \varepsilon_\theta  \mathcal{P}_\zeta^2 \frac{1}{p^6} \hspace{2pt} . 
\end{equation}
Notice that this term can be enhanced if the ratio $f^\prime / f$ is sufficiently large.
The slow-roll parameters are usually of order $10^{-3} -10^{-1}$. We suppose that $\varepsilon_\varphi$ and $\xi_\theta = {f^\prime}^2 f^{-2} \varepsilon_\theta$ are of this order of  magnitude. Hence, the slow-roll parameters suppress this amplitude by a factor $10^{-\frac{9}{2}} - 10^{-\frac{3}{2}}$. To compensate this, the ratio $f^\prime / f$ should satisfy 
\begin{equation}
\frac{f^\prime}{f} \sim 10^{\frac{3}{2}} -  10^{\frac{9}{2}} \hspace{1pt} .
\end{equation}
This is a rather tight condition, since it forces the function $f$ to have a nearly exponential behaviour. Indeed, assume that $f^\prime / f$ is a constant, then the previous condition becomes
\begin{equation}
f \, \propto \, e^{A \varphi},
\end{equation}
where $A \sim 30 - 30\cdot 10^3$. Remember that one has to integrate over the whole background trajectory. If we want to produce large non-Gaussianities, we need to require that the previous condition is satisfied for a sufficiently long period during inflation. 
In addition to this, the ratio $f^\prime / f$ enters into the background equation of motion~\eqref{eq:theta-background-W-1} and can break the slow-roll regime, accelerating the field $\theta$. 

Notice that the third derivative $f^{\prime\prime\prime}$ is not constrained by~\eqref{eq:NSF-f-slow-roll-generic}.
Hence, a function with $f^{\prime\prime\prime} / f \gg 1$ can enhance some of the previous terms, e.g. the contribution~\eqref{eq:NSF-bispectrum-1-A1-f} or~\eqref{eq:NSF-bispectrum-3-A1-f}. This is a viable way to obtain large non-Gaussianities for a nearly straight trajectory. Obviously, the function $f$ should be chosen properly (but the 
field-space metric is usually fixed \textit{a priori} for a given model, and we can modify only the potential term, as in our example). In conclusion, if we want to build a model with Lagrangian~\eqref{eq:tgh-lagrangian} that can produce large non-Gaussianities with a nearly straight trajectory, we need to consider a function $f$ with a high third derivative.

An analysis of multi-field models of inflation in the context supergravity-based multifield $\alpha$-attractors leading to a regime of almost-straight trajectories has been performed in~\cite{Achucarro:2017ing}. In these models, the angular component of the scalar field is frozen during inflation, and the evolution continues to be effectively one-dimensional (for a reason which is completely different from that of ~\cite{Kallosh:2013daa}). In this case the fields roll along the ridges of the potential without falling to the valleys. As a result, the authors of~\cite{Achucarro:2017ing} find that only  a low-level of primordial non-Gaussianity can be generated. There are some overlaps with the analysis performed here, given similar starting actions,  similarities but also differences (for example they use the $\Delta N$ formalism while her we use the {\it in-in} formalism). Indeed, as we already mentioned, the precise result and level of primordial non-Gaussianity in our case depend on the precise background evolution, and hence also on the precise initial conditions, and on the steepness of the potential in the angular direction. These are indeed possible exceptions to the result of small non-Gaussianity of~\cite{Achucarro:2017ing}. In our analysis we have been as general as possible, considering also these exceptions, see, e.g., our discussion in Section 2.4.4. and Table 1. Another possibility where the multi-field alpha-attractor regime may lead to non-gaussianities is cascade inflation described in ~\cite{Kallosh:2017wnt} (see also \cite{Kallosh:2017ced}). It would be interesting to investigate whether our analysis can be applied to such scenarios. 
Finally note that some of these exceptions might also be related to the previous analysis we have done in Section 2.3, where inflation is driven by the angular direction, possibly leading to natural inflation in the context of $\alpha$-attractor supergravity \cite{Linde:2018hmx, Yamada:2018nsk}.  In this case there might be an intermediate stage between two regimes driven by the radial component of the inflaton field~\cite{Linde:2018hmx} which might lead to large primordial non-Gaussianity according to our analysis of Section 2.3.

\subsubsection{Massive isocurvaton}
\label{2.4.5}

Let us suppose that the potential $V$ depends on both $\varphi$ and $\theta$. The Hamiltonian has the general expression~\eqref{eq:H-0-H-INT}. The new contributions to $f_{\rm NL}$, due to the dependence on $\theta$, are estimated in table~\ref{tab:bispectrum-contributions}. 

\begin{table}[t!]
\centering
\resizebox{\textwidth}{!}{%
\begin{tabular}{l | c c c c}
\toprule 
Couplings & $Z_2 B_1$ & $Z_2 B_2 $   &   $Z_1 B_1$ & $Z_2 Z_2 A_1$   \\
\midrule
$f_{\rm NL}$  &$\frac{V_{\varphi\theta}}{H^2} \frac{{f^\prime}^2 }{f^{5/2}} \sqrt{\varepsilon_\theta} \sqrt{\varepsilon_\varphi} $ & $\frac{V_{\varphi\theta}}{H^2} \frac{V_{\varphi\varphi\theta}}{H} \frac{1}{f} \sqrt{\varepsilon_\varphi}$   & $\frac{V_{\varphi\varphi\theta}}{H} \frac{f^\prime}{f^{3/2} }  \sqrt{\varepsilon_\theta} \sqrt{\varepsilon_\varphi} $ & $\frac{V_{\varphi\theta}^2}{H^4} \frac{{f^\prime}^3 }{f^3} \frac{1}{f} \varepsilon_\theta   \sqrt{\varepsilon_\varphi}$   \\
\midrule
\midrule
Couplings & $Z_2 Z_2 A_2 $ & $Z_2 Z_1 A_1 $   &   $Z_2 Z_1 A_2 $ & $Z_2 Z_2 C_1$   \\
\midrule
$f_{\rm NL}$ & $\frac{V_{\varphi\theta}^2}{H^4} \frac{V_{\varphi\varphi\varphi}}{H} \frac{1}{f} \sqrt{\varepsilon_\varphi}$  & $\frac{V_{\varphi\theta}}{H^2}  \frac{ {f^\prime}^4 }{f^4 } \frac{1}{\sqrt{f}} \varepsilon_\theta^{3/2}  \sqrt{\varepsilon_\varphi}  $ & $ \frac{V_{\varphi\theta}}{H^2}  \frac{V_{\varphi\varphi\varphi}}{H} \frac{f^\prime}{f^{3/2}}   \sqrt{\varepsilon_\theta}   \sqrt{\varepsilon_\varphi} $  & $\frac{V_{\varphi\theta}^2}{H^4} \frac{f^\prime}{f} \frac{1}{f}    \sqrt{\varepsilon_\varphi}$ \\
\midrule
\midrule
Couplings & $Z_2 Z_2 C_2 $ & $Z_2 Z_1 C_1 $   &   $Z_2 Z_1 C_2 $ & $ Z_1 Z_1 Z_1 B_1 $   \\
\midrule
$f_{\rm NL}$ & $\frac{V_{\varphi\theta}^2}{H^4} \frac{V_{\varphi\theta\theta}}{H} \frac{1}{f^2}   \sqrt{\varepsilon_\varphi} $  & $\frac{V_{\varphi\theta}}{H^2} \frac{{f^\prime}^2}{f^2} \frac{1}{\sqrt{f}}  \sqrt{\varepsilon_\theta}   \sqrt{\varepsilon_\varphi} $ & $ \frac{V_{\varphi\theta}}{H^2} \frac{V_{\varphi\theta\theta}}{H} \frac{f^\prime}{f^{5/2}}      \sqrt{\varepsilon_\theta} \sqrt{\varepsilon_\varphi} $  & $ \frac{{f^\prime}^5}{f^5} \varepsilon_\theta^2 \sqrt{\varepsilon_\varphi} $ \\
\midrule
\midrule
Couplings & $Z_1 Z_1 Z_2 D_1 $ & $Z_1 Z_2 Z_2 B_1 $   &   $Z_2 Z_2 Z_2 B_2 $ & $Z_2 Z_2 Z_2 D_1$   \\
\midrule
$f_{\rm NL}$ & $\frac{V_{\theta\theta\theta}}{H} \frac{V_{\varphi\theta}}{H^2} \frac{ {f^\prime}^2 }{f^4}  \varepsilon_\theta \sqrt{\varepsilon_\varphi} $  & $\frac{V_{\varphi\theta}^2}{H^4} \frac{{f^\prime}^3}{f^3} \frac{1}{f}  \varepsilon_\theta   \sqrt{\varepsilon_\varphi} $ & $ \frac{V_{\varphi\varphi\theta}}{H} \frac{V_{\varphi\theta}^3}{H^6} \frac{1}{f^2}  \sqrt{\varepsilon_\varphi} $  & $ \frac{V_{\theta\theta\theta}}{H} \frac{V_{\varphi\theta}^3}{H^6} \frac{1}{f^3} \sqrt{\varepsilon_\varphi} $ \\
\bottomrule
\end{tabular} }
\caption{Contribution to $f_{\rm NL}$ for different terms of the expansion of the bispectrum, using the \textit{in-in} formalism. The couplings $Z,A,B,C,D$ are defined in equation~\eqref{eq:H-0-H-INT}. The first entry $Z_2 B_1$ should be interpreted as the contribution of the correlator $\braket{\zeta^3 H^{(2)}_{Z_2} H^{(3)}_{B_1} }$ to the non-linearity parameter $f_{\rm NL}$. We have reported all the terms up to third order and some terms at fourth order. }
\label{tab:bispectrum-contributions}
\end{table}

Notice that there are terms that do not contain the slow-roll parameter $\varepsilon_\theta$. All these terms contain the coupling $Z_2$ and the corresponding Hamiltonian terms describe an interaction between $\delta\varphi$ and $\delta\theta$, mediated by the potential $V$. In addition to this, notice that many of the terms are multiplied by an inverse power of $f$. For the function~\eqref{eq:function-f} these are suppression factors, but for another metric these could lead to an enhancement, even if the ratio $f^\prime /f$ is of order unity. For example one can consider an exponential function like $\exp(-\varphi)$. Moreover, all the terms with coupling $A_1$ contain the third derivative $f^{\prime\prime\prime}$ and hence can be enhanced with an appropriate function $f$, as discussed in the previous subsection.

If we introduce the parameter $\alpha$, some additional terms which can be enhanced appear. For example, the terms $Z_2 Z_2 A_2$ and $Z_2 Z_2 Z_2 B_2$ scale as $\alpha^{-2}$ and hence large enhancement are possible. Obviously, the introduction of a potential for $\theta$ can alter significantly the dynamics of the system and the trajectory could be no longer nearly straight. These terms can nonetheless increase $f_{\rm NL}$, after a rotation of the basis $(\delta\varphi, \delta\theta)$, into that of the curvature and isocurvature perturbations.

Finally, let us summarize the results of our analysis. If we suppose that the dependence of $V$ on $(\varphi,\theta)$ is given by~(\ref{eq:two-alpha-model-2}), then terms that contains $V_\varphi$ are suppressed by $\exp \bigl(-\sqrt{ 2/3 } \varphi \bigl)$. Therefore, it is very difficult to produce large non-Gaussianities if the trajectory is nearly straight and we consider a model like~\eqref{eq:NSF-potential-phi}.
However, thanks to the parameter $\alpha$ one can produce non-negligible non-Gaussianities even in a model where there is a nearly single-field background trajectory, thanks to the term $\braket{\zeta^3}^{(3)}_{A_2}$~\eqref{eq:NSF-bispectrum-3-A2} (or other terms if the potential $V$ depends on both $\varphi$ and $\theta$). It is worth noticing that this enhancement relies on two facts. First, the parameter $\alpha$ enters into the potential $V$ and into the non-canonical kinetic term $f$ of the $\theta$ field. Secondly, the parameter $\alpha$ is not simply a rescaling of the field $\varphi$. This type of enhancement is one of the main results we have obtained in this section. If we let $f$ be a generic function and consider an appropriate potential $V$, we can obtain the production of large non-Gaussianities, even if the trajectory is practically of single-field type, through interactions between the fields (in particular thanks to the third derivative $f^{\prime\prime\prime}$ that is not directly constrained by the slow-roll conditions). 

\subsection{$f_{\rm NL}$ estimated via the $\delta N$ formalism: contribution to the local shape for a nearly straight trajectory.}
\label{finalsec}

As an example of the cases discussed in the previous section~\ref{2.4.5} let us consider the following potential
\begin{equation}
V(\varphi,\theta) =\bigl( 1 + M^2 \sin^2 \theta \bigl) \tanh^2 \left( \varphi / \sqrt{6} \hspace{1pt} \right)\, .
\end{equation}
If the field $\theta$ starts near a minimum of $\sin^2(\theta)$ then the background trajectories are nearly straight. If one wants to estimate the value of $f_{\rm NL}$ on superhorizon scales in the local configuration one can also use the $\delta N$ formalism. We rely on the analysis and the formulae derived by Ref.~\cite{Choi:2007su}, since they studied a Lagrangian similar to~\eqref{eq:two-alpha-model-2}. In particular, we adapt their equation (4.10) to our non-canonical kinetic term. 
This choice allows us to estimate the magnitude of local non-Gaussianities produced during the super-horizon evolution of the perturbations. We will not provide details of this calculation as it is tedious but fairly straightforward and we will simply focus on the results.

\begin{figure}[!h]
\centering
\includegraphics[width=.7\textwidth]{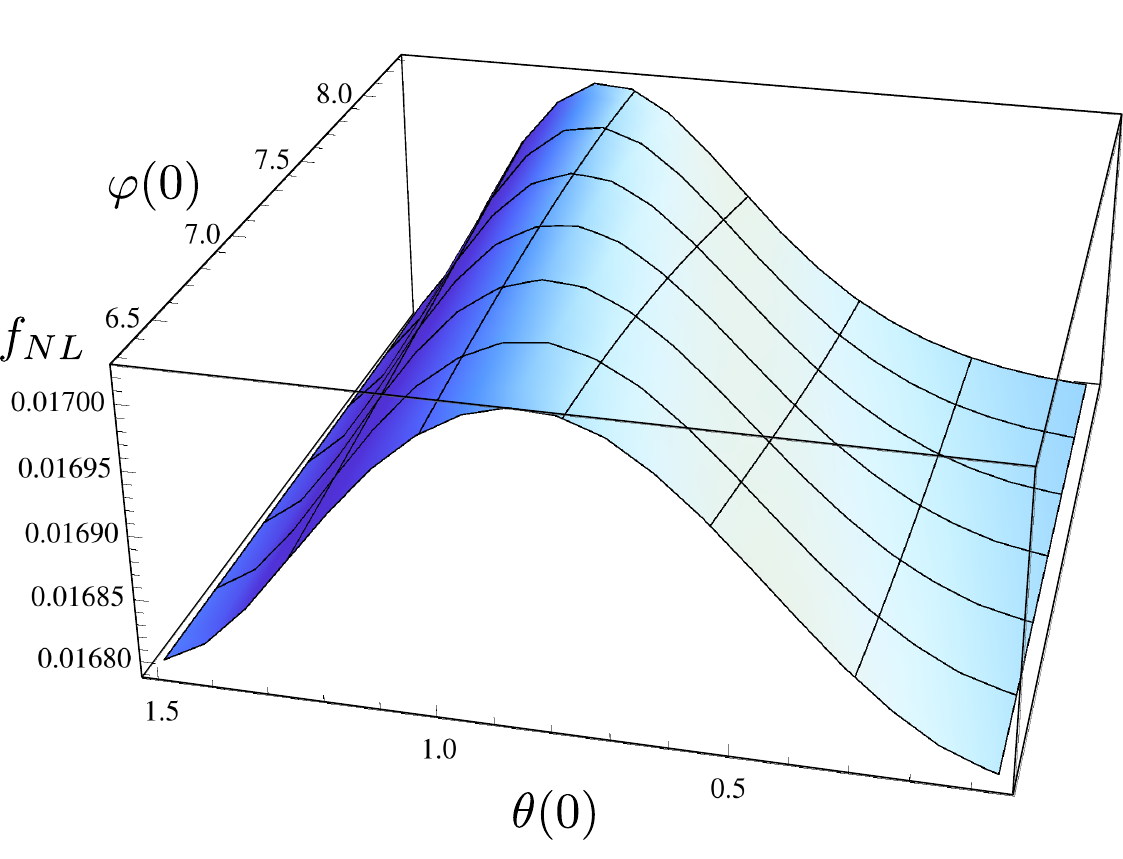}
\caption{$f_{\rm NL}$ for the model $V(\varphi,\theta) =\bigl( 1 + M^2 \sin^2 \theta \bigl) \tanh^2 \left( \varphi / \sqrt{6} \hspace{1pt} \right) $ estimated via the $\delta N$ formalism for $M^2=1$ and different initial configurations of the fields $\theta$ and $\phi$. The initial conditions are: $\varphi^\prime(0)=0$ and $\theta^\prime(0)=0$. The value of $f_{\rm NL}$ , in its local shape, is computed for a scale that exits the horizon 50 $e$-folds before the end of inflation.}
\label{fig:tanh-sin-W-fNL}
\end{figure}

\begin{figure}
\centering
\includegraphics[width=.7\textwidth]{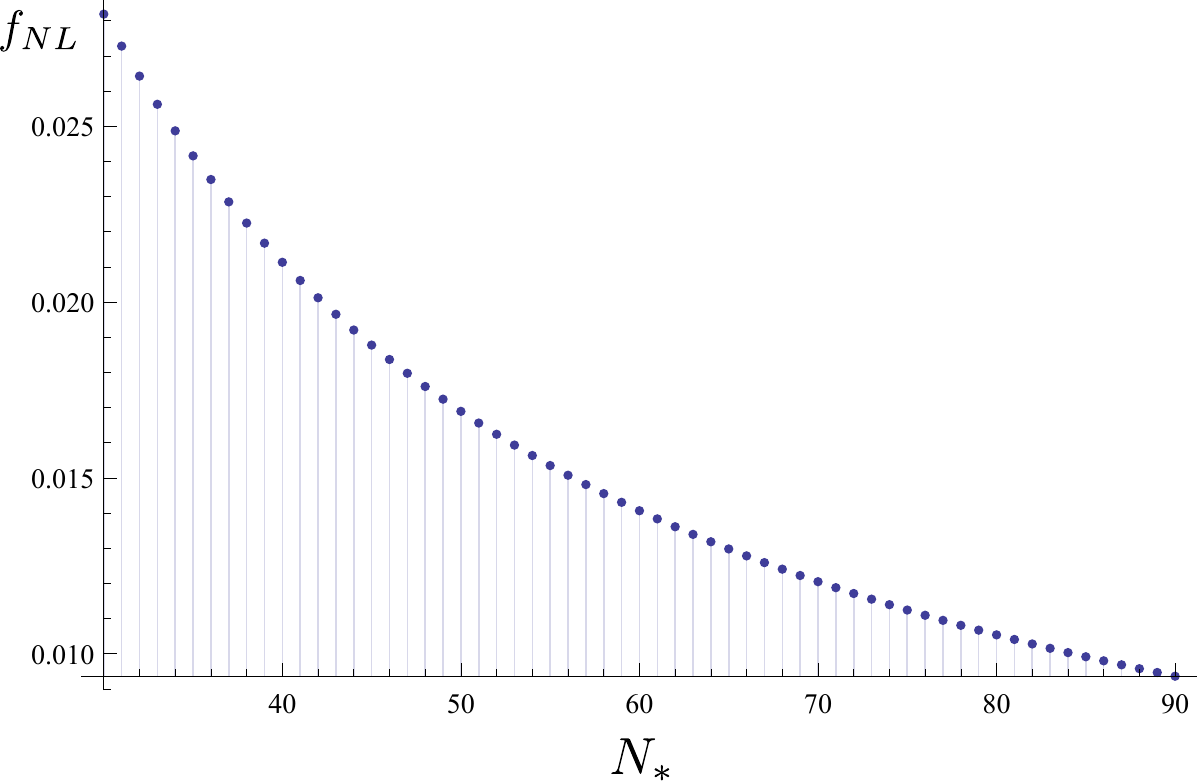}
\caption{$f_{\rm NL}$ for the model $V^{(2)}(\varphi,\theta) =\bigl( 1 + M^2 \sin^2 (\theta) \bigl) \tanh^2 \left( \varphi / \sqrt{6 \alpha} \hspace{1pt} \right) $ estimated using the $\delta N$ formalism (same as in Fig.~\ref{fig:tanh-sin-W-fNL} but introducing the parameter $\alpha$). In this case $\alpha=10^{-1}$, $M^2=1$, $\varphi(0)=7$, $\theta(0)=1.3$, $\varphi^\prime(0)=0$ and $\theta^\prime(0)=0$ as a function of the scale. $N_*$ is the number of $e$-folds from the end of inflation at which the scale crosses the horizon. }
\label{fig:tanh-sin-W-fNL-2}
\end{figure}

Let us focus on figures~\ref{fig:tanh-sin-W-fNL} and \ref{fig:tanh-sin-W-fNL-2}. We show values of $f_{\rm NL}$ in its local shape for a nearly straight trajectory. We investigate the effect of adding $\alpha$ to the potential. We show the case for $M^2=1$, but we have verified that the values of $f_{\rm NL}$ do not change much by changing this parameter, at least as far as $M^2=0.01$. The fields start from rest and inflation starts immediately in the region considered. We choose null initial velocities since their effect on the estimations of $f_{\rm NL}$ is marginal if inflation occurs. The initial velocity $\varphi^\prime$ has practically no effect on the inflationary dynamics because it is suppressed in a few $e$-folds before reaching the slow-roll regime. Conversely, the initial velocity $\theta^\prime$ affects mainly the viable initial conditions and only slightly the inflationary trajectory (remember that inflation occurs only if $\theta^\prime(0) \ll 1$). Therefore, considering the fields initially at rest is a good approximation. Varying $\alpha$ does not help to increase the amount of $f_{\rm NL}$. Hence, at least on super-horizon scales and for local non-Gaussianities, the parameter $\alpha$ cannot enhance $f_{\rm NL}$. It can be proven that the magnitude of $f_{\rm NL}$ is of order $\sqrt{\varepsilon_\varphi}$ evaluated at the horizon crossing for our model. Indeed, this factor gives the most important contribution to the expression of $f_{\rm NL}$ (4.10) in~\cite{Choi:2007su}. However, at least in the limit when $M^2$ vanishes to produce an almost straight trajectory (with the second field $\theta$ not participating to inflation) there is indeed the possibility to generate a higher level of primordial non-Gaussianity.

 \section{Conclusions}
 \label{sec:conclusions}
 
We have explored physical models that modify the laws of gravity at the scale of inflation to investigate what observable predictions can be obtained. In the context of inflation, any modification of gravity can be seen as an extra field that interacts with the inflaton. If the interaction is non-negligible, then non-Gaussian features may be generated during the slow-roll phase. 
Our motivation is to probe deviations from Einstein gravity at the highest energies available in cosmological observations, which are not far from the Planck energy scale. 

We have focused on inflationary models that arise in supergravity, the low-energy limit of string theory, which remains a most promising candidate to unify the laws of
Nature, but, to do so, it introduces modifications of Einstein gravity at high (possibly inflation-scale) energies.

To this aim, we investigated a fairly general model of inflation that can be mapped into supergravity theories and studied $\alpha$-attractor models. In particular, we focused on the two-field $\alpha$-attractor models described by the Lagrangian~\ref{eq:two-alpha-model-2}.
A first analysis of the background dynamics was performed in section~\ref{sec:W-model-initials}. We concentrated especially on the behaviour of the system in the region where inflation is most likely to occur. 
The most stringent condition to satisfy in order to have inflation is that the initial velocity of the angular field $\theta$ should be small. We estimated the magnitude of this smallness using both an analytical approximation and a numerical analysis. We showed that it is not difficult to obtain an inflationary period that lasts for at least $50-70$ $e$-folds.

In section~\ref{sec:CT}, we considered a simple curved trajectory in field space following~\cite{Chen:2009zp}. Indeed, we showed that for a representative model large primordial non-Gaussianities can be easily generated thanks to the $\alpha$ parameter (see equations~\ref{eq:CT-chens-fNL} and~\ref{eq:chen-term-enhancement-alpha}). These non-Gaussianities are produced by an interaction term between the fields (corresponding to transfer of non-Gaussianities from the isocurvature to the adiabatic mode), while we proved that other contributions should be negligible. 

In section~\ref{sec:NSF}, we studied another example of trajectory, a nearly straight one. This will be the most pessimistic situation in which the generation of non-Gaussianities is minimized. This type of trajectory is representative (at least in certain regions of the field space and under some assumptions) of the behaviour of the fields described via the Lagrangian~\eqref{eq:two-alpha-model-2}. In this case we found that 
there is the possibility for large non-Gaussianities to be obtained still by varying the parameter $\alpha$, thanks to interactions between the fields. The  interesting finding is that such 
non-Gaussianities can be produced even if the trajectory is almost straight and the second field does not participate actively to inflation (but its presence is nonetheless fundamental for this enhancement).
In addition to this, we found indications that it is possible to enhance the level of non-Gaussianity considering a model different from~\eqref{eq:two-alpha-model-2}, if the third derivative of the function $f$, defined in equation~\eqref{eq:tgh-lagrangian}, is high. Finally we considered even more general cases, summarized in Table~\ref{tab:bispectrum-contributions} (for some of which primordial non-Gaussianity can still be enhanced) and we gave a simple concrete example in Sec~\ref{finalsec}.

The main findings of our work is that we find no obvious obstacle in having models where a non-Gaussian signal can be large, even with a value $|f_{\rm NL}| \sim 1$ (of the equilateral or local type). 
This signal could be potentially measured with future surveys (see, e.g. \cite{Kamion,Carbone, Karagiannis,Castorina:2018zfk}), thus probing possible modifications of gravity at high energies. This opens up a window to physics at energy scales not reachable on Earth by any foreseeable future experiment.\\

\textit{Acknowledgements.---} Funding for this work was partially provided by the Spanish MINECO under projects AYA2014-58747-P  AEI/FEDER, UE, and MDM-2014-0369 of ICCUB (Unidad de Excelencia `Mar{\'\i}a de Maeztu'). NB and SM  We acknowledge partial financial support by ASI Grant No. 2016-24-H.0. Ê LV acknowledges support by  European Union's Horizon 2020 research and innovation programme ERC (BePreSySe, grant agreement 725327). N.B. and S.M. thank Gianguido Dall'Agata and Augusto Sagnotti for interesting discussions and for their valuable comments on the draft. R.J. thanks Joan Simon for useful comments on the draft.  We also would like to thank Andrei Linde for very useful correspondence over the preprint and on the results of this work, which helped us to improve it.

\appendix
\section{Appendix}
\label{AppA}
\subsection{Conformal invariance and conformal gauge}
\label{sec:conformal-invariance-gauge}

In the following we summarize for convenience some of the results of Ref.~\cite{Kallosh:2013hoa}  and some of the main properties of the so called ``T''-models for single-field inflation. Let us consider a simple but instructive toy-model with two scalar fields $\phi$ and $\chi$. 
Suppose the Lagrangian has the form~\cite{Kallosh:2013hoa}
\begin{equation}
\mathcal{L}_{toy} = \sqrt{-g} \biggl[ \frac{1}{2} \partial_\mu \chi \partial^\mu \chi + \frac{\chi^2}{12} R - \frac{1}{2} \partial_\mu \phi \partial^\mu \phi - \frac{\phi^2}{12} R - \frac{\lambda}{4} \bigl( \phi^2 - \chi^2 \bigl)^2 \biggl],
\label{eq:lagrangian-toy}
\end{equation}
where $\lambda$ is a dimensionless constant and $R$ is the Ricci curvature. Notice that the kinetic term for $\chi$ has the wrong sign, but this is not a problem as it will be clear below. The Lagrangian is locally conformally invariant, i.e. invariant under the transformations
\begin{subequations}
\begin{align}
&\widetilde{g}_{\mu\nu} = e^{-2\sigma (x)} g_{\mu\nu} \hspace{1pt} , \\
&\widetilde{\chi} = e^{\sigma (x) } \chi \hspace{1pt}, \\ 
&\widetilde{\phi} = e^{\sigma (x) } \phi \hspace{1pt}.
\end{align}
\end{subequations}
It is also invariant under a $\text{SO} (1,1)$ global symmetry, that acts like a boost between the fields. The field $\chi$ is called ``conformon". 

We can preserve conformal invariance and break $\text{SO}(1,1)$ symmetry, by introducing a function $F(\phi / \chi)$ into~\eqref{eq:lagrangian-toy}~\cite{Kallosh:2013hoa},
\begin{equation}
\mathcal{L} = \sqrt{-g} \biggl[ \frac{1}{2} \partial_\mu \chi \partial^\mu \chi + \frac{\chi^2}{12} R - \frac{1}{2} \partial_\mu \phi \partial^\mu \phi - \frac{\phi^2}{12} R - \frac{F(\phi / \chi)}{36} \left( \phi^2 - \chi^2 \right)^2 \biggl].
\label{eq:lagrangian-linde}
\end{equation}
In the limit $F(\phi / \chi) \rightarrow \text{const}$, we restore $\text{SO}(1,1)$ symmetry.
After having fixed the gauge $\chi = \sqrt{6}$ (which therefore brings to a breaking of conformal invariance), our Lagrangian in the Jordan frame becomes
\begin{equation}
\mathcal{L} = \sqrt{-g} \biggl[ \frac{R}{2} \biggl( 1- \frac{\phi^2}{6} \biggl)  - \frac{1}{2} \partial_\mu \phi \partial^\mu \phi - F\biggl(\frac{\phi}{\sqrt{6}}\biggl) \biggl( 1- \frac{\phi^2}{6} \biggl)^{\hspace{-3pt} 2 \hspace{1pt}}  \biggl].
\label{eq:T-model-jordan}
\end{equation}
Notice the presence of a UV cutoff at $\phi = \sqrt{6} M_{Pl}$ where $M_{Pl}$ denotes the reduced Planck mass. This is due to the fact that, if the field exceeded the value $\sqrt{6}$ (in Planck units), then the theory would describe antigravity instead of gravity (due to the change of the sign in front of $R$).
After the conformal transformation
\begin{equation}
g_{\mu\nu} \rightarrow \biggl( 1-\frac{\phi^2}{6}\biggl)^{\hspace{-3pt} -1} g_{\mu\nu} \hspace{1pt} ,
\label{eq:conformal-transformation-a} 
\end{equation}
and the introduction of a new field $\varphi$ that satisfies
\begin{equation}
\frac{d \varphi}{d\phi} = \biggl( 1-\frac{\phi^2}{6} \biggl)^{\hspace{-3pt} -1} ,
\label{eq:redefinition-a} 
\end{equation} 
the starting Lagrangian~\eqref{eq:lagrangian-linde} takes the form
\begin{equation}
\mathcal{L} = \sqrt{-g} \biggl[ \frac{R}{2} - \frac{1}{2} \partial_\mu \varphi \partial^\mu \varphi - F\biggl(\tanh \frac{\varphi}{\sqrt{6\alpha}}\biggl)  \biggl],
\label{eq:T-model-lagrangian}
\end{equation}
where we have introduced a dimensionless parameter $\alpha$, that allows us to shift the UV cutoff~\cite{Kallosh:2014rga}; this parameter arises naturally in supergravity theories and is related to the K\"{a}hler curvature of the manifold. For our purposes, it is sufficient to think of $\alpha$ as a parameter influencing the UV cutoff\footnote{The interested reader can look at Ref.~\cite{Kallosh:2013yoa}.}. We usually consider $\alpha = 1$, but it will be simple to reintroduce it. In fact, it is sufficient to rescale all the occurrences $\varphi / M_{Pl} \rightarrow \varphi / (\sqrt{\alpha} M_{Pl}) $ (notice that the kinetic term of $\varphi$ is not rescaled).

\begin{figure}
\centering
{\includegraphics[width=.475\textwidth]{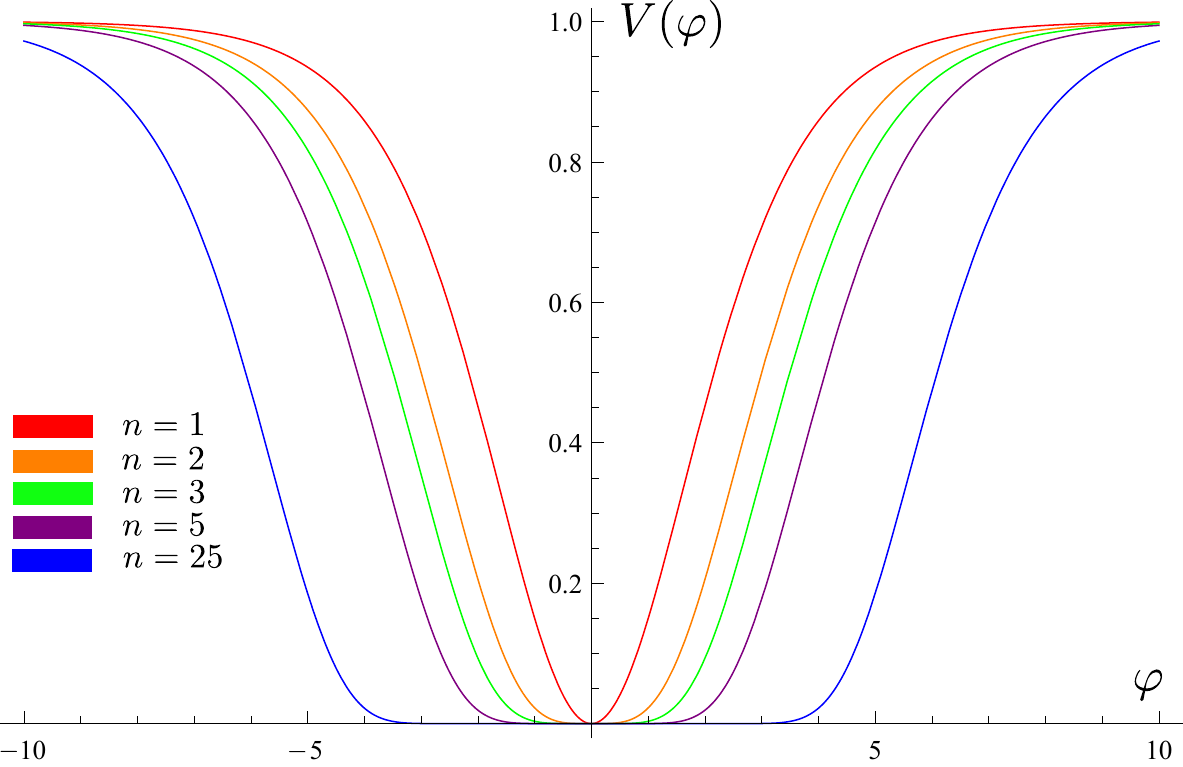}
\caption{Effects of the variation of $n$ with $\alpha=1$. Note that the potentials are very different for small values of $\varphi$, but they asymptotically coincide for $\varphi \rightarrow \infty$.}
\label{fig:t-model}}
\end{figure}

\begin{figure}
\centering
{\includegraphics[width=.475\textwidth]{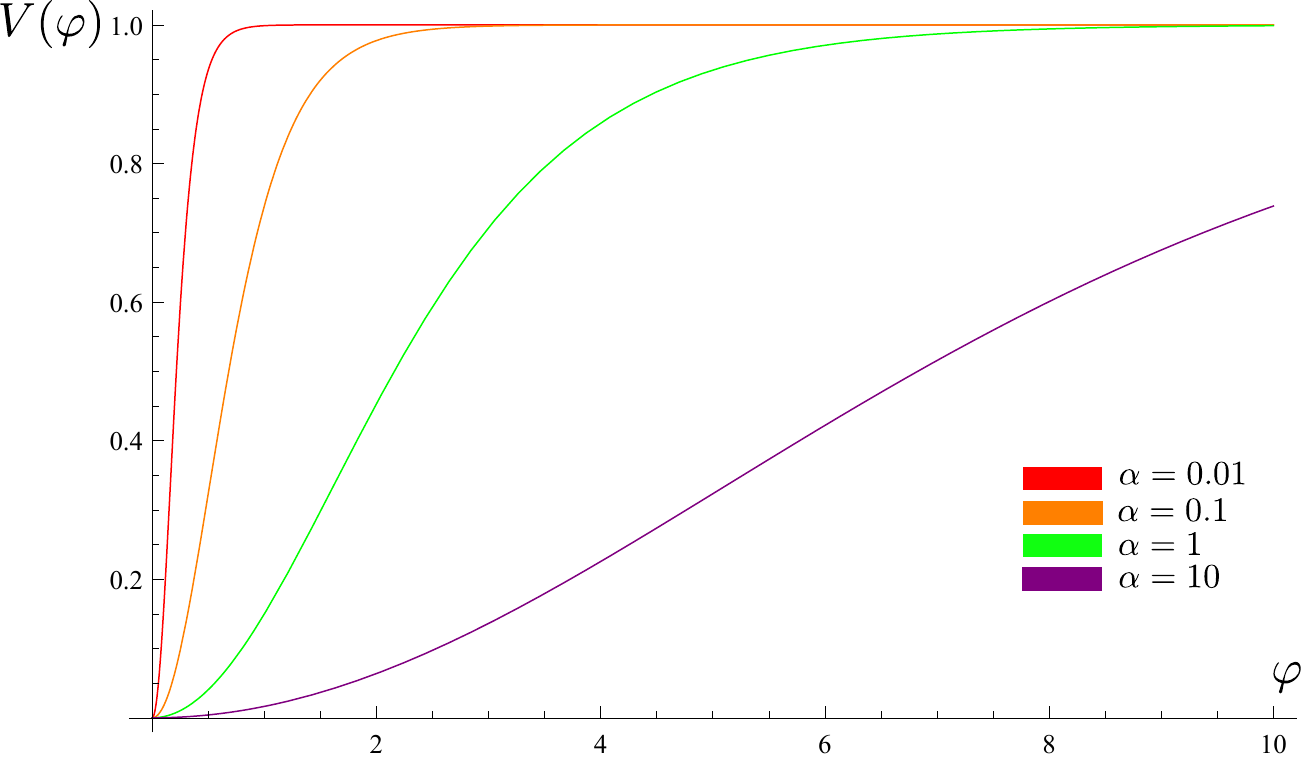}
\label{fig:t-model-alpha}}
\caption{Effects of the variation of $\alpha$ with $n=1$. Notice the effects of the variation of $\alpha$ on the inflationary region (i.e. the plateau of the potential).}
\label{fig:t-model-potentials}
\end{figure}

The hyperbolic tangent has the asymptotic behaviour $\tanh \varphi \rightarrow \pm 1$ in the limit $\varphi\rightarrow\pm\infty$ and then $F\rightarrow \text{const}$ in the same limit.
The function $F(\phi / \chi)$ is arbitrary, therefore it is possible to mimic an arbitrary chaotic inflation potential. We point out that we can obtain a generic potential starting from a conformal theory with spontaneously broken conformal invariance. On the other hand, this procedure looks sometimes a little bit artificial. Indeed, to obtain the simple term $\varphi^2$ we need to choose $F(\tanh \frac{\varphi}{\sqrt{6}}) \sim ( \tanh^{-1} \tanh \varphi )^2$. Alternatively, the function $F$ may be thought as a deviation from an inflationary theory driven by a cosmological constant. Thus, it is interesting to study the simplest case of monomial functions $F(\phi / \chi) = \lambda (\phi / \chi)^{2n}$.
In this case the transformed potential of~\eqref{eq:T-model-lagrangian} has the form
\begin{equation}
V(\varphi) = \lambda_n \tanh^{2n} \biggl( \frac{\varphi}{\sqrt{6\alpha}} \biggl) ,
\label{eq:t-model-potential}
\end{equation}
and these theories are known as T-models. The potential for different values of $n$ and $\alpha$ is shown in figure~\ref{fig:t-model-potentials}.

Inflation occurs in the limit $\varphi \gg 1$, where the potential is flat enough. In this limit, the potential is \begin{equation}
 V(\varphi) = \lambda \left( \frac{1- e^{-\sqrt{2/(3\alpha)}\varphi}}{1+ e^{-\sqrt{2/(3\alpha)}\varphi}} \right)^{\hspace{-3pt} 2n} = \lambda \left( 1- 4n e^{-\sqrt{\frac{2}{3\alpha}}\varphi} + \mathcal{O} \left( e^{-2\sqrt{\frac{2}{3\alpha}}\varphi} \right) \right).
 \label{eq:T-models-potential-expansion}
\end{equation}
Deviations from flatness are exponentially suppressed in the limit $\varphi \gg 1$.
Using
\begin{equation}
\frac{d\varphi}{dN} =  \frac{V^\prime}{V} =  n \sqrt{\frac{2}{3\alpha}} \left( \sinh\frac{\varphi}{\sqrt{6\alpha}} \cosh\frac{\varphi}{\sqrt{6\alpha}} \right) ^{\hspace{-3pt} -1}   ,
\end{equation}
and after integrating with respect to $dN$, we obtain the number of $e$-folds of inflation during slow-roll
\begin{equation}
N = \frac{3\alpha}{2n} \sinh^2 \frac{\varphi(N)}{\sqrt{6\alpha}} \hspace{1pt} .
\label{eq:T-models-efoldings}
\end{equation}

\begin{figure}[!t]
  \centering
    \includegraphics[width=0.7\textwidth]{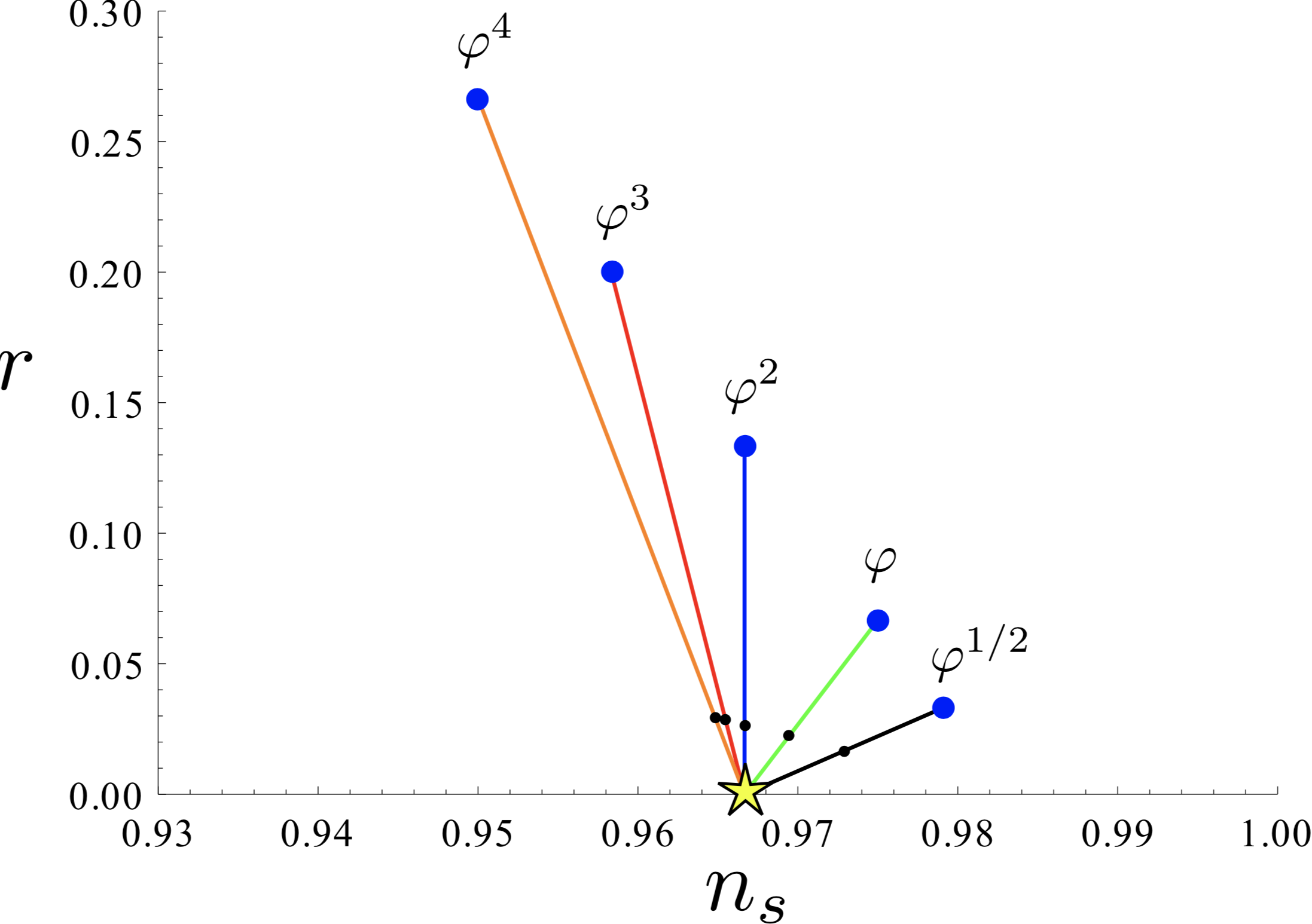}
      \caption{The $r-n_s$ plane for different potentials~\eqref{eq:t-model-potential} with $2n = \frac{1}{2},1,2,3,4$ for $N=60$. This plot can be obtained from equation~\eqref{eq:T-models-relationship} with $n$ and $N$ fixed and varying $\alpha$ (the full expressions for $n_s$~\eqref{eq:T-models-ns-full} and $r$~\eqref{eq:T-models-r-full} should be used). The yellow star is the attractor point $(1-2/N,0)$ for $\alpha\rightarrow 0$. In the limit $\alpha\rightarrow \infty$, we recover the predictions $(1-(n+1)/N,8n/N)$ of the monomial potential $\varphi^{2n}$ (see eq.~\eqref{eq:monomial-potentials-predictions-2}), which are represented by blue dots. Black dots correspond to $\alpha = 10$. The predictions of different potentials merge when $\alpha \lesssim 1$. \textit{Planck} data constrain the value of $\alpha$ to $\alpha \lesssim 14$ $95 \%$ C.L.~\cite{Ade:2015lrj}. }
      \label{fig:alpha-attractors}
\end{figure}

The slow-roll parameters are given by
\begin{subequations}
\begin{align}
\varepsilon &= \frac{1}{2} \biggl( \frac{V^\prime}{V} \biggl)^{\hspace{-3pt} 2} = \frac{n^2}{3\alpha} \left( \sinh\frac{\varphi}{\sqrt{6\alpha}} \cosh\frac{\varphi}{\sqrt{6\alpha}} \right) ^{\hspace{-3pt} -2} , \\
\eta &=  \frac{V^{\prime\prime}}{V}  = - \frac{2 n}{3\alpha}    \left( 1 - \frac{2n -1}{2} \sinh^{-2}\frac{\varphi}{\sqrt{6\alpha}}   \right) \left( \cosh \frac{\varphi}{\sqrt{6\alpha}} \right)^{\hspace{-3pt} -2}  ,
\end{align}
\end{subequations}
and it is immediate to calculate
\begin{subequations}
\begin{align}
&n_s - 1 = 2 \eta - 6 \varepsilon = -\frac{8n}{3\alpha} \left( \sinh \sqrt{\frac{2}{3\alpha}} \varphi \right)^{\hspace{-3pt} -2} \left( n+  \cosh \sqrt{\frac{2}{3\alpha}} \varphi\right) , \label{eq:T-models-ns-full} \\
&r = 16 \varepsilon = \frac{16 n^2}{3\alpha} \left( \sinh\frac{\varphi}{\sqrt{6\alpha}} \cosh\frac{\varphi}{\sqrt{6\alpha}} \right) ^{\hspace{-3pt} -2}  . \label{eq:T-models-r-full}
\end{align}
\end{subequations}
After some algebra, one can obtain the relation
\begin{equation}
r = \left( \frac{2n-1}{8n} - \frac{N}{6\alpha} - \frac{3}{8} \right)^{\hspace{-3pt} -1} \left( n_s -1 \right) .
\label{eq:T-models-relationship}
\end{equation}
This equation relates the predictions of the $\alpha$-attractor models in the $r-n_s$ plane. Varying $\alpha$ or $n$ it is possible to have different predictions. The previous relationship is shown in figure~\ref{fig:alpha-attractors} for some values of $n$ as a function of $\alpha$. Notice the attractor point and the behaviour for different potentials~\eqref{eq:t-model-potential} in the limit $\alpha\rightarrow 0$. The $\alpha$-attractor models owe their name to this $\alpha$-dependent behaviour.

If we consider the limit $\varphi \gg 1$, then we can approximate~\cite{Kallosh:2013hoa}
\begin{subequations}
\begin{align}
N &\simeq \frac{3\alpha}{8n} e^{\sqrt{\frac{2}{3\alpha}} \varphi(N) }  , \\
\varepsilon &\simeq \frac{3\alpha}{4 N^2} \hspace{2pt},\\
\eta &\simeq - \frac{1}{N} \hspace{1pt} , 
\end{align}
\label{eq:T-models-high-N-approximation}
\end{subequations}
and the predictions reduce to
\begin{subequations}
\begin{align}
&n_s - 1 \simeq - \frac{2}{N} \hspace{2pt}  , \label{eq:n-s-alpha} \\
&r  \simeq\frac{12 \alpha}{N^2} \hspace{2pt} , \label{eq:r-alpha}
\end{align}
\label{eq:T-models-predictions}
\end{subequations}
at lowest order in $1/N$. Let us suppose that the observable window during inflation consists of 60 $e$-foldings, then the models predict $n_s = 0.967$ which is in perfect agreement with the latest \textit{Planck} satellite result $n_s = 0.9645 \pm 0.0049$~\cite{Ade:2015lrj}. A variation of $\alpha$ corresponds to a change in $r$ and that insofar we have only an upper limit $r<0.07$ (95\% CL)~\cite{Array:2015xqh}.

For $\alpha =1$, the predictions coincide with that of the Starobinsky model~\cite{Starobinsky:1980te,Starobinsky:1983zz,Whitt:1984pd,Kofman:1985aw} and the Higgs inflation~\cite{Salopek:1988qh, Bezrukov:2007ep} at leading order in $1/N$. 

In the limit of large $\alpha$, the predictions coincide with those of chaotic inflation with potential $V\sim \varphi^{2n}$, that is~\cite{Kallosh:2013yoa,Kallosh:2014rga}
\begin{subequations}
\begin{align}
&n_s - 1 = -\frac{n+1}{N} \hspace{2pt} ,  \\
&r = \frac{8n}{N} \hspace{2pt}.
\end{align}
\label{eq:monomial-potentials-predictions-2}
\end{subequations}
This behaviour is easy to understand by looking at the expression for the inflaton potential~\eqref{eq:t-model-potential}, in the limit $\alpha\rightarrow \infty$.
 
\subsection{Two-field attractors}
\label{sec:two-field-attractors}

The Lagrangian~\eqref{eq:lagrangian-linde} can be trivially extended~\cite{Kallosh:2013daa} to include $n$ fields $\phi_I$, together with the conformon field $\chi$,
\begin{equation}
\mathcal{L} = \sqrt{-g} \biggl[ \frac{1}{2} \partial_\mu \chi \partial^\mu \chi + \frac{\chi^2}{12} R - \frac{1}{2} \partial_\mu \phi_I \partial^\mu \phi_I - \frac{\phi_I^2}{12} R - \frac{V(\phi_I / \chi)}{36} \bigl(\chi^2 - \phi_I^2 \bigl)^2 \biggl],
\label{eq:lagrangian-two-1}
\end{equation}
where a sum over $I$ is implicit.
This Lagrangian is locally conformally invariant and, in the limit $V\rightarrow \text{const} $, it has also a global $\text{SO} (n,1)$ symmetry.

We are interested only in the two-field case, then $I=1,2$. Hence, the previous Lagrangian explicitly reads
\begin{equation}
\mathcal{L} = \sqrt{-g} \biggl[ \frac{1}{2} \partial_\mu \chi \partial^\mu \chi + \frac{\chi^2}{12} R - \frac{1}{2} \partial_\mu \phi_1 \partial^\mu \phi_1 - \frac{1}{2} \partial_\mu \phi_2 \partial^\mu \phi_2  - \frac{\phi_1^2 + \phi_2^2}{12} R - \frac{V(\phi_1 / \chi, \phi_2 / \chi)}{36} \bigl(\chi^2 - \phi_1^2 -  \phi_2^2 \bigl)^2 \biggl]\, .
\end{equation}
Observing the form of this Lagrangian, it is natural to parametrize the fields as
\begin{subequations}
\begin{align}
\phi_1 &= \rho \cos \theta \hspace{1pt} , \\
\phi_2 &= \rho \sin \theta \hspace{1pt} .
\end{align}
\end{subequations}
Then we get
\begin{equation}
\mathcal{L} = \sqrt{-g} \biggl[\frac{\chi^2-\rho^2}{12} R + \frac{1}{2} \partial_\mu \chi \partial^\mu \chi - \frac{1}{2} \partial_\mu \rho \partial^\mu \rho - \frac{1}{2} \rho^2 \partial_\mu \theta \partial^\mu \theta - \frac{\left(\chi^2-\rho^2 \right)^2 }{36}  V \left( \frac{\rho}{\chi} \cos \theta , \frac{\rho}{\chi} \sin \theta \right) \biggl],
\end{equation}
and we can fix the gauge $\chi^2 - \rho^2 = 6$ to obtain
\begin{equation}
\mathcal{L} = \sqrt{-g} \biggl[\frac{R}{2} - \frac{1}{2} \frac{ \partial_\mu \rho \partial^\mu \rho}{1+ \frac{\rho^2}{6}} - \frac{1}{2} \rho^2 \partial_\mu \theta \partial^\mu \theta -  V \biggl( \frac{\rho}{\sqrt{\rho^2+6}} \cos \theta , \frac{\rho}{\sqrt{\rho^2+6}} \sin \theta \biggl) \biggl].
\end{equation}
Since there is a non-canonically normalized kinetic term for $\rho$, we introduce a new field $\varphi$ such that 
\begin{equation}
\rho = \sqrt{6} \sinh \frac{\varphi}{\sqrt{6}} \hspace{2pt} ,
\end{equation}
in order to get the following expression for the Lagrangian
\begin{equation}
\mathcal{L} = \sqrt{-g} \biggl[\frac{R}{2} - \frac{1}{2} \partial_\mu \varphi \partial^\mu \varphi -  3 \sinh^2 \frac{\varphi}{\sqrt{6}} \; \partial_\mu \theta \partial^\mu \theta \,  - V \biggl( \tanh\frac{\varphi}{\sqrt{6}} \cos \theta , \tanh\frac{\varphi}{\sqrt{6}} \sin \theta \biggl) \biggl].
\label{eq:two-alpha-model}
\end{equation}
Note that now $\varphi$ has a canonical kinetic term as opposed to $\theta$. Furthermore it is impossible to obtain a standard kinetic term for $\theta$ since its redefinition involves the field $\varphi$ and then a new non-canonical kinetic term for $\varphi$.  

Before concluding, let us relate explicitly the dependence on the initial variables $\phi_1$ and $\phi_2$ of the potential $V$ into Lagrangian~\eqref{eq:lagrangian-two-1} with that on $\phi$ and $\theta$ of the potential in~\eqref{eq:two-alpha-model}.
After the gauge fixing, the dependence on $\phi_1$ and $\phi_2$ of the original potential of Lagrangian~\eqref{eq:lagrangian-two-1} is given by
\begin{equation}
V = V\biggl( \frac{\phi_1}{\sqrt{6+\phi^2_1+\phi^2_2}} , \frac{\phi_2}{\sqrt{6+\phi^2_1+\phi^2_2}} \biggl) \equiv V(z_1,z_2) \hspace{1pt} ,
\end{equation}
where in the last equality we have defined the variables $z_1$ and $z_2$. Passing from Lagrangian~\eqref{eq:lagrangian-two-1} to~\eqref{eq:two-alpha-model} corresponds to
\begin{subequations}
\begin{align}
& z_1 \rightarrow \tanh \frac{\varphi}{\sqrt{6}} \; \cos\theta \hspace{1pt} , \\
& z_2 \rightarrow \tanh \frac{\varphi}{\sqrt{6}} \; \sin\theta \hspace{1pt} .
\end{align}
\end{subequations}

Finally, let us rewrite, for convenience, the full dependence of the potential $V$,
\begin{equation}
 V \biggl( \tanh\frac{\varphi}{\sqrt{6}} \, \cos \theta \hspace{2pt} , \hspace{2pt} \tanh\frac{\varphi}{\sqrt{6}} \, \sin \theta \biggl) .
\label{eq:F-full-dependence}
\end{equation}
Given this dependence, it is natural for the field $\varphi$ to have a T-model-like potential (see equation~\eqref{eq:t-model-potential}).

\section{Setup of the \textit{in-in} formalism}
\label{sec:setup-of-in-in}

At first, we write the Hamiltonian associated with the Lagrangian~\eqref{eq:tgh-lagrangian}. The conjugate momenta are
\begin{subequations}
\begin{align}
\pi_\varphi &= \frac{\partial \mathcal{L}}{\partial \dot{\varphi}} = a^3 \dot{\varphi} \hspace{1pt},  \\
\pi_\theta &= \frac{\partial \mathcal{L}}{\partial \dot{\theta}} =  a^3 f(\varphi) \dot{\theta} \hspace{1pt}, 
\end{align}
\label{eq:in-in-pi-momenta-W}
\end{subequations}
and thus the Hamiltonian is given by
\begin{equation}
\mathcal{H} = \frac{\pi_\varphi^2}{2 a^3} + \frac{\pi_\theta^2}{2 a^3 f} + \frac{a}{2} \left( \partial_i \varphi \partial_i \varphi + f \, \partial_i \theta \partial_i \theta \right) + a^3 V \hspace{1pt} . \label{eq:in-in-hamiltonian-W}
\end{equation}
One can easily check that the Hamilton equations of motion reproduce equations~\eqref{eq:W-eq-motion-phi} and~\eqref{eq:W-eq-motion-theta}. 

Following the \textit{in-in} prescriptions (see, e.g,~\cite{Weinberg:2006ac}) the fields and their conjugate momenta should be perturbed. Formally, the perturbed Hamiltonian can be written as
\begin{equation}
\mathcal{H} = \mathcal{H}^{(0)} + \sum_{i=1}^{+\infty} \delta \mathcal{H}^{(i)},
\end{equation}
where $\mathcal{H}^{(0)}$ is the background term and $\mathcal{H}^{(i)}$ contains terms of $i$-th order of the fields' perturbations.
For our purposes, it is sufficient to consider terms up to third order. 
Through a straightforward calculation, the different contributions are given by
\begin{align}
\mathcal{H}^{(0)} &= \frac{\pi_\varphi^2}{2 a^3} + \frac{\pi_\theta^2}{2 a^3 f}  + a^3 V \hspace{1pt} , \\
\delta\mathcal{H}^{(1)} &= \frac{\pi_\varphi}{a^3} \delta\pi_\varphi + \frac{\pi_\theta^2}{2 a^3} \left(\frac{1}{f} \right)^{\hspace{-0.3em} \prime} \delta\varphi + \frac{\pi_\theta}{a^3} \left(\frac{1}{f} \right) \delta\pi_\theta + a^3 \left( V_\varphi \delta\varphi + V_\theta \delta\theta \right) , \\
\delta\mathcal{H}^{(2)} &= \frac{\delta\pi_\varphi^2}{2 a^3} +  \frac{\pi_\theta^2}{4 a^3} \left(\frac{1}{f} \right)^{\hspace{-0.3em} \prime\prime} \delta\varphi^2 + \frac{\pi_\theta}{a^3} \left(\frac{1}{f} \right)^{\hspace{-0.3em} \prime} \delta\varphi \delta\pi_\theta + \frac{1}{2 a^3} \left(\frac{1}{f} \right)  \delta\pi_\theta^2 \nonumber \\
&\phantom{=}+ \frac{a}{2} \left( (\partial_i \delta\varphi)^2 + f (\partial_i \delta\theta)^2 \right) + a^3 \left( \frac{1}{2} V_{\varphi\varphi} \delta\varphi^2 + V_{\varphi\theta} \delta\varphi\delta\theta + \frac{1}{2} V_{\theta\theta} \delta\theta^2  \right) , \\ 
\delta\mathcal{H}^{(3)} &=  \frac{\pi_\theta^2}{12 a^3} \left(\frac{1}{f} \right)^{\hspace{-0.3em} \prime\prime\prime} \delta\varphi^3 + \frac{\pi_\theta}{2 a^3} \left(\frac{1}{f} \right)^{\hspace{-0.3em} \prime\prime} \delta\pi_\theta \delta\varphi^2 + \frac{1}{2 a^3} \left(\frac{1}{f} \right)^{\hspace{-0.3em} \prime} \delta\pi_\theta^2 \delta\varphi +  \frac{a}{2} f^\prime \delta\varphi \left(\partial_i \delta\theta \right)^2 \nonumber \\
&\phantom{=}+ \frac{a^3}{6} \left( V_{\varphi\varphi\varphi} \delta\varphi^3 + 3 V_{\varphi\varphi\theta} \delta\varphi^2 \delta\theta + 3 V_{\varphi\theta\theta} \delta\varphi \delta\theta^2 + V_{\theta\theta\theta} \delta\theta^3 \right) .
\end{align}
Obtaining equations~\eqref{eq:W-perturbations-phi} and~\eqref{eq:W-perturbations-theta} from this perturbed Hamiltonian is rather tedious. One can neglect the linear terms in this calculation since they do not contribute to the equations for the fluctuations. 

Notice that, for simplicity, we have not considered the perturbations of the metric.  Indeed, since our main goal is to study the non-Gaussianities produced through the interactions between the 
fields, we expect that the contribution from metric perturbations is subdominant.

The next step is the separation of the Hamiltonian into a kinetic part and an interaction part. After that, one can substitute the perturbations of the momenta with those of the fields via the relation 
\begin{equation}
\dot{\delta\phi}_I
 = \frac{\partial \mathcal{H}^{(0)}}{\partial \delta\pi_I} \hspace{1pt} .
\end{equation}
The resulting terms are given by
\begin{subequations}
\begin{align}
\mathcal{H}_0 &=  \frac{a^3}{2} \dot{\delta\varphi}^2 +  \frac{a^3 f^2 \dot{\theta}^2}{4} \left(\frac{1}{f} \right)^{\hspace{-0.3em} \prime\prime} \delta\varphi^2 + \frac{ a^3 f}{2} \dot{\delta\theta}^2  + \frac{a}{2} \left( (\partial_i \delta\varphi)^2 + f (\partial_i \delta\theta)^2 \right) \nonumber  \\
&\phantom{=}+ \frac{a^3}{2} \left(  V_{\varphi\varphi} \delta\varphi^2  +  V_{\theta\theta} \delta\theta^2  \right), \label{eq:H-0-kinetic} \\ 
\mathcal{H}^{(2)}_{INT} &=  -  a^3  f^\prime \dot{\theta}  \delta\varphi \dot{\delta\theta} + a^3  V_{\varphi\theta} \delta\varphi\delta\theta \label{eq:H-2-INT}  \\
&\equiv  \; a^3 Z_1  \delta\varphi \dot{\delta\theta} + a^3 Z_2 \delta\varphi\delta\theta \hspace{1pt} , \nonumber  \\ 
\mathcal{H}^{(3)}_{INT} &=  \frac{a^3 f^2 \dot{\theta}^2 }{12} \left(\frac{1}{f} \right)^{\hspace{-0.3em} \prime\prime\prime} \delta\varphi^3 + \frac{a^3 f^2 \dot{\theta} }{2} \left(\frac{1}{f} \right)^{\hspace{-0.3em} \prime\prime} \dot{\delta\theta} \delta\varphi^2 + \frac{a^3 f^2 }{2 } \left(\frac{1}{f} \right)^{\hspace{-0.3em} \prime} \dot{\delta\theta}^2 \delta\varphi +  \frac{a}{2} f^\prime \delta\varphi (\partial_i \delta\theta)^2 \nonumber \\
&\phantom{=}+ \frac{a^3}{6} \left( V_{\varphi\varphi\varphi} \delta\varphi^3 + 3 V_{\varphi\varphi\theta} \delta\varphi^2 \delta\theta + 3 V_{\varphi\theta\theta} \delta\varphi \delta\theta^2 + V_{\theta\theta\theta} \delta\theta^3 \right) \label{eq:H-3-INT}  \\
&\equiv  \; a^3 \Bigl[  A_1 \delta\varphi^3 + B_1 \dot{\delta\theta} \delta\varphi^2 + C_1 \bigl( \dot{\delta\theta}^2 \delta\varphi - a^{-2} \delta\varphi (\partial_i \delta\theta)^2 \bigl) \nonumber \\
&\phantom{=} + A_2 \delta\varphi^3 + B_2 \delta\varphi^2 \delta\theta + C_2 \delta\varphi \delta\theta^2 + D  \delta\theta^3 \Bigl] ,  \nonumber
\end{align}
\label{eq:H-0-H-INT}
\end{subequations}
where we have defined the coupling constants $Z_i,A_i,B_i,C_i,D$. The term $\mathcal{H}_0$ describes the kinetic part, $\mathcal{H}^{(2)}_{INT}$ and $\mathcal{H}^{(3)}_{INT}$ are respectively the second and third order interaction terms. All the fields are written in the interaction picture but, for brevity, we will not employ any distinctive notation. Notice that there are derivatives of the field perturbation $\delta\theta$ in the interaction terms, as opposed to $\delta\varphi$. 

Before proceeding, let us make some comments on the coupling constants (that indeed are not constant). The couplings can be divided into two classes, those which depend on $f$ (related to the non-canonical kinetic term) and those which depend on the derivatives of the potential. 
The latter couple the field perturbations directly without derivatives. Their magnitude is constrained by the slow-roll conditions that require a sufficiently flat potential (at least along the direction of motion of the inflaton).
The other terms, that depend on $f$, contain the background field velocity $\dot{\theta}$ except for two of them. 
Let us remind that $\dot{\theta} / H$ should be small during inflation in order to have slow-roll (see section~\ref{sec:W-model-initials}).
Note that, if $\dot{\theta} = 0$, some of the couplings vanish. The coupling of the first term of $\mathcal{H}^{(2)}_{INT}$ goes as $f^\prime \dot{\theta}$. Notice that in our model with $f$ given by~\eqref{eq:function-f}, we have $f \sim f^\prime \sim f^{\prime\prime}$ and then the couplings of the first two terms of $\mathcal{H}^{(3)}_{INT}$ behave respectively as $f \dot{\theta}^2$ and $f \dot{\theta}$. We are left with two terms that do not depend on $\dot{\theta}$. The first one is $ - \frac{a^3}{2}  f^{\prime} \dot{\delta\theta}^2 \delta\varphi$ and its coupling behaves as $f^\prime$. This is rather interesting, since there are no tight constraints\footnote{We are refering here to the constraints that derive from slow-roll conditions. For example, from the condition $\frac{1}{2} \dot{\varphi}^2 + \frac{f}{2} \dot{\theta}^2 \ll V$ it is clear that $f$ can be large if $\dot{\theta}$ is sufficiently small. On the other hand, if $f$ is small enough, $\dot{\theta}$ may be high (however $\ddot{\theta}$ should be small in order to have slow-roll). \label{fn:previous}} on the derivatives of $f$. The coupling of the last term $ \frac{a}{2} f^\prime \delta\varphi (\partial_i \delta\theta)^2$ seems to behave as $f^\prime$, but the spatial derivatives should be kept in consideration. In fact, in Fourier space, the coupling goes like $f^\prime k^2$ and, on large scales (when $f^\prime$ is large), $k$ is small and therefore one has to study the relative amplitude of the factors.
For a generic function $f$ the previous arguments should be reconsidered, accounting for the specific expression of $f$ (some terms with the derivatives might be enhanced or suppressed) and the velocity $\dot{\theta}$ might be very large (similarly to what disscussed in footnote~\ref{fn:previous}).

The equations of motion for the free fields (in the interaction picture), following from the kinetic Hamiltonian~\eqref{eq:H-0-kinetic}, are
\begin{align}
&\ddot{\delta\varphi} + 3 H \dot{\delta\varphi}  - \frac{1}{a^2} \nabla^2 \delta\varphi +  \frac{ \dot{\theta}^2}{2} f^2 \biggl(\frac{1}{f} \biggl)^{\hspace{-0.3em} \prime\prime} \delta\varphi +  V_{\varphi\varphi} \delta\varphi  = 0 \hspace{1pt}, \label{eq:in-in-W-phi} \\
&\ddot{\delta\theta} + 3 H \dot{\delta\theta}  - \frac{1}{a^2} \nabla^2 \delta\theta + \frac{f^\prime}{f} \dot{\varphi} \dot{\delta\theta} + \frac{1}{f} V_{\theta\theta} \delta\theta = 0 \hspace{1pt}. \label{eq:in-in-W-theta}
\end{align}
Interactions between the perturbations are absent unlike in equations~\eqref{eq:W-perturbations-phi} and~\eqref{eq:W-perturbations-theta}. 
The reason is that the previous equations derive from the free Hamiltonian and hence there are no interactions between the perturbations. 

As previously done, we quantize the interaction picture fields as
\begin{subequations}
\begin{align}
\delta \varphi_{\textbf{k}}^I &= u_\textbf{k} a_\textbf{k} + u_{-\textbf{k}}^* a_{-\textbf{k}}^\dagger \hspace{1pt} , \label{eq:W-delta-phi-in-in} \\
\delta \theta_{\textbf{k}}^I &= v_\textbf{k} b_\textbf{k} + v_{-\textbf{k}}^* b_{-\textbf{k}}^\dagger \hspace{1pt}, \label{eq:W-delta-theta-in-in}
\end{align}
\end{subequations}
with the standard commutation relations. We will assume that the Universe expansion is quasi-de Sitter-like. 

Let us start with the equation~\eqref{eq:in-in-W-phi} for $\delta\varphi$, which is the simplest one to solve. After having defined $q \equiv a u$ and having switched to conformal time, this equation becomes
\begin{equation}
q^{\prime\prime} - (\mathscr{H}^2+ \mathscr{H}^\prime ) q + k^2 q + a^2 \biggl(\frac{{f^\prime}^2}{f} - \frac{f^{\prime\prime}}{2} \biggl) \dot{\theta}^2 q + a^2  V_{\varphi\varphi} q = 0 \hspace{1pt} \, , 
\label{eq:in-in-W-phi-partial}
\end{equation}
where notice that here a prime denotes a derivative w.r.t. conformal time (except for $f^{\prime}$ which stands for $df/d\varphi$)  and $\mathscr{H}=a^{\prime}/a$. Let us define the following set of parameters,
\begin{subequations}
\begin{align}
\varepsilon &= \frac{ {\varphi^\prime}^2 +f {\theta^\prime}^2 }{2 \mathscr{H}^2} \equiv \varepsilon_\varphi + \varepsilon_\theta \hspace{1pt} , \label{eq:epsilon-phi-theta} \\
\eta_\varphi &\equiv \frac{a^2 V_{\varphi\varphi}}{3 \mathscr{H}^2} \hspace{1pt}, \label{eq:eta-phi} \\
\eta_\theta &\equiv \frac{a^2 V_{\theta\theta}}{3 f \mathscr{H}^2} \hspace{1pt}, \label{eq:eta-theta} \\
\xi_\varphi &\equiv  \frac{f^\prime \varphi^\prime}{f \mathscr{H}} = \frac{f^\prime}{f} \sqrt{2 \varepsilon_\varphi}  \hspace{1pt}, \label{eq:xi-phi} \\
\xi_\theta &\equiv  \frac{ {f^\prime}^2 {\theta^\prime}^2}{ 2  f \mathscr{H}^2 } =  \frac{ {f^\prime}^2 }{f^2} \varepsilon_\theta \hspace{1pt}, \label{eq:xi-theta} \\
\gamma_{\theta} &\equiv  \frac{f^{\prime\prime} {\theta^\prime}^2}{2 \mathscr{H}^2} = \frac{f^{\prime\prime}}{f} \varepsilon_\theta \hspace{1pt} . \label{eq:gamma-theta}
\end{align}
\label{eq:my-slow-roll}
\end{subequations}
In the next section, we will prove that all these parameters are actually slow-roll parameters for our model. For the moment, we are interested only in their definitions.

The previously defined parameters~\eqref{eq:my-slow-roll}, equation~\eqref{eq:in-in-W-phi-partial} can be recast in the form
\begin{equation}
q^{\prime\prime} + \left( k^2 - \frac{\nu^2 - \frac{1}{4}}{\tau^2} \right) q = 0 \hspace{1pt} ,
\end{equation}
where
\begin{equation}
\nu^2 = \frac{9}{4} + 3 \varepsilon -3 \eta_\varphi + \gamma_\theta -2\xi_\theta \hspace{1pt} ,
\label{eq:nu-phi}
\end{equation}
to first order in the slow-roll parameters.
Notice that the non-canonical kinetic term is responsible for the correction $\gamma_\theta -2\xi_\theta$. Finally, the mode function $u_k$ associated to $\delta\varphi$ is
\begin{equation}
u_k (\tau) = \frac{\sqrt{\pi}}{2} e^{i(\nu + \frac{1}{2}) \frac{\pi}{2}} (-\tau)^{\frac{3}{2}} H \, H^{(1)}_\nu (-k\tau) \hspace{1pt} .
\label{eq:u-k}
\end{equation}

Now, let us consider $\delta\theta$. Defining $p \equiv a v$, equation~\eqref{eq:in-in-W-theta} becomes
\begin{equation}
 p^{\prime\prime} +   \frac{f^{\prime}}{f} \varphi^\prime p^\prime + \biggl( k^2 - \left( \mathcal{H}^\prime + \mathcal{H}^2 \right) + \frac{a^2}{f} V_{\theta\theta}  -  \frac{f^{\prime}}{f} \varphi^\prime \mathcal{H} \biggl) p   = 0  \hspace{1pt} .
\end{equation}
The non-canonical kinetic term is responsible for the contribution $\frac{f^{\prime}}{f} \varphi^\prime p^\prime$. Its presence spoils the standard strategy to obtain the mode functions we have followed so far. In order to proceed, we need to introduce the rescaled function
\begin{equation}
\tilde{p} \equiv \sqrt{f} p \hspace{1pt} .
\end{equation}
We do not write down all the calculations (apart ffrom the non-trivial redefinition, they are similar to the former ones), but we are comforted to get once again the familiar result
\begin{equation}
\tilde{p}^{\prime\prime} + \left( k^2 - \frac{\tilde{\nu}^2 - \frac{1}{4}}{\tau^2} \right) \tilde{p}   = 0 \hspace{1pt}  ,
\end{equation}
where now
\begin{equation}
\tilde{\nu}^2 = \frac{9}{4} + 3 \varepsilon + \frac{f^{\prime\prime}}{f}  \varepsilon   - 3 \eta_\theta + \frac{3}{2} \xi_\varphi  + \frac{1}{2} \xi_\theta \hspace{1pt} ,
\label{eq:nu-theta}
\end{equation}
to first order in the slow-roll parameters.
At the end, the mode function $v_k$ has the form
\begin{equation}
v_k (\tau) = \frac{1}{\sqrt{f}} \frac{\sqrt{\pi}}{2} e^{i(\tilde{\nu} + \frac{1}{2}) \frac{\pi}{2}} (-\tau)^{\frac{3}{2}} H \, H^{(1)}_{\tilde{\nu}} (-k\tau) \hspace{1pt} .
\label{eq:v-k}
\end{equation}
Notice that the factor $1/ \sqrt{f}$ is a suppression factor in our model, since $f$ is given by~\eqref{eq:function-f} and inflation occurs usually for $\varphi \gg 1$. 

\section{Non-canonical kinetic term and slow-roll}
\label{sec:non-canonical-kinetic-slow-roll}

The standard definition of the slow-roll parameters should be suitably  modified when there is a non-canonical kinetic term. 
In this section, we will prove that the parameters in~\ref{eq:my-slow-roll} are slow-roll parameters in the sense that they are small during the slow-roll regime. 

It is instructive to begin with a more general situation. Let us start with the multi-field action
\begin{equation}
S = \int d^4 x \sqrt{-g} \biggl( \frac{R}{2} -\frac{1}{2} G_{IJ} \, g^{\mu\nu} \partial_\mu \phi^I \partial_\nu \phi^J - V(\phi_1, \phi_2)  \biggl),
\label{eq:S-multi}
\end{equation}
where $G_{IJ} = G_{IJ} (\phi_1, \phi_2)$ and capital Latin indices refer to field space ($I=\{1,2\}$). Notice that we allow the kinetic terms to be non-canonical (they are canonical if $G_{IJ} \equiv \delta_{IJ}$). In addition, notice that now we deal with two metrics: $g_{\mu\nu}$ is the metric in coordinate space and $G_{IJ}$ is the one in field space.
Equations of motion follow from the variation of the action~\eqref{eq:S-multi} with respect to the fields. This calculation is tedious but straightforward and the result in the FRW spacetime is~\cite{Lahiri:2005xj}
\begin{equation}
\ddot{\phi}^{I} - \Gamma^{I}_{JK} \partial_\mu \phi^J \partial^\mu \phi^K + 3 H \dot{\phi}^I + V^{,I} = 0 \hspace{1pt},
\label{eq:two-field-motion}
\end{equation}
where the connection coefficients (or Christoffel symbols) are defined as
\begin{equation}
\Gamma^{I}_{JK} = \frac{1}{2} G^{IL} \left(  G_{LK,J} + G_{LJ,K} - G_{JK,L}   \right) .
\label{eq:christoffel-definition}
\end{equation}
It is worth noticing that the term with Christoffel symbols $ \Gamma^{I}_{JK}$ accounts for the non-trivial structure of field space. The non-trivial field metric introduces additional couplings between the fields into the equations of motion.

The background equations of motion are
\begin{subequations}
\begin{align}
&\ddot{\phi}^I +\Gamma^{I}_{JK} \dot{\phi}^J \dot{\phi}^K + 3H \dot{\phi}^I + G^{IJ} V_{,J} = 0 \hspace{1pt}, \\
&H^2 = \frac{1}{3} \biggl( \frac{1}{2} G_{IJ} \dot{\phi}^I \dot{\phi}^J + V \biggl), \\
&\dot{H} = -  \frac{1}{2} G_{IJ} \dot{\phi}^I \dot{\phi}^J .
\end{align}
\end{subequations}
In order to have slow-roll inflation $\varepsilon\equiv - \dot{H} / H^2 \ll 1$, and hence the energy density of the universe should be dominated by the potential energy of the fields, 
\begin{equation}
\frac{1}{2} G_{IJ} \dot{\phi}^I \dot{\phi}^J \ll V \hspace{2pt}.
\label{eq:small-constraint}
\end{equation}
It is natural to define the slow-roll parameter
\begin{equation}
\varepsilon \equiv - \frac{\dot{H}}{H^2} = \frac{\mathcal{H}^2 - \mathcal{H}^\prime}{\mathcal{H}^2} = \frac{G_{IJ} {\phi^I}^\prime {\phi^J}^\prime}{2 \mathcal{H}^2} \hspace{1pt} ,
\end{equation}
which quantifies the variation over time of $H$.
Let us make a comment on the construction of the slow-roll parameters. These parameters are dimensionless and should be small. The first condition can be satisfied with appropriate powers of $H$. Indeed, the expansion rate $H$ provides both a mass scale and a time scale, hence every time derivative should be paired with $H^{-1}$ and also every field. The smallness condition can be 
obtained imposing constraints like~\eqref{eq:small-constraint}.

Other constraints to impose are 
\begin{align}
\lvert \ddot{\phi}^I \lvert \ll 3H \lvert \dot{\phi}^I \lvert \hspace{1pt},\label{eq:small-constraint-2}\\
\lvert \Gamma^{I}_{JK} \dot{\phi}^J \dot{\phi}^K \lvert \ll 3H \lvert \dot{\phi}^I \lvert \hspace{1pt}. \label{eq:small-constraint-3}
\end{align}
Using the first condition, it can be proven that 
\begin{equation}
\eta_{IJ} \equiv \frac{1}{3} \frac{V_{,IK} G^{KJ} }{H^2} = \frac{a^2}{3} \frac{V_{,IK} G^{KJ}}{\mathcal{H}^2}
\end{equation}
are slow-roll parameters (notice the analogy with the usual slow-roll parameter for single-field models od inflation).
Instead, from the second condition, one can define the slow-roll parameters
\begin{equation}
\Upsilon_{I} \equiv \frac{\Gamma^{I}_{JK} {\phi^J}^\prime {\phi^K}^\prime }{\mathscr{H}^2} \hspace{2pt}.
\end{equation} 

Now, let us return to our model with the metric~\eqref{eq:metric-f-matrix}.
The parameters~\eqref{eq:epsilon-phi-theta}-\eqref{eq:eta-theta} follow from the definitions just mentioned. Hence, they are small during slow-roll.
We are left with the parameters~\eqref{eq:xi-phi}-\eqref{eq:gamma-theta}. They are small during slow-roll because they can be rewritten in terms of $\varepsilon_\varphi$ and $\varepsilon_\theta$ and in our model $f\sim f^\prime \sim f^{\prime\prime}$. 
Notice that $\xi_\varphi$~\eqref{eq:xi-phi} is only the square root of a slow-roll parameter in our model.

Let us comment on the case of a generic function $f$. The slow-roll condition $\frac{1}{2} \dot{\varphi}^2 + \frac{f}{2} \dot{\theta}^2 \ll V$ does not forbid $\dot{\theta}$ to be large if $f$ is sufficiently small. On the other hand, $\frac{f {\theta^\prime}^2 }{2 \mathscr{H}^2} \ll 1$ and $\frac{f^\prime {\theta^\prime}^2 }{2 \mathscr{H}^2} \ll 1$, due to the conditions~\eqref{eq:small-constraint} and~\eqref{eq:small-constraint-3}, but there are no constraints on $\frac{f^{\prime\prime} {\theta^\prime}^2 }{2 \mathscr{H}^2}$ that might be large. As an example, suppose that the trajectory is a circle in field space with constant radius $\varphi(0)$ (this circumstance can be obtained with a confining potential along the radial direction). Moreover, suppose that $f(\varphi) = (\varphi -\varphi(0) )^2$. Notice that the velocity $\dot{\theta}$ is not limited by slow-roll conditions. Therefore $f(\varphi(0)) = f^\prime (\varphi(0)) = 0$, but $f^{\prime\prime} = 2$. Hence $\gamma_{\theta}$ is not a slow-roll parameter for this model\footnote{Many authors require that the field space is nearly flat and hence it has little curvature. With this assumption, one can show that $\xi_\varphi$, $\xi_\theta$ and $\gamma_{\theta}$ (defined in~\eqref{eq:xi-phi}-\eqref{eq:gamma-theta}) are indeed small. Let us stress the fact that this condition is independent of slow-roll and slow-turn conditions, but it can affect them. See for example the appendix of Ref.~\cite{Nakamura:1996da}.}. 


\begin{thebibliography}{100}

\bibitem[Bartolo et al.(2014)]{MG} 
N.~Bartolo, D.~Cannone, R.~Jimenez, S.~Matarrese and L.~Verde,
  ``Mild quasilocal non-Gaussianity as a signature of modified gravity during inflation,''
  Phys.\ Rev.\ Lett.\  {\bf 113} (2014) no.16,  161303
    [arXiv:1407.6719 [astro-ph.CO]].

\bibitem{strings}
  For some reviews, see, e.g.: M.~B.~Green, J.~H.~Schwarz and E.~Witten, ``Superstring Theory'', 2 vols., Cambridge, UK: Cambridge Univ. Press (1987); J.~Polchinski, ``String theory'', 2 vols. Cambridge, UK: Cambridge Univ. Press (1998);  C.~V.~Johnson, ``D-branes,'' USA: Cambridge Univ. Press (2003); B.~Zwiebach, ``A first course in string theory''
Cambridge, UK: Cambridge Univ. Press (2004); K.~Becker, M.~Becker and J.~H.~Schwarz,
``String theory and M-theory: A modern introduction'' Cambridge, UK: Cambridge Univ.
Press (2007); E.~Kiritsis, ``String theory in a nutshell'', Princeton, NJ: Princeton Univ. Press (2007).

\bibitem[Gangui et al.(1994)]{Gangui} 
A.~Gangui, F.~Lucchin, S.~Matarrese and S.~Mollerach,
  ``The Three point correlation function of the cosmic microwave background in inflationary models,''
  Astrophys.\ J.\  {\bf 430} (1994) 447
  [astro-ph/9312033].

\bibitem{Acquaviva:2002ud}
V.~Acquaviva, N.~Bartolo, S.~Matarrese and A.~Riotto,
  ``Second order cosmological perturbations from inflation,''
  Nucl.\ Phys.\ B {\bf 667} (2003) 119
    [astro-ph/0209156].

\bibitem{Maldacena:2002vr}
J.~M.~Maldacena,
  ``Non-Gaussian features of primordial fluctuations in single field inflationary models,''
  JHEP {\bf 0305} (2003) 013
    [astro-ph/0210603].
  
\bibitem[Mu{\~n}oz et al.(2015)]{Kamion} 
 J.~B.~Mu–oz, Y.~Ali-Ha•moud and M.~Kamionkowski,
  ``Primordial non-gaussianity from the bispectrum of 21-cm fluctuations in the dark ages,''
  Phys.\ Rev.\ D {\bf 92} (2015) no.8,  083508
  [arXiv:1506.04152 [astro-ph.CO]].

\bibitem{Pillepich:2006fj} 
 A.~Pillepich, C.~Porciani and S.~Matarrese,
  ``The bispectrum of redshifted 21-cm fluctuations from the dark ages,''
  Astrophys.\ J.\  {\bf 662} (2007) 1
  [astro-ph/0611126].
   
\bibitem{Cooray:2006km} 
A.~Cooray,
  ``21-cm Background Anisotropies Can Discern Primordial Non-Gaussianity,''
  Phys.\ Rev.\ Lett.\  {\bf 97} (2006) 261301
   [astro-ph/0610257].
  
\bibitem{Pajer:2012vz} 
E.~Pajer and M.~Zaldarriaga,
  ``A New Window on Primordial non-Gaussianity,''
  Phys.\ Rev.\ Lett.\  {\bf 109} (2012) 021302
  [arXiv:1201.5375 [astro-ph.CO]].
    
\bibitem{Ganc:2012ae} 
 J.~Ganc and E.~Komatsu,
  ``Scale-dependent bias of galaxies and mu-type distortion of the cosmic microwave background spectrum from single-field inflation with a modified initial state,''
  Phys.\ Rev.\ D {\bf 86} (2012) 023518
    [arXiv:1204.4241 [astro-ph.CO]].
    
\bibitem{Emami:2015xqa} 
R.~Emami, E.~Dimastrogiovanni, J.~Chluba and M.~Kamionkowski,
  ``Probing the scale dependence of non-Gaussianity with spectral distortions of the cosmic microwave background,''
  Phys.\ Rev.\ D {\bf 91} (2015) no.12,  123531
  [arXiv:1504.00675 [astro-ph.CO]].

\bibitem{Bartolo:2015fqz} 
N.~Bartolo, M.~Liguori and M.~Shiraishi,
  ``Primordial trispectra and CMB spectral distortions,''
  JCAP {\bf 1603} (2016) no.03,  029
  [arXiv:1511.01474 [astro-ph.CO]].
  
\bibitem{Berkin:1991nm} 
 A.~L.~Berkin and K.~I.~Maeda,
  ``Inflation in generalized Einstein theories,''
  Phys.\ Rev.\ D {\bf 44} (1991) 1691.
 
\bibitem{Mollerach:1991qx} 
S.~Mollerach and S.~Matarrese,
  ``Nonscale invariant density perturbations from chaotic extended inflation,''
  Phys.\ Rev.\ D {\bf 45} (1992) 1961.

\bibitem[{\'A}lvarez-Gaum{\'e} et al.(2010)]{Minimal} 
L.~Alvarez-Gaume, C.~Gomez and R.~Jimenez,
  ``Minimal Inflation,''
  Phys.\ Lett.\ B {\bf 690} (2010) 68
   [arXiv:1001.0010 [hep-th]].

\bibitem[{\'A}lvarez-Gaum{\'e} et al.(2011)]{MinimalII} 
L.~Alvarez-Gaume, C.~Gomez and R.~Jimenez,
  ``A Minimal Inflation Scenario,''
  JCAP {\bf 1103} (2011) 027
   [arXiv:1101.4948 [hep-th]].

\bibitem{Maldacena:2011nz} 
J.~M.~Maldacena and G.~L.~Pimentel,
  ``On graviton non-Gaussianities during inflation,''
  JHEP {\bf 1109} (2011) 045
  [arXiv:1104.2846 [hep-th]].

\bibitem{Soda:2011am} 
J.~Soda, H.~Kodama and M.~Nozawa,
  ``Parity Violation in Graviton Non-gaussianity,''
  JHEP {\bf 1108} (2011) 067
  [arXiv:1106.3228 [hep-th]].
    
\bibitem{Kallosh:2013wya}
R.~Kallosh and A.~Linde,
  ``Superconformal generalization of the chaotic inflation model $\frac{\lambda}{4} \phi^{4} - \frac{\xi}{2} \phi^{2}R$,''
  JCAP {\bf 1306} (2013) 027
    [arXiv:1306.3211 [hep-th]].

\bibitem{Kallosh:2013xya}
R.~Kallosh and A.~Linde,
  ``Superconformal generalizations of the Starobinsky model,''
  JCAP {\bf 1306} (2013) 028
    [arXiv:1306.3214 [hep-th]].

\bibitem{Kallosh:2013hoa}
R.~Kallosh and A.~Linde,
  ``Universality Class in Conformal Inflation,''
  JCAP {\bf 1307} (2013) 002
  [arXiv:1306.5220 [hep-th]].
  
\bibitem{Starobinsky:1980te}
A.~A.~Starobinsky,
  ``A New Type of Isotropic Cosmological Models Without Singularity,''
  Phys.\ Lett.\ B {\bf 91} (1980) 99

\bibitem{Choudhury:2014uxa}
  S.~Choudhury, ``Constraining N = 1 supergravity inflation with non-minimal Kaehler operators using $\delta$N formalism,''
  JHEP {\bf 1404} (2014) 105
  [arXiv:1402.1251 [hep-th]].


\bibitem{Ellis:2014opa}
  J.~Ellis, M.~A.~G.~Garc'a, D.~V.~Nanopoulos and K.~A.~Olive, ``Two-Field Analysis of No-Scale Supergravity Inflation,''
  JCAP {\bf 1501} (2015) 010
   [arXiv:1409.8197 [hep-ph]].

\bibitem{Kawai:2014gqa} 
  S.~Kawai and J.~Kim, ``Testing supersymmetric Higgs inflation with non-Gaussianity,''
  Phys.\ Rev.\ D {\bf 91}, no. 4, 045021 (2015)
[arXiv:1411.5188 [hep-ph]].
   

\bibitem{Arkani-Hamed:2015bza}
  N.~Arkani-Hamed and J.~Maldacena, ``Cosmological Collider Physics,''
  arXiv:1503.08043 [hep-th].


\bibitem{Baumann:2015xxa} 
  D.~Baumann, H.~Lee and G.~L.~Pimentel, ``High-Scale Inflation and the Tensor Tilt,''
  JHEP {\bf 1601}, 101 (2016)
    [arXiv:1507.07250 [hep-th]].
  



\bibitem{Hetz:2016ics}
  A.~Hetz and G.~A.~Palma,``Sound Speed of Primordial Fluctuations in Supergravity Inflation,''
  Phys.\ Rev.\ Lett.\  {\bf 117} (2016) no.10,  101301
[arXiv:1601.05457 [hep-th]].



\bibitem{Lee:2016vti}
  H.~Lee, D.~Baumann and G.~L.~Pimentel, ``Non-Gaussianity as a Particle Detector,''
  JHEP {\bf 1612} (2016) 040
  [arXiv:1607.03735 [hep-th]].


\bibitem{Bartolo:2017szm}
  N.~Bartolo and G.~Orlando, ``Parity breaking signatures from a Chern-Simons coupling during inflation: the case of non-Gaussian gravitational waves,''
  JCAP {\bf 1707} (2017) 034
   [arXiv:1706.04627 [astro-ph.CO]].

  
  

\bibitem{Carbone}
C.~Carbone, L.~Verde and S.~Matarrese,
  ``Non-Gaussian halo bias and future galaxy surveys,''
  Astrophys.\ J.\  {\bf 684} (2008) L1
  [arXiv:0806.1950 [astro-ph]].
  
\bibitem[Karagiannis et al.(2018)]{Karagiannis} 
D.~Karagiannis, A.~Lazanu, M.~Liguori, A.~Raccanelli, N.~Bartolo and L.~Verde,
  ``Constraining Primordial non-Gaussianity with Bispectrum and Power Spectum from Upcoming Optical and Radio Surveys,''
  arXiv:1801.09280 [astro-ph.CO].

\bibitem{Castorina:2018zfk} 
E.~Castorina, Y.~Feng, U.~Seljak and F.~Villaescusa-Navarro,
  ``Primordial non-Gaussianities and zero bias tracers of the Large Scale Structure,''
  arXiv:1803.11539 [astro-ph.CO].
  
  
  \bibitem{supergravity1}
D.~Z.~Freedman, P.~van Nieuwenhuizen and S.~Ferrara,``Progress Toward a Theory of Supergravity,''
  Phys.\ Rev.\ {\bf D 13} (1976) 3214;

\bibitem{supergravity2}
S.~Deser and B.~Zumino, ``Consistent Supergravity,''
  Phys.\ Lett.\ {\bf B 62} (1976) 335.\\

\bibitem{supergravityreviews}
For a comprehensive review see, e.g., 
 D.~Z.~Freedman and A.~Van Proeyen, ``Supergravity,''
  Cambridge, UK: Cambridge Univ. Pr. (2012) 607 p.
A recent elementary review is:
S.~Ferrara and A.~Sagnotti, ``Supergravity at 40: Reflections and Perspectives,''
  Riv.\ Nuovo Cim.\  {\bf 40} (2017) no.6,  1
   [J.\ Phys.\ Conf.\ Ser.\  {\bf 873} (2017) no.1,  012014]
  [arXiv:1702.00743 [hep-th]].
 

\bibitem{Ferrara:2013rsa}
S.~Ferrara, R.~Kallosh, A.~Linde and M.~Porrati,
  ``Minimal Supergravity Models of Inflation,''
  Phys.\ Rev.\ D {\bf 88} (2013) no.8,  085038
  [arXiv:1307.7696 [hep-th]].


\bibitem{Kallosh:2013yoa}
R.~Kallosh, A.~Linde and D.~Roest,
  ``Superconformal Inflationary $\alpha$-Attractors,''
  JHEP {\bf 1311} (2013) 198
  [arXiv:1311.0472 [hep-th]].
  
  \bibitem{Dall'Agata:2014oka}
G.~Dall'Agata and F.~Zwirner,
  ``On sgoldstino-less supergravity models of inflation,''
  JHEP {\bf 1412} (2014) 172
  [arXiv:1411.2605 [hep-th]].


\bibitem{Kallosh:2015lwa}
  R.~Kallosh and A.~Linde, ``Planck, LHC, and $\alpha$-attractors,''
  Phys.\ Rev.\ D {\bf 91} (2015) 083528
   [arXiv:1502.07733 [astro-ph.CO]].


\bibitem{Roest:2015qya}
  D.~Roest and M.~Scalisi, ``Cosmological attractors from ?-scale supergravity,''
  Phys.\ Rev.\ D {\bf 92} (2015) 043525
    [arXiv:1503.07909 [hep-th]].
 
 
 \bibitem{DiMarco:2017sqo}
  A.~Di Marco, P.~Cabella and N.~Vittorio, ``Reconstruction of $\alpha$-attractor supergravity models of inflation,''
  Phys.\ Rev.\ D {\bf 95} (2017) no.2,  023516
    [arXiv:1703.06472 [astro-ph.CO]].

\bibitem{DiMarco:2017zek}
  A.~Di Marco, P.~Cabella and N.~Vittorio, ``Constraining the general reheating phase in the $\alpha$-attractor inflationary cosmology,''
  Phys.\ Rev.\ D {\bf 95} (2017) no.10,  103502
 [arXiv:1705.04622 [astro-ph.CO]].


\bibitem{Kallosh:2017wku}
  R.~Kallosh, A.~Linde, D.~Roest, A.~Westphal and Y.~Yamada, ``Fibre Inflation and $\alpha$-attractors,''
  JHEP {\bf 1802} (2018) 117
 [arXiv:1707.05830 [hep-th]].

\bibitem{Kallosh:2013daa}
 R.~Kallosh and A.~Linde,
  ``Multi-field Conformal Cosmological Attractors,''
  JCAP {\bf 1312} (2013) 006
  [arXiv:1309.2015 [hep-th]].


\bibitem{Kallosh:2015zsa}
  R.~Kallosh and A.~Linde, ``Escher in the Sky,''
  Comptes Rendus Physique {\bf 16} (2015) 914
 [arXiv:1503.06785 [hep-th]].


\bibitem{Cecotti:1987sa}
  S.~Cecotti, ``Higher Derivative Supergravity Is Equivalent To Standard Supergravity Coupled To Matter. 1.,''
  Phys.\ Lett.\ B {\bf 190} (1987) 86.
  doi:10.1016/0370-2693(87)90844-6



\bibitem[Matarrese et al.(2000)]{MF} 
 S.~Matarrese, L.~Verde and R.~Jimenez,
  ``The Abundance of high-redshift objects as a probe of non-Gaussian initial conditions,''
  Astrophys.\ J.\  {\bf 541} (2000) 10
  [astro-ph/0001366].

\bibitem{Komatsu:2009kd} 
E.~Komatsu {\it et al.},
 ``Non-Gaussianity as a Probe of the Physics of the Primordial Universe and the Astrophysics of the Low Redshift Universe,''
 arXiv:0902.4759 [astro-ph.CO].

\bibitem[Dalal et al.(2008)]{Dalal} 
N.~Dalal, O.~Dore, D.~Huterer and A.~Shirokov,
  ``The imprints of primordial non-gaussianities on large-scale structure: scale dependent bias and abundance of virialized objects,''
  Phys.\ Rev.\ D {\bf 77} (2008) 123514
  [arXiv:0710.4560 [astro-ph]].

\bibitem[Matarrese \& Verde(2008)]{HaloBiasI} 
S.~Matarrese and L.~Verde,
  ``The effect of primordial non-Gaussianity on halo bias,''
  Astrophys.\ J.\  {\bf 677} (2008) L77
  [arXiv:0801.4826 [astro-ph]].

\bibitem[Verde \& Matarrese(2009)]{HaloBiasII} 
L.~Verde and S.~Matarrese,
  ``Detectability of the effect of Inflationary non-Gaussianity on halo bias,''
  Astrophys.\ J.\  {\bf 706} (2009) L91
  [arXiv:0909.3224 [astro-ph.CO]].

\bibitem{Ade:2015ava} 
  P.~A.~R.~Ade {\it et al.} [Planck Collaboration], 
  ``Planck 2015 results. XVII. Constraints on primordial non-Gaussianity,''
  Astron.\ Astrophys.\  {\bf 594} (2016) A17
  [arXiv:1502.01592 [astro-ph.CO]].


\bibitem{Choi:2007su}
K.~Y.~Choi, L.~M.~H.~Hall and C.~van de Bruck,
  ``Spectral Running and Non-Gaussianity from Slow-Roll Inflation in Generalised Two-Field Models,''
  JCAP {\bf 0702} (2007) 029
  [astro-ph/0701247].

\bibitem{Ellis:2013nxa}
J.~Ellis, D.~V.~Nanopoulos and K.~A.~Olive,
  ``Starobinsky-like Inflationary Models as Avatars of No-Scale Supergravity,''
  JCAP {\bf 1310} (2013) 009
  [arXiv:1307.3537 [hep-th]].

\bibitem{Yamaguchi:2011kg}
M.~Yamaguchi,
  ``Supergravity based inflation models: a review,''
  Class.\ Quant.\ Grav.\  {\bf 28} (2011) 103001
  [arXiv:1101.2488 [astro-ph.CO]].


\bibitem{Antoniadis:2014oya}
  I.~Antoniadis, E.~Dudas, S.~Ferrara and A.~Sagnotti, ``The VolkovÐAkulovÐStarobinsky supergravity,''
  Phys.\ Lett.\ B {\bf 733} (2014) 32
    [arXiv:1403.3269 [hep-th]].



\bibitem{Ferrara:2014kva}
  S.~Ferrara, R.~Kallosh and A.~Linde, ``Cosmology with Nilpotent Superfields,''
  JHEP {\bf 1410} (2014) 143
  [arXiv:1408.4096 [hep-th]].




\bibitem{Ferrara:2015cwa}
S.~Ferrara and A.~Sagnotti,
  ``Supersymmetry and Inflation,''
  Int.\ J.\ Mod.\ Phys.\  {\bf 1} (2017) 29
  [arXiv:1509.01500 [hep-th]].
  
  \bibitem{S}
  S.~Sugimoto, ``Anomaly cancellations in type I D9-D9-bar system and the USp(32)  string theory,''
Prog.\ Theor.\ Phys.\  {\bf 102} (1999) 685 [arXiv:hep-th/9905159];

\bibitem{ADS}
I.~Antoniadis, E.~Dudas and A.~Sagnotti, ``Brane supersymmetry breaking,''
Phys.\ Lett.\ {\bf B 464} (1999) 38 [arXiv:hep-th/9908023];

\bibitem{A}
C.~Angelantonj, ``Comments on open-string orbifolds with a non-vanishing B(ab),''
Nucl.\ Phys.\ {\bf B 566} (2000) 126 [arXiv:hep-th/9908064];

\bibitem{Scalisi1}
M.~Scalisi , ``Cosmological $\alpha$ -attractors and de Sitter landscape""
JHEP 1512 (2015) 134 [arXiv:1506.01368]

\bibitem{carrasco}
J.J.M.~ Carrasco, R.~Kallosh, A.~Linde, ``Minimal supergravity inflation'',
Phys.Rev. D93 (2016) no.6, 061301 [arXiv:1512.00546] 


\bibitem{Mourad:2017rrl}
  J.~Mourad and A.~Sagnotti, ``An Update on Brane Supersymmetry Breaking,''
  arXiv:1711.11494 [hep-th].
   
  

\bibitem{Ade:2015lrj}
P.~A.~R.~Ade {\it et al.} [Planck Collaboration],
  ``Planck 2015 results. XX. Constraints on inflation,''
  Astron.\ Astrophys.\  {\bf 594} (2016) A20
  [arXiv:1502.02114 [astro-ph.CO]].

\bibitem{0067-0049-208-2-20}
C.~L.~Bennett {\it et al.} [WMAP Collaboration],
  ``Nine-Year Wilkinson Microwave Anisotropy Probe (WMAP) Observations: Final Maps and Results,''
  Astrophys.\ J.\ Suppl.\  {\bf 208} (2013) 20
  [arXiv:1212.5225 [astro-ph.CO]].

\bibitem{Kallosh:2013tua}
R.~Kallosh, A.~Linde and D.~Roest,
  ``Universal Attractor for Inflation at Strong Coupling,''
  Phys.\ Rev.\ Lett.\  {\bf 112} (2014) no.1,  011303
  [arXiv:1310.3950 [hep-th]].

\bibitem{Giudice:2014toa}
G.~F.~Giudice and H.~M.~Lee,
  ``Starobinsky-like inflation from induced gravity,''
  Phys.\ Lett.\ B {\bf 733} (2014) 58
  [arXiv:1402.2129 [hep-ph]].

\bibitem{Pallis:2013yda}
C.~Pallis,
  ``Linking Starobinsky-Type Inflation in no-Scale Supergravity to MSSM,''
  JCAP {\bf 1404} (2014) 024
   Erratum: [JCAP {\bf 1707} (2017) no.07,  E01]
  [arXiv:1312.3623 [hep-ph]].

\bibitem{Kallosh:2014rga}
R.~Kallosh, A.~Linde and D.~Roest,
  ``Large field inflation and double $\alpha$-attractors,''
  JHEP {\bf 1408} (2014) 052
  [arXiv:1405.3646 [hep-th]].

\bibitem{Kallosh:2014laa}
R.~Kallosh, A.~Linde and D.~Roest,
``The double attractor behavior of induced inflation,''
   JHEP {\bf 1409} (2014) 062
  [arXiv:1407.4471 [hep-th]].
  
  \bibitem{Linde:2017pwt}
  A.~Linde,
 ``On the problem of initial conditions for inflation,''
  arXiv:1710.04278 [hep-th].
  
  \bibitem{Achucarro:2017ing}
  A.~Achœcarro, R.~Kallosh, A.~Linde, D.~G.~Wang and Y.~Welling,
  `Universality of multi-field $\alpha$-attractors,''
  JCAP {\bf 1804} (2018) no.04,  028
  doi:10.1088/1475-7516/2018/04/028
  [arXiv:1711.09478 [hep-th]].
  
\bibitem{Kallosh:2017wnt}
  R.~Kallosh, A.~Linde, D.~Roest and Y.~Yamada,
  JHEP {\bf 1707} (2017) 057
  doi:10.1007/JHEP07(2017)057
  [arXiv:1705.09247 [hep-th]].
  
  
\bibitem{Kallosh:2017ced} 
  R.~Kallosh, A.~Linde, T.~Wrase and Y.~Yamada,
  JHEP {\bf 1704}, 144 (2017)
  doi:10.1007/JHEP04(2017)144
  [arXiv:1704.04829 [hep-th]].
  
  \bibitem{Kallosh:2017wnt}
  R.~Kallosh, A.~Linde, D.~Roest and Y.~Yamada,
  ``$ \overline{D3} $ induced geometric inflation,''
  JHEP {\bf 1707} (2017) 057
  doi:10.1007/JHEP07(2017)057
  [arXiv:1705.09247 [hep-th]].
  
    \bibitem{Kallosh:2017ced}
  R.~Kallosh, A.~Linde, T.~Wrase and Y.~Yamada,
  ``Maximal Supersymmetry and B-Mode Targets,''
  JHEP {\bf 1704} (2017) 144
  doi:10.1007/JHEP04(2017)144
  [arXiv:1704.04829 [hep-th]].
  
  \bibitem{Akrami:2017cir}
  Y.~Akrami, R.~Kallosh, A.~Linde and V.~Vardanyan,
  ``Dark energy, $\alpha$-attractors, and large-scale structure surveys,''
  JCAP {\bf 1806} (2018) no.06,  041
  doi:10.1088/1475-7516/2018/06/041
  [arXiv:1712.09693 [hep-th]].
  
  
  \bibitem{Mosk:2014cba}
B.~Mosk and J.~P.~van der Schaar,
  ``Chaotic inflation limits for non-minimal models with a Starobinsky attractor,''
  JCAP {\bf 1412} (2014) no.12,  022
  [arXiv:1407.4686 [hep-th]].


\bibitem{Dias:2018pgj}
  M.~Dias, J.~Frazer, A.~Retolaza, M.~Scalisi and A.~Westphal, ``Pole N-flation,''
  arXiv:1805.02659 [hep-th].
  
\bibitem{Linde:2017pwt} 
  A.~Linde,
  arXiv:1710.04278 [hep-th].


\bibitem{Dias:2015rca}
M.~Dias, J.~Frazer and D.~Seery,
  ``Computing observables in curved multifield models of inflationÑA guide (with code) to the transport method,''
  JCAP {\bf 1512} (2015) no.12,  030
   [arXiv:1502.03125 [astro-ph.CO]].
   
   
\bibitem{Lahiri:2005xj}
  J.~Lahiri and G.~Bhattacharya, ``Perturbative analysis of multiple-field cosmological inflation,''
  Annals Phys.\  {\bf 321} (2006) 999
[astro-ph/0507630].


\bibitem{Peterson:2010np}
  C.~M.~Peterson and M.~Tegmark,
  ``Testing Two-Field Inflation,''
  Phys.\ Rev.\ D {\bf 83} (2011) 023522
  [arXiv:1005.4056 [astro-ph.CO]].


\bibitem{vandeBruck:2014ata} 
  C.~van de Bruck and M.~Robinson, 
  ``Power Spectra beyond the Slow Roll Approximation in Theories with Non-Canonical Kinetic Terms,''
  JCAP {\bf 1408}, 024 (2014)
[arXiv:1404.7806 [astro-ph.CO]].

\bibitem{Chen:2009zp}
 X.~Chen and Y.~Wang,
  ``Quasi-Single Field Inflation and Non-Gaussianities,''
  JCAP {\bf 1004} (2010) 027
  [arXiv:0911.3380 [hep-th]].

\bibitem{Bartolo:2001cw}
N.~Bartolo, S.~Matarrese and A.~Riotto,
  ``Nongaussianity from inflation,''
  Phys.\ Rev.\ D {\bf 65} (2002) 103505
  [hep-ph/0112261].

\bibitem{Bartolo:2001vw}
N.~Bartolo, S.~Matarrese and A.~Riotto,
  ``Oscillations during inflation and cosmological density perturbations,''
  Phys.\ Rev.\ D {\bf 64} (2001) 083514
  [astro-ph/0106022].

\bibitem{Weinberg:2006ac}
  S.~Weinberg, ``Quantum contributions to cosmological correlations. II. Can these corrections become large?,''
  Phys.\ Rev.\ D {\bf 74} (2006) 023508
    [hep-th/0605244].


\bibitem{Freese:1990rb}
K.~Freese, J.~A. Frieman, and A.~V. Olinto, ``{Natural inflation with pseudo -
  Nambu-Goldstone bosons}'', {\em Phys. Rev. Lett.}, vol.~65, pp.~3233--3236,
  1990.

\bibitem{Adams:1992bn}
F.~C.~Adams, J.~R.~Bond, K.~Freese, J.~A.~Frieman and A.~V.~Olinto,
  ``Natural inflation: Particle physics models, power law spectra for large scale structure, and constraints from COBE,''
  Phys.\ Rev.\ D {\bf 47} (1993) 426
  [hep-ph/9207245].

\bibitem{Linde:2018hmx}
  A.~Linde, D.~G.~Wang, Y.~Welling, Y.~Yamada and A.~Achœcarro,
  JCAP {\bf 1807} (2018) no.07,  035
  doi:10.1088/1475-7516/2018/07/035
  [arXiv:1803.09911 [hep-th]].

\bibitem{Yamada:2018nsk}
  Y.~Yamada,
  JHEP {\bf 1804} (2018) 006
  doi:10.1007/JHEP04(2018)006
  [arXiv:1802.04848 [hep-th]].


\bibitem{Allen:1987vq}
  T.~J.~Allen, B.~Grinstein and M.~B.~Wise,``Nongaussian Density Perturbations in Inflationary Cosmologies,''
  Phys.\ Lett.\ B {\bf 197} (1987) 66.
  doi:10.1016/0370-2693(87)90343-1
  
  

\bibitem{Bernardeau:2003nx}
  F.~Bernardeau, T.~Brunier and J.~P.~Uzan, ``High order correlation functions for self interacting scalar field in de Sitter space,''
  Phys.\ Rev.\ D {\bf 69} (2004) 063520
  [astro-ph/0311422].

\bibitem{Seery:2007wf}
  D.~Seery, ``One-loop corrections to the curvature perturbation from inflation,''
  JCAP {\bf 0802} (2008) 006
  doi:10.1088/1475-7516/2008/02/006
  [arXiv:0707.3378 [astro-ph]].



\bibitem{Array:2015xqh}
P.~A.~R.~Ade {\it et al.} [BICEP2 and Keck Array Collaborations],
  ``Improved Constraints on Cosmology and Foregrounds from BICEP2 and Keck Array Cosmic Microwave Background Data with Inclusion of 95 GHz Band,''
  Phys.\ Rev.\ Lett.\  {\bf 116} (2016) 031302
  [arXiv:1510.09217 [astro-ph.CO]].

\bibitem{Whitt:1984pd}
B.~Whitt,
  ``Fourth Order Gravity as General Relativity Plus Matter,''
  Phys.\ Lett.\  {\bf 145B} (1984) 176.

\bibitem{Starobinsky:1983zz}
 A.~A.~Starobinsky,
  ``The Perturbation Spectrum Evolving from a Nonsingular Initially De-Sitter Cosmology and the Microwave Background Anisotropy,''
  Sov.\ Astron.\ Lett.\  {\bf 9} (1983) 302.

\bibitem{Kofman:1985aw}
 L.~A.~Kofman, A.~D.~Linde and A.~A.~Starobinsky,
  ``Inflationary Universe Generated by the Combined Action of a Scalar Field and Gravitational Vacuum Polarization,''
  Phys.\ Lett.\  {\bf 157B} (1985) 361.

\bibitem{Salopek:1988qh}
D.~S.~Salopek, J.~R.~Bond and J.~M.~Bardeen,
  ``Designing Density Fluctuation Spectra in Inflation,''
  Phys.\ Rev.\ D {\bf 40} (1989) 1753.
  
 \bibitem{Bezrukov:2007ep}
  F.~L.~Bezrukov and M.~Shaposhnikov,
  ``The Standard Model Higgs boson as the inflaton,''
  Phys.\ Lett.\ B {\bf 659} (2008) 703
  [arXiv:0710.3755 [hep-th]].


\bibitem{Nakamura:1996da}
T.~T.~Nakamura and E.~D.~Stewart,
  ``The Spectrum of cosmological perturbations produced by a multicomponent inflaton to second order in the slow roll approximation,''
  Phys.\ Lett.\ B {\bf 381} (1996) 413
  [astro-ph/9604103].	

\end{thebibliography}
\end{document}